\documentclass[12pt]{article}
\usepackage{times}
\usepackage{epsfig}
\usepackage{amssymb,amsmath,amsthm}
\usepackage{amsfonts}%,amscd}

\newtheorem{defn}{Definition}[section]

\newtheorem{lemma}[defn]{Lemma}

\newtheorem{thm}[defn]{Theorem}

\newcommand{\be}{\begin{equation}}
\newcommand{\ee}{\end{equation}}
\newcommand{\bea}{\begin{eqnarray}}
\newcommand{\eea}{\end{eqnarray}}
\newcommand{\beas}{\begin{eqnarray*}}
\newcommand{\eeas}{\end{eqnarray*}}

\newcommand{\goto}{\rightarrow}

\newcommand{\ink}{\rule{.5\baselineskip}{.55\baselineskip}}
\newcommand{\ts}{\textstyle}

\newcommand{\noi}{\noindent}

\newcommand{\ve}{\varepsilon}

\newcommand{\skp}{\vspace{\baselineskip}}

\newcommand{\R}{{\mathbb R}}

\newcommand{\N}{\mathbb N}

\newcommand{\ebetak}{{\cal E}_{\beta,K}}

\newcommand{\tc}{\tilde{c}}
\newcommand{\tz}{\tilde{z}}

\newcommand{\bn}{\beta_n}
\newcommand{\kn}{K_n}
\newcommand{\gbk}{G_{\beta,K}}
\newcommand{\gn}{G_{\bn,\kn}}
\newcommand{\bc}{\beta_c}
\newcommand{\kc}{K_c}
\newcommand{\kcbc}{K_c(\beta_c)}
\newcommand{\pnbnkn}{P_{n,\bn,\kn}}
\newcommand{\nv}{n^{-v}}
\newcommand{\nw}{n^{-w}}

\newcommand{\xng}{x/n^\gamma}
\newcommand{\rng}{R n^\gamma}
\newcommand{\xixng}{\xi(\xng)}

\newcounter{bean}
\newcommand{\benuma}{\setlength{\labelwidth}{.25in}
\begin{list}%
{(\alph{bean})}{\usecounter{bean}}}
\newcommand{\eenuma}{\end{list}}

\newcommand{\beginsec}{\setcounter{equation}{0}}

\begin{document}

% \pagewiselinenumbers
% use with "\usepackage{lineno}"

\title{
Multiple Critical Behavior \\
of Probabilistic Limit Theorems \\
in the Neighborhood of a Tricritical Point}
\author{Marius Costeniuc,\normalsize{$\,^1$} \vspace{-.1in}\\  
\small{marius@mis.mpg.de} \vspace{-.125in} \\ \\
Richard S.\ Ellis,\normalsize{$\,^2$} \vspace{-.1in} \\
\small{rsellis@math.umass.edu} \vspace{-.125in} \\ \\
Peter Tak-Hun Otto\normalsize{$\,^3$} \vspace{-.1in}\\
\small{potto@willamette.edu} \vspace{-.125in}\\ \\
\normalsize{$^1$ Max Planck Institute for Mathematics in the Sciences} \vspace{-.05in}  \\
\normalsize{Inselstrasse 22--26} \vspace{-.05in}  \\
\normalsize{D-04103 Leipzig, Germany} \vspace{-.1in} \\ \\
\normalsize{$^2$ Department of Mathematics and Statistics} \vspace{-.05in} \\ 
\normalsize{University of Massachusetts} \vspace{-.05in} \\ 
\normalsize{Amherst, MA 01003} \vspace{-.1in}\\ \\
\normalsize{$^3$ Department of Mathematics} \vspace{-.05in}\\ 
\normalsize{Willamette University} \vspace{-.05in}\\ 
\normalsize{Salem, OR 97301}}
\maketitle

\begin{abstract}
We derive probabilistic limit theorems that reveal
the intricate structure of the phase transitions in a mean-field version of the
Blume-Emery-Griffiths model \cite{BluEmeGri}. 
These probabilistic limit theorems consist of scaling limits for the total 
spin and moderate deviation principles (MDPs) for the total spin. 
The model under study is defined by a probability distribution that depends
on the parameters $n$, $\beta$, and $K$, which represent, respectively, the number of
spins, the inverse temperature, and the interaction strength.  
The intricate structure of the phase transitions 
is revealed by the existence of 18 scaling limits and 18 MDPs
for the total spin.
These limit results are obtained as $(\beta,K)$ converges along appropriate 
sequences $(\bn,\kn)$ to points belonging to various subsets of the phase diagram,
which include a curve of second-order points and a tricritical point. 
The forms of the limiting densities in the 
scaling limits and of the rate functions
in the MDPs reflect the influence of one or more sets that lie in neighborhoods 
of the critical points and the tricritical point.  
Of all the scaling limits, the structure of those near the tricritical point 
is by far the most complex, exhibiting new types of critical behavior
when observed in a limit-theorem phase diagram 
in the space of the two parameters that parametrize the scaling limits.  
\end{abstract}

\noi
{\it American Mathematical Society 2000 Subject Classifications.}  Primary 60F10, 60F05, Secondary 82B20
\skp

\noi

\pagestyle{myheadings}
\markboth{M.~Costeniuc, R.~S.~Ellis, and P.~T.~Otto}{Costeniuc, Ellis, and Otto: 
Critical Behavior of Probabilistic Limit Theorems}
% Need next line for line numbers.
% \pagewiselinenumbers

\section{Introduction}
\beginsec
\label{section:intro}

The purpose of this paper is to analyze a new set of phenomena
associated with the critical behavior 
of probabilistic limit theorems for a mean-field version of an important
lattice-spin model due to Blume, Emery, and Griffiths \cite{BluEmeGri}. 
These probabilistic limit theorems consist of scaling limits for the total 
spin and moderate deviation principles (MDPs) for the total spin. 

We will refer to the mean-field model studied in this paper as the BEG
model; it is equivalent to the Blume-Emery-Griffiths model on 
the complete graph on $n$ vertices.  In contrast to the mean-field version of the Ising
model known as the Curie-Weiss model,
whose only phase transition is a continuous, second-order phase transition 
at the critical inverse temperature \cite[\S IV.4]{Ellis}, the BEG model 
exhibits both a curve of continuous, second-order 
points; a curve of discontinuous, first-order points; and 
a tricritical point, which separates the two curves \cite{EllOttTou,NagBon}.  It is
one of the few models, and certainly one of the simplest, that 
exhibit this intricate phase-transition structure. 
 
Applications of the Blume-Emery-Griffiths model to a diverse 
range of physical systems are discussed in \cite[\S 1]{EllOttTou} 
and in \cite[\S 3.3]{NagBon}, where the model is called the Blume-Emery-Griffiths-Rys model.  
As the latter reference points out, the model studied in the present paper 
is actually a mean-field version of a precursor of the Blume-Emery-Griffiths-Rys 
model due to Blume \cite{Blu} and Capel \cite{Cap1,Cap2,Cap3}. 
With apologies to these authors, we follow the nomenclature of our earlier paper
\cite{EllOttTou} by referring to this mean-field version 
as the BEG model.

The BEG model is defined by a probability distribution $P_{n,\beta,K}$, 
where $n$ equals the number of spins, $\beta$ is the inverse temperature, 
and $K$ is the interaction strength.  We investigate the complex structure
of the phase transitions in the model 
by deriving 36 different limit results for the total 
spin $S_n$ as $(\beta,K)$ converges along appropriate 
sequences $(\bn,\kn)$ to points belonging to three separate classes: (1) the tricritical point, 
(2) the curve
of second-order points, and (3) the single-phase region lying under that curve. 
In case 1, we obtain 13 scaling limits and 13 MDPs;
in case 2, 4 scaling limits and 4 MDPs; and in case 3, 1 scaling limit
and 1 MDP.  As we will see, the numbers 13, 4, and 1 represent natural and exhaustive enumerations
of three classes of polynomials that arise in the related settings of the 
scaling limits and the MDPs.

The existence of $18 = 13 + 4 + 1$ scaling limits and 18 MDPs reflects the 
intricate structure of the phase transitions in the BEG model.  It is hoped that our insights can also be applied to other statistical mechanical models that exhibit other types of phase transitions and critical phenomena and 
thus, presumably, other possibilities for scaling limits of macroscopic random variables like the total spin
in the BEG model \cite{EllMac}.

Before saying more about the limit theorems in the BEG model and their critical behavior, 
we summarize a number of 
facts concerning the phase-transition structure of the model \cite{EllOttTou}.
For $\beta > 0$ and $K >0$ we denote by $\ebetak$ the set of equilibrium macrostates
of the model corresponding to the macroscopic variable of the spin per site.
In \cite{EllOttTou} it is proved that there exists a critical
inverse temperature $\beta_c = \log 4$ and that for $\beta > 0$ there exists 
a critical value $K_c(\beta) > 0$ having the following properties.
\begin{enumerate}
\item For $\beta > 0$ and $0 < K < K_c(\beta)$, $\ebetak$ consists of the unique
pure phase 0.
\item For $\beta > 0$ and $K > K_c(\beta)$, $\ebetak$ consists of two distinct, nonzero
phases.
\item For $0 < \beta \leq \beta_c$, as $K$ increases through $K_c(\beta)$, $\ebetak$
undergoes a continuous bifurcation, which corresponds to a second-order
phase transition.
\item For $\beta > \beta_c$, as $K$ increases through $K_c(\beta)$, $\ebetak$
undergoes a discontinuous bifurcation, which corresponds to a first-order
phase transition.
\item The point $(\beta_c,K_c(\beta_c)) = (\log 4, 3/[2 \log 4])$ in the positive quadrant
of the $\beta$-$K$ plane separates the second-order phase transition noted in item 2
from the first-order phase transition noted in item 4.  The point 
$(\beta_c,K_c(\beta_c))$ is called the tricritical point.
\end{enumerate}

The limit theorems to be considered in the present paper focus
on the values of $\beta$ and $K$ in items 1, 3, and 5.  For each such $(\beta,K)$,
$\ebetak$ consists of the unique pure phase 0. 
Figure 1 shows the corresponding portion of the phase diagram, which exhibits
three sets $A$, $B$, and $C$.  $C$ is the singleton set containing the
tricritical point $(\beta_c,K_c(\beta_c))$,
$B$ is the curve of second-order points defined by
\be
\label{eqn:definesetb}
B = \{(\beta,K) \in \R^2 : 0 < \beta < \beta_c, K = K_c(\beta)\},
\ee
and $A$ is the single-phase region lying under $B \cup C$ and defined by
\be
\label{eqn:defineseta}
A = \{(\beta,K) \in \R^2 : 0 < \beta \leq \beta_c, 0 < K < K_c(\beta)\}.
\ee

\begin{figure}[h]
\begin{center}
\epsfig{file=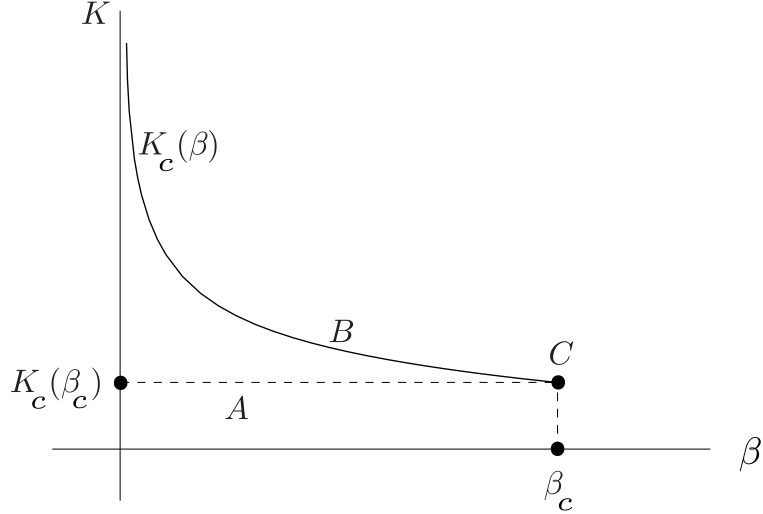,width=10cm}
\caption{\small The sets $A$, $B$, and $C$}
\end{center}
\end{figure}

In the remainder of this introduction we focus on the scaling limits for the total
spin $S_n$ when $(\beta,K)$ converges to the tricritical point $(\beta_c,K_c(\beta_c))$
along appropriate sequences $(\bn,\kn)$.  
These scaling limits describe the limiting distribution
of $S_n/n^{1-\gamma}$ with respect to $\pnbnkn$ for appropriate choices
of $\gamma \in (0,{1}/{2})$.  The simplest sequences
for which the full range of scaling limits appear are 
defined in terms of parameters $\alpha > 0$, $\theta > 0$, $b \not = 0$,
and $k \not = 0$ by
\be
\label{eqn:betanknfirst}
\beta_n = \log(e^{\beta_c} - {b}/{n^\alpha}) \ \mbox{ and }
\ K_n = K(\beta_n) - {k}/{n^\theta},
\ee  
where $K(\beta) = (e^\beta + 2)/(4\beta)$ for $\beta > 0$.
$K(\beta)$ coincides with $K_c(\beta)$ for $0 < \beta \leq \bc$
and satisfies $K(\beta) > K_c(\beta)$ for $\beta > \bc$ \cite[Thms.\
3.6, 3.8]{EllOttTou}. A detailed overview of all
the limit theorems in the paper, including those 
discussed here, is given in the next section.

In each of the scaling limits the form of the limiting density 
reflects the influence of one or more of the sets $A$, $B$, and $C$ that
lie in a neighborhood of the tricritical point.  The influence of those sets,
which depends only on $\alpha$ and $\theta$ and not on $b$ or $k$ in (\ref{eqn:betanknfirst}), 
is shown in Figure 2.  In that figure the positive quadrant
of the $\alpha$-$\theta$ plane is partitioned into the following sets.
\begin{enumerate} 
\item Three open sets labeled $A$, $B$, and $C$.
\item Three line segments labeled $A+B$, $A+C$, and $B+C$ that separate the three
open sets in item 1.
\item The point equal to $({1}/{3},{2}/{3})$
and labeled $A+B+C$ at which the three line segments in item 2 meet.
\end{enumerate}

\begin{figure}[h]
\begin{center}
\epsfig{file=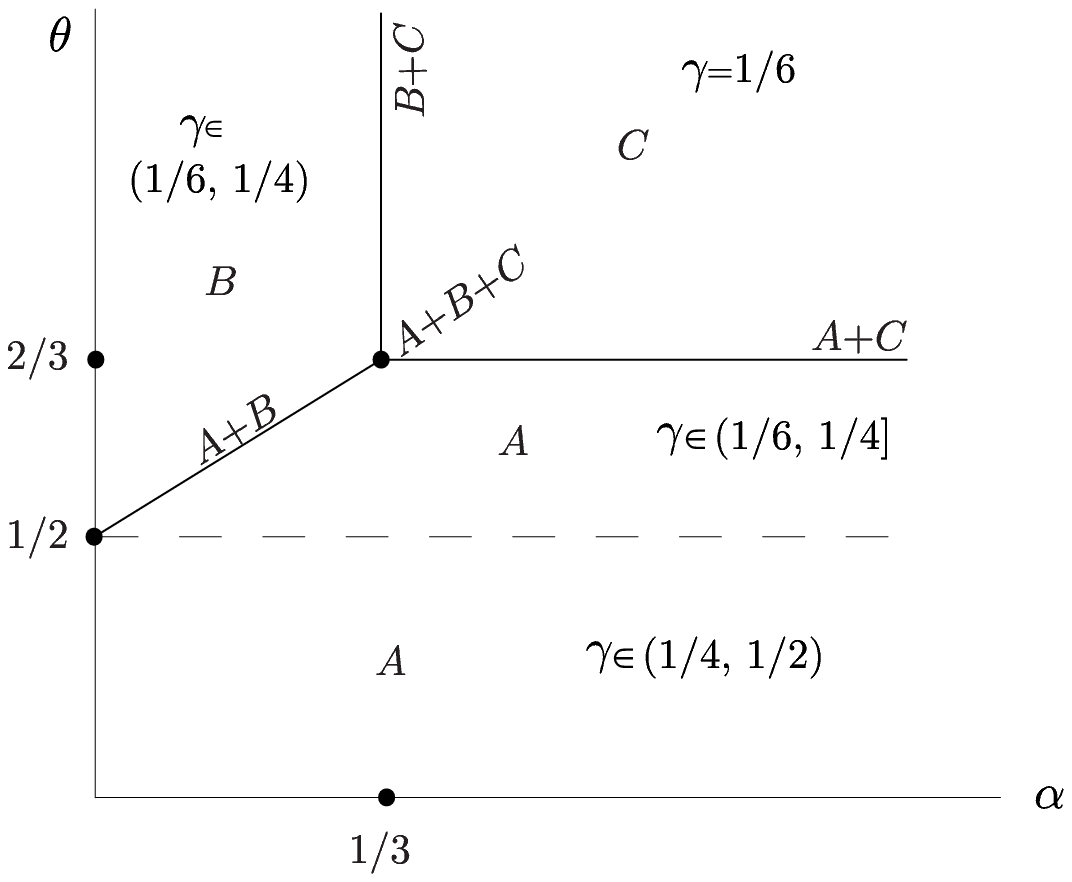,width=8cm}
\caption{\small Influence of $C$, $B$, and $A$ when $(\bn,\kn) \goto (\bc,\kcbc)$}
\end{center}
\end{figure} 

Figure 2 is a limit-theorem phase diagram that summarizes the critical behavior
of the scaling limits in a neighborhood of the tricritical point.  This critical
behavior consists of the following phenomena, which can be verified by examining the
statement of the scaling limits in Theorem \ref{thm:betakinc}.  

\begin{enumerate}
\item  When $(\alpha,\theta)$ lies in one of the open sets labeled $A$, $B$, or $C$, then the 
limiting density in the corresponding scaling limit shows the influence only 
of that single set.  Hence these three open sets correspond to the pure
phases of the scaling limits.
\item When $(\alpha,\theta)$ lies in one
of the line segments labeled $A+B$, $A+C$, or $B+C$, then the limiting density shows
the influence of both sets, $A$ and $B$, $A$ and $C$, or $B$ and $C$, respectively.  Hence these
three line segments correspond to the coexistence of the pure-phase scaling limits noted in item 1.  
\item  When $(\alpha,\theta)$ equals the point labeled $A+B+C$, then the limiting
density shows the influence of all three sets $A$, $B$, and $C$.   
This point is
the analogue of the tricritical point in the standard phase diagram, a portion
of which is shown in Figure 1.
Indeed, any neighborhood of the tricritical point in the $\beta$-$K$ plane
contains values of $\beta$ and $K$ corresponding to all the different phase-transition
behaviors of the model.  Similarly, any neighborhood of the analogue
of the tricritical point in the limit-theorem phase diagram contains values of $\alpha$ and $\theta$ corresponding
to all the different forms of the scaling limits, which number 13. 
\item As $(\alpha,\theta)$ crosses any of the line segments labeled $A+B$, $A+C$, or $B+C$, 
the values of $\gamma$ in the scaling limits change
continuously, which corresponds to a second-order phase transition; by contrast,
the forms of the limiting densities change discontinuously, which 
corresponds to a first-order phase transition.
\end{enumerate}

As noted in items 1, 2, and 3, the influence of the sets upon the forms of the limiting densities 
reveals a fascinating geometric feature of the BEG model.  This feature is completely unexpected
because the model has no geometric
structure. In fact, each spin interacts equally with all the other spins via a mean-field
Hamiltonian, and so the model is independent of dimension. 
The discussion of the scaling limits given here, 
including the notion of the influence of a set on the form of the limiting
density, will be greatly amplified in the next section. 

The scaling limits of $S_n/n^{1-\gamma}$ corresponding to the choices of $\alpha$
and $\theta$ in Figure 2 are derived in Theorem \ref{thm:betakinc}, where we determine
the values of $\alpha$, $\theta$, and $\gamma$ leading to the various forms
of the limit.  In Figure 2 the value or range of values of $\gamma$ are also 
shown for $(\alpha,\theta)$ lying in the sets labeled $A$, $B$, and $C$.  The 
set labeled $A$ is divided into two subsets by the line $\theta = 1/2$;
the ranges of values of $\gamma$ are different in the two subsets.

The three seeds from which the present paper grew are \cite{EllOttTou}, 
\cite{EllNew}, and \cite{EicLow}. In the first paper the phase-transition 
structure of the BEG model is analyzed.  In the second paper
scaling limits are proved for a class of models that includes the 
Curie-Weiss model as a special case.  In the third paper 4 different MDPs are 
obtained for the Curie-Weiss model when the inverse temperature converges
to the critical inverse temperature in the model along appropriate sequences
$\beta_n$. The results derived in the present paper greatly extend both the scaling 
limits in \cite{EllNew} and the MDPs in \cite{EicLow}.  This is the case because
the BEG model has a much more intricate structure of phase transitions
than the Curie-Weiss model
and so exhibits a much richer class both of scaling limits and of MDPs.  As we will outline
near the end of the next section, both the scaling limits and the MDPs are proved by a unified method. 

This unified method is based, in part, on properties of a function $G_{\beta,K}$ defined in 
(\ref{eqn:introgbetak}). This function plays a central role in every aspect of the analysis of the BEG 
model considered in the present paper as well as in its prequel \cite{EllOttTou}.
In summary these are the following. 

\begin{itemize}
\item The set $\ebetak$ of equilibrium macrostates for the BEG model
is defined as the set of zeroes of the rate function in the LDP for the $P_{n,\beta,K}$-distributions
of $S_n/n$ given in Theorem \ref{thm:ldppnbetak}.  In turn, this set
coincides with the set of global minimum points of $G_{\beta,K}$ [see (\ref{eqn:ebetak})].
This characterization of $\ebetak$ allowed us to carry out the detailed analysis of
the phase-transition structure of the model in \cite{EllOttTou}.
\item The canonical free energy $\varphi(\beta,K)$ equals the global minimum value
of $G_{\beta,K}$ [see item 2 after (\ref{eqn:introgbetak})].  
\item The distribution of $S_n/n^{1-\gamma}$ can be expressed directly in terms of
$G_{\beta,K}$ [Lem.\ \ref{lem:G}].
\item $\gbk$ is the rate function in a second LDP involving $S_n/n$ given in 
part (b) of Lemma \ref{lem:awayfrom0}.  The estimates derived from this LDP and given in 
parts (c) and (d) of the lemma are the key estimates needed to control error terms
in the proofs of the scaling limits and the MDPs for $S_n/n^{1-\gamma}$.
Lemma \ref{lem:awayfrom0} is the main technical innovation in the paper.
\item  When a certain quantity $w$ defined in terms of
$\alpha$, $\theta$, and $\gamma$ equals 0, the 13 different forms of the Taylor 
expansion of $nG_{\beta_n,K_n}(x/n^\gamma)$ for appropriate
sequences $(\bn,\kn)$ and $\gamma \in (0,1/2)$
yield the 13 different forms of the scaling limits of $S_n/n^{1-\gamma}$
[Thm.\ \ref{thm:betakinc}].
\item When $w < 0$, the 13 different forms of the Taylor expansion of 
$n^{1+w}G_{\beta_n,K_n}(x/n^\gamma)$ for appropriate
sequences $(\bn,\kn)$ and $\gamma \in (0,1/2)$
yield the 13 different forms of the MDPs of $S_n/n^{1-\gamma}$ [Thm.\ \ref{thm:mdpc}].
\end{itemize}

This discussion shows that all the magic is in the function $G_{\beta,K}$. 
The fact that the wide variety of phenomena derived in the present paper and in
\cite{EllOttTou} can be obtained via properties of a single function
is an appealing feature of the BEG model.
Besides the Curie-Weiss model and generalizations studied in
\cite{EicLow,EllNew,EllNewRos,Pap} and numerous other papers, this feature
is shared with a number of other mean-field models, including a mean-field
version of the nearest neighbor Potts model known as the
Curie-Weiss-Potts model \cite{EllWan}, the mean-field XY Heisenberg model \cite{AntRuf},
and the Hopfield model of spin glasses and neural networks \cite{PasFig}. 
These mean-field models have in common the fact that the interaction
terms in their Hamiltonians can be written as a quadratic function.
Scaling limits and MDPs for these models have either
been proved, or in principle could be proved, by
techniques similar to those used in the present paper.
Some of these techniques are generalized in \cite{ChaSet}, in which
the quadratic term in the Hamiltonian is replaced by the moment
generating function of suitable random variables.  
Other generalizations are given in \cite{Bol1,Bol2,EllRos1,EllRos2}.  
The analysis of the 
equilibrium macrostates and the associated phase transitions in the BEG model,
which underlies the present paper, 
is carried out in \cite{EllOttTou} using
large deviation techniques.  While this is an elegant method that
provides exact, analytical results, it has the restriction that it works most
efficiently in models with long-range interactions, as explained in
\cite{BarBouDauRuf}.

The Hopfield model of spin glasses and neural networks
has received a great deal of attention, and
limit theorems for this model have been actively studied.  The
Hamiltonian in the Hopfield model can be
written as a quadratic function of the overlap parameter, a feature that
it shares with the Curie-Weiss model and the BEG model, in which 
the Hamiltonian can be written as a quadratic function of the spin per site.
For the Hopfield model both central limit theorems and non-classical scaling limits for
the overlap parameter are studied in 
\cite{BovGay,Gen1,Gen2,GenLow}, and MDPs are studied 
in \cite{EicLow2}.  These limit theorems include the cases when the
inverse temperature is constant and when the inverse
temperature parameter converges to the critical inverse temperature
at an appropriate rate \cite{EicLow2,GenLow}.  

We next preview the contents of the present paper.
In section \ref{section:overview} a detailed
overview is given of the scaling limits and the MDPs that will be derived.
In section \ref{section:phasetr} 
 we summarize the results in \cite{EllOttTou} on the structure of
the set of equilibrium macrostates of the BEG model and the
associated phase transitions.  In section \ref{section:G} we introduce the function
$G_{\beta,K}$, properties of which are integral to the proofs of the scaling limits
and MDPs.  These properties include a formula
for the distribution of the total spin in terms of $G_{\beta,K}$ [Lem.\ \ref{lem:G}],
several forms of the Taylor expansions
of $G_{\beta,K}$ that will be used to derive the limit theorems [Thm.\ \ref{thm:taylor}],
and two estimates in Lemma \ref{lem:awayfrom0} for controlling error terms 
in the proofs of the scaling limits and the MDPs.

In sections \ref{section:scalinga}--\ref{section:mdp} we apply
the results in the previous sections to derive the scaling limits and the MDPs.
Sections \ref{section:scalinga} and \ref{section:scalingb}
are devoted to scaling limits for $S_n/n^{1-\gamma}$ when appropriate
sequences $(\bn,\kn)$ converge to points $(\beta,K) \in A$ and to 
points $(\beta,K_c(\beta)) \in B$, where $A$ and $B$ are the sets defined
in (\ref{eqn:defineseta}) and (\ref{eqn:definesetb}).  When $(\bn,\kn) \goto (\beta,K) \in A$
we obtain only 1 scaling limit, which is independent of the sequence $(\bn,\kn)$ [Thm.\ \ref{thm:betakina}].
The situation for $(\beta,K_c(\beta)) \in B$ is much more interesting; 
for appropriate choices of $(\bn,\kn) \goto (\beta,K) \in B$,
4 different forms of the scaling limits arise [Thm.\ \ref{thm:betakinc}].  
The scaling limits proved in these two sections are warm-ups for the even more complicated
scaling limits proved in section \ref{section:scalingc}.  
In that section, for appropriate choices of $(\bn,\kn)$ converging
to the tricritical point $(\bc,\kcbc)$
 we obtain 13 different forms of the scaling limits [Thm.\ \ref{thm:betakinc}].
Finally, in section \ref{section:mdp} we obtain 1 MDP for 
$S_n/n^{1-\gamma}$ when $(\bn,\kn) \goto (\beta,K) \in A$ [Thm.\ \ref{thm:mdpa}],
4 MDPs when $(\bn,\kn) \goto (\beta,K_c(\beta)) \in B$ [Thm.\ \ref{thm:mdpb}], and 
13 MDPs when $(\bn,\kn) \goto (\bc,K_c(\bc))$ [Thm.\ \ref{thm:mdpc}]. 
The MDPs are proved by showing the equivalent Laplace principles, which is carried
out by a method closely related to that used to prove the scaling limits in the earlier sections.  Being able
to prove both classes of limit theorems via a unified method is one of the  
attractive features of this paper. 

\skp
\noi
{\bf Acknowledgements.}  We would like to thank an anonymous referee of 
\cite{EllOttTou} who suggested studying scaling limits for $S_n/n^{1-\gamma}$
in the BEG model using sequences $(\bn,\kn)$ converging to various
points $(\beta,K)$.  We would also like to thank Jonathan Machta
for useful discussions on the material of the present paper.
The research of Richard S.\ Ellis is supported 
in part by a grant from the National Science Foundation (NSF-DMS-0604071). 
The research of Peter Otto was supported in part
by a grant from the Faculty Research Fund at Union College.

\section{Overview of the Limit Theorems}
\beginsec
\label{section:overview}

\iffalse
Particularly in the case of the tricritical point, the limit theorems have a fascinating
structure, which is summarized in Figure 2 in the next section.  We focus now
on the scaling limits for the total spin $S_n$; 
though somewhat analogous, the structure of the MDPs is actually more complicated. 
The sequences $(\bn,\kn)$ that converge to the tricritical point and that yield
the 13 different scaling limits are defined
in terms of two positive parameters, $\alpha$ and $\theta$.  As we explain in 
detail in the next section,
the positive quadrant of the $\alpha$-$\theta$ plane is divided into three open sets
labeled $A$, $B$, and $C$ in Figure 2.  For $\alpha$ and $\theta$ in each of these
three sets, the limit results take three separate forms.  As one crosses the line segment
labeled $A+B$, the form of the scaling limits changes abruptly.
\fi

This paper is devoted to scaling limits and MDPs for the total spin in the BEG
model.  In order to highlight the novelty of these results, we introduce some notation.
The BEG model is a lattice-spin model defined on 
the complete graph on $n$ vertices $1,2,\ldots,n$.  The
spin at site $j \in \{1,2,...,n\}$ is denoted by $\omega_j$, a
quantity taking values in $\Lambda = \{-1,0,1\}$.  The joint distribution 
of the spins $\omega_j$ is defined by a probability measure  $P_{n,\beta,K}$
on the configuration space $\Lambda^n$ [see (\ref{eqn:pnbetak})].  The sequence $P_{n,\beta,K}$ 
for $n \in \N$ defines the canonical ensemble for the BEG model.  

Through the particular form of the interactions among the spins, 
the measures $P_{n,\beta,K}$ incorporate an alignment effect that underlies
the phase-transition structure of the model.
As $\beta \goto 0$, $P_{n,\beta,K}$ converges
weakly to the product measure on $\Lambda^n$ with marginals equal to the uniform
measure on $\Lambda$.  Similarly, as $K \goto 0$, $P_{n,\beta,K}$
converges weakly to another product measure on $\Lambda^n$.
By contrast, as $K \goto \infty$, $P_{n,\beta,K}$ concentrates
on the configurations $\omega^+$ and $\omega^-$ in which the spins are all $1$ or $-1$;
by symmetry, as $K \goto \infty$, $P_{n,\beta,K}$ converges weakly 
to the sum of point masses $\frac{1}{2}(\delta_{\omega^+} +
\delta_{\omega^-}$).  The phase-transition structure
of the model reflects the persistence of this alignment effect in the limit $n \goto \infty$. 

We define $S_n = \sum_{j=1}^n \omega_j$, which represents the total spin.
In this paper we will consider numerous weak limits of the distributions of 
$S_n/n^{1-\gamma}$, where $\gamma \in [0,1)$.  The distributions are with respect
to $P_{n,\beta,K}$ for fixed $\beta > 0$ and $K > 0$ and, more generally, with
respect to $P_{n,\beta_n,K_n}$, where $(\beta_n,K_n)$ are appropriate sequences converging
to specific values of $(\beta,K)$.  The use of $P_{n,\beta_n,K_n}$ to study
weak limits in place of $P_{n,\beta,K}$ is the basic innovation of this paper, which will reveal the 
intricate phase-transition structure of the model.   
If $\nu$ is a probability measure on $\R$, then the notation $P_{n,\beta_n,K_n}\{S_n/n^{1-\gamma} \in dx\}
\Longrightarrow \nu$ means 
that the distributions of $S_n/n^{1-\gamma}$ with respect to $P_{n,\beta_n,K_n}$ 
converge weakly to $\nu$ as $n \goto \infty$. 
If $f$ is a nonnegative integrable function on $\R$, 
then the notation $\pnbnkn\{S_n/n^{1-\gamma} \in dx\} \Longrightarrow f \, dx$ 
means that the distributions of $S_n/n^{1-\gamma}$ 
converge weakly to the 
probability measure on $\R$ having a density proportional to $f$ with respect to Lebesgue measure. 

The first hint of the intricacy of the phase-transition 
structure of the BEG model can be seen 
by examining the law of large numbers and its breakdown, which we
consider with respect to $P_{n,\beta,K}$ for fixed $\beta > 0$ and $K > 0$.  
The intuition is that for sufficiently small $K > 0$ the interactions among the spins are sufficiently
weak so that the analogue of the classical law of large numbers holds.  However, for sufficiently large
$K > 0$ the interactions among the spins are sufficiently strong to cause the classical law of large
numbers to break down.  This intuition is in fact correct.  
In \cite{EllOttTou} it is proved that there exist $K_c(\beta) > 0$,
defined for $\beta > 0$, 
and $z(\beta,K)$, defined for $\beta > 0$ and $K \geq K_c(\beta)$,
in terms of which the following limits hold.  
The form of the limits for $K = K_c(\beta)$ is given
in (\ref{eqn:smallbetakcbeta}) and (\ref{eqn:largebetakcbeta}).
\begin{itemize}
\item For any $\beta > 0$ and $0 < K < K_c(\beta)$
\be
\label{eqn:smallk}
P_{n,\beta,K}\{{S_n}/{n} \in dx\} \Longrightarrow \delta_0.
\ee
\item For any $\beta > 0$ and $K > K_c(\beta)$ we have $z(\beta,K) > 0$ and 
\be
\label{eqn:largek}
P_{n,\beta,K}\{{S_n}/{n} \in dx\} \Longrightarrow \ts\frac{1}{2} \!\left(\delta_{z(\beta,K)}
+ \delta_{-z(\beta,K)}\right).
\ee
\end{itemize}
The proofs of these two limits are indicated at the end of section \ref{section:phasetr},
where they are derived from the LDP given in part (a) of Theorem 
\ref{thm:ldppnbetak}.

As we explain in section \ref{section:phasetr}, for each $\beta > 0$ and
$K > 0$ the sets of mass points of the limiting measures represent
the sets of equilibrium macrostates of the BEG model, which we denote by
$\ebetak$.  Thus, for 
$\beta > 0$ and $0 < K < K_c(\beta)$, $\ebetak = \{0\}$ while for $\beta > 0$ and 
$K > K_c(\beta)$, $\ebetak = \{\pm z(\beta,K)\}$.  
The quantity $z(\beta,K)$ is a positive, increasing, continuous function
for $K > K_c(\beta)$.  The limit of $z(\beta,K)$ as 
$K \goto K_c(\beta)^+$ depends on whether $\beta \leq \beta_c$
or $\beta > \beta_c$, where $\beta_c = \log 4$ represents the critical inverse
temperature of the model.  For $\beta > \beta_c$ we have 
$z(\beta,K_c(\beta)) > 0$, and
\[
\lim_{K \goto K_c(\beta)^+} z(\beta,K) = \left\{
\begin{array}{ll} 0 & \ \mbox{ if } 0 < \beta \leq \beta_c  \\
z(\beta,K_c(\beta))& \ \mbox{ if } \beta > \beta_c   .
\end{array}  \right.
\]
Consistent with this limit behavior is the fact that 
$\mathcal{E}_{\beta,K_c(\beta)}$ equals $\{0\}$ for $0 < \beta \leq \bc$
and equals $\{0, \pm z(\beta,K_c(\beta))\}$ for $\beta > \bc$. 
The limit behavior of $z(\beta,K)$ exhibited in the last display
shows that the sets $\ebetak$ undergo
a continuous bifurcation at $K = K_c(\beta)$ for $0 < \beta \leq \beta_c$
and a discontinuous bifurcation at $K = K_c(\beta)$ for $\beta > \beta_c$.
From the viewpoint of statistical mechanics, the dual bifurcation behavior
of the model corresponds to 
a continuous, second-order phase transition at $(\beta,K_c(\beta))$ for 
$0 < \beta \leq \beta_c$
and a discontinuous, first-order phase transition at $(\beta,K_c(\beta))$
for $\beta > \bc$.
The point $(\beta_c,K_c(\beta_c)) = (\log 4, 3/[2 \log 4])$ 
separates the second-order phase
transition from the first-order phase transition and is called the tricritical point.

The different behavior of the two phase transitions is reflected in the form of 
the limits of $S_n/n$ when $K = K_c(\beta)$.  For $0 < \beta \leq \beta_c$, 
we have the law of large numbers 
\be
\label{eqn:smallbetakcbeta}
P_{n,\beta,K_c(\beta)}\{{S_n}/{n} \in dx\} \Longrightarrow \delta_0,
\ee
while for $\beta > \beta_c$ the limit is expressed in terms of a measure
supported at the three points in ${\cal E}_{\beta,K_c(\beta)}$:
\be
\label{eqn:largebetakcbeta}
P_{n,\beta,K_c(\beta)}\{{S_n}/{n} \in dx\} \Longrightarrow 
\lambda_0 \delta_0 + \lambda_1 \!\left(\delta_{z(\beta,K_c(\beta))}
+ \delta_{-z(\beta,K_c(\beta))}\right).
\ee
In the last limit $\lambda_0$ and $\lambda_1$ are positive numbers satisfying
$\lambda_0 + 2 \lambda_1 =1$ and given explicitly in (\ref{eqn:lambdakappa}).  
As we point out at the end of section \ref{section:phasetr}, 
(\ref{eqn:smallbetakcbeta}) follows immediately from the LDP
given in part (a) of Theorem \ref{thm:ldppnbetak}. However, the proof of (\ref{eqn:largebetakcbeta}) is 
more subtle and is postponed until after Theorem \ref{thm:secondordertype}.

Further evidence of the intricacy of the phase-transition structure of the model
can be seen if one jumps from the context of the law of large numbers and its breakdown to
the context of scaling limits for $S_n$ that are related to the central
limit theorem and its breakdown.   We consider three cases, in all of which
$\ebetak = \{0\}$.  Case 1 is defined by $\beta > 0$
and $0 < K < K_c(\beta)$.  For these values of $\beta$ and $K$
the interactions among the spins are sufficiently weak, and the analogue of
the classical central limit theorem holds.  As we prove in 
Theorem \ref{thm:betakina} when $0 < \beta \leq \beta_c$,
\be
\label{eqn:2ndorder}
P_{n,\beta,K}\{{S_n}/{n^{1/2}} \in dx\} \Longrightarrow \exp(-c_2 x^2) \, dx,
\ee
where $c_2 = c_2(\beta,K)$ is defined
in (\ref{eqn:c2betak}).  The same limit holds when $\beta > \bc$ and $0 < K < K_c(\beta)$.

Case 2 is defined by $0 < \beta < \beta_c$ 
and $K = K_c(\beta)$.  In this case the central limit scaling
$n^{1/2}$ in (\ref{eqn:2ndorder}) must be replaced by $n^{3/4}$, which 
reflects the onset of long-range order represented
by the second-order phase transition at $(\beta,K_c(\beta))$.  We have the 
nonclassical limit
\be
\label{eqn:4thorder}
P_{n,\beta,K_c(\beta)}\{{S_n}/{n^{3/4}} \in dx\} \Longrightarrow \exp(-c_4 x^4) \, dx,
\ee
where $c_4 = c_4(\beta,K) > 0$ is defined in (\ref{eqn:c4betak}).  The limit in the last
display is a special case of one of the limits proved in Theorem \ref{thm:betakinb} [see the note
after the statement of the theorem].

Case 3 focuses on the 
tricritical point $(\beta_c,K_c(\beta_c))$.  Not only is there an onset
of long-range order represented by the second-order phase transition
at this point, but also this point separates
the second-order phase transition for $\beta < \beta_c$
and the first-order phase transition for
$\beta > \beta_c$.  This more intricate phase-transition behavior in 
a neighborhood of the tricritical point is reflected in the replacement
of the scaling $n^{3/4}$ for $0 < \beta < \beta_c$ by $n^{5/6}$.  In this case
\be
\label{eqn:6thorder}
P_{n,\beta_c,K_c(\beta_c)}\{{S_n}/{n^{5/6}} 
\in dx\} \Longrightarrow \exp(-c_6 x^6) \, dx,
\ee
where $c_6 = 9/40$. The limit
in the last display is a special case of one of the limits proved in Theorem \ref{thm:betakinc}
[see the note after the statement of the theorem].

For all other values of $\beta > 0$ and $K > 0$ ---
those satisfying $0 < \beta \leq \beta_c$, $K >  K_c(\beta)$
and $\beta > \beta_c$, $K \geq K_c(\beta)$ --- the limit theorems
have different forms because the set $\ebetak$
of equilibrium macrostates consists of more than one point.
\iffalse
For all $\beta > 0$ and $K > K_c(\beta)$, 
we have $\ebetak = \{\pm z(\beta,K)\}$, while
for all $\beta > \beta_c$ and $K = K_c(\beta)$, we have
$\mathcal{E}_{\beta,K_c(\beta)} = \{0, \pm z(\beta,K_c(\beta))\}$.  
\fi
In both of these cases, 
for any equilibrium macrostate $\tilde{z}$, $(S_n - n \tilde{z})/n^{1/2}$
satisfies a central-limit-type limit when $S_n/n$ is conditioned to lie 
in a sufficiently small neighborhood of $\tilde{z}$.  The explicit
form of the limit is given in part (b) of Theorem 6.6 in \cite{EllOttTou}.

We are now ready to outline the main contribution of this paper, which is 
to exhibit the intricate probabilistic behavior of the 
BEG model in neighborhoods of the tricritical point $(\beta_c,K_c(\beta_c))$,
second-order points $(\beta,K_c(\beta))$ for $0 < \beta < \beta_c$,
and points $(\beta,K)$ for $0 < \beta \leq \bc$ and $0 < K < K_c(\beta)$.
We do this by studying scaling limits and MDPs for 
$S_n/n^{1-\gamma}$ with respect to $P_{n,\beta_n,K_n}$ for appropriate
sequences $(\beta_n,K_n)$ that converge to points belonging
to these three classes and for appropriate
choices of $\gamma \in (0,\frac{1}{2}]$.  In order to facilitate
the discussion, we denote by $C$ the singleton set containing the
tricritical point $(\beta_c,K_c(\beta_c))$,
by $B$ the curve of second-order points defined by 
\[
\label{eqn:defineB}
B = \{(\beta,K) \in \R^2 : 0 < \beta < \beta_c, K = K_c(\beta)\},
\]
and by $A$ the single-phase region lying under $B \cup C$ and defined by
\[
A = \{(\beta,K) \in \R^2 : 0 < \beta \leq \beta_c, 0 < K < K_c(\beta)\}.
\]
The sets $A$, $B$, and $C$ are shown in Figure 1 in the introduction.
In the rest of this section we focus mainly
on the scaling limits and MDPs for $S_n/n^{1-\gamma}$ when $(\beta_n,K_n)$ is an appropriate
sequence that converges to $(\beta_c,K_c(\beta_c))$.  Scaling limits and MDPs when 
$(\beta_n,K_n)$ converges to $(\beta,K_c(\beta)) \in B$ and to $(\beta,K) \in A$ are 
treated, respectively, in Theorems \ref{thm:betakinb} and \ref{thm:mdpb} and in 
Theorems \ref{thm:betakina} and \ref{thm:mdpa}.

Corresponding to each $(\beta,K) \in A \cup B \cup C$ 
there exists a unique equilibrium macrostate at
0. We do not consider scaling limits and MDPs in the neighborhoods
of other points corresponding to which there exist nonunique
equilibrium macrostates.  In all or most cases of
nonunique equilibrium macrostates, we expect that the scaling
limits and MDPs are conditioned limits as in \cite[Thm.\ 6.6(b)]{EllOttTou}
and \cite[Thm.\ 1.1]{EicLow}; however, we have not worked out the
details.

Through the limits (\ref{eqn:2ndorder}), (\ref{eqn:4thorder}), 
and (\ref{eqn:6thorder}), each of the sets $A$, $B$, and $C$ is associated, respectively, 
with the term $x^2$, $x^4$, and $x^6$. 
Specifically, for fixed $(\beta,K)$
\be
\label{eqn:c2c4c6}
P_{n,\beta,K}\{{S_n}/{n^{1-\gamma}} \in dx\} \Longrightarrow \left\{
\begin{array}{ll}  
\exp(-c_2 x^2) \, dx & \ \mbox{ with } \gamma = 1/2 \mbox{ if } (\beta,K) \in A \\
\exp(-c_4 x^4) \, dx & \ \mbox{ with } \gamma = 1/4 \mbox{ if } (\beta,K) \in B \\
\exp(-c_6 x^6) \, dx & \ \mbox{ with } \gamma = 1/6 \mbox{ if } (\beta,K) \in C,
\end{array}
\right.
\ee
where $c_2$ and $c_4$ are positive and depend on $\beta$ and $K$, and $c_6 = 9/40$.
Theorem \ref{thm:betakinc} shows that for appropriate sequences 
$(\beta_n,K_n)$ converging to $(\beta_c,K_c(\beta_c)) $, for appropriate choices of 
$\gamma \in (0,1/2)$, and for appropriate coefficients 
$\tc_2$, $\tc_4$, and $\tc_6$
\be
\label{eqn:13cases}
P_{n,\beta_n,K_n}\{S_n/n^{1-\gamma} \in dx\} 
\Longrightarrow \exp(-\tc_2 x^2 - \tc_4 x^4 - \tc_6 x^6) \, dx.
\ee 
As we show in Table 2.1, $G(x) = \tc_2 x^2 + \tc_4 x^4 + \tc_6 x^6$ 
takes all of the 13 possible forms
of an even polynomial of degree 6, 4, or 2 satisfying $G(0) = 0$ and $G(x) \goto \infty$
as $|x| \goto \infty$.  Each of the 13 cases shows the influence of one or more of the sets
$C$, $B$, and $A$ through the presence of the term $x^6$, $x^4$,
or $x^2$ associated with that set by the limit (\ref{eqn:c2c4c6}).
The coefficient $c_6 = 9/40$ is the same as
in (\ref{eqn:6thorder}), $\bar{c}_4 = 3/16$, and $b$ and $k$ are
any nonzero real numbers subject only to the requirement that $\exp(-G)$ is 
integrable. Because in every case $\gamma \in (0,1/2)$, 
the scaling of $S_n$ by $n^{1-\gamma}$ is non-classical.    

\begin{center}
% \label{table:13cases}
\begin{tabular}{||c|c|l||} \hline \hline
{\em {\bf case}} & {\em {\bf influence}} & \boldmath $\pnbnkn\{S_n/n^{1-\gamma} \in dx\} \Longrightarrow
\exp[-G(x)] \, dx$ \unboldmath \\ 
\hline \hline 
1 & $C$ & $G(x) = c_6 x^6$, $\ c_6 > 0$ \\ \hline
2 & $B$ & $G(x) = b \bar{c}_4 x^4$, $\ b > 0$, $\bar{c}_4 > 0$ \\ \hline
3 & $A$ & $G(x) = k \bc x^2$, $\ k > 0$ \\ \hline
4--5 & $B+C$ & $G(x) = b \bar{c}_4 x^4 + c_6 x^6$, $\ b \not = 0$ \\ \hline
6--7 & $A+C$ & $G(x) = k \bc x^2 + c_6 x^6$, $\ k \not = 0$ \\ \hline
8--9 & $A+B$ & $G(x) = k \bc x^2 + b \bar{c}_4 x^4$, $\ k \not = 0$, $b > 0$ \\ \hline
10--13 & $A+B+C$ & $G(x) = k \bc x^2 + b \bar{c}_4 x^4 + c_6 x^6$, $k \not = 0$, $b \not = 0$ 
\\ \hline
\end{tabular}

\vspace{.05in}
Table 2.1: {\small 13 cases of the scaling limits
in (\ref{eqn:13cases}) for $(\bn,\kn)$ in (\ref{eqn:betankn})
and $\gamma \in (0,1/2)$}
\end{center}

The forms of the scaling limits in (\ref{eqn:13cases}) 
depend crucially on the appropriate choices of the sequences $(\beta_n,K_n)$
converging to $(\beta_c,K_c(\beta_c))$.   The simplest sequences
for which all 13 cases of the limit (\ref{eqn:13cases}) arise are 
defined in terms of parameters $\alpha > 0$, $\theta > 0$, $b \not = 0$,
and $k \not = 0$
\be
\label{eqn:betankn}
\beta_n = \log(e^{\beta_c} - {b}/{n^\alpha}) \ \mbox{ and }
\ K_n = K(\beta_n) - {k}/{n^\theta},
\ee  
where $K(\beta) = (e^\beta + 2)/(4\beta)$ for $\beta > 0$.
$K(\beta)$ coincides with $K_c(\beta)$ for $0 < \beta \leq \bc$
and satisfies $K(\beta) > K_c(\beta)$ for $\beta > \bc$ \cite[Thms.\
3.6, 3.8]{EllOttTou}. Since $\beta_n \goto \beta_c$
and since $K(\cdot)$ is continuous,
we have $K(\beta_n) \goto K_c$; thus the convergence $(\beta_n,K_n) 
\goto (\beta_c,K_c(\beta_c))$ is valid.
In section \ref{section:scalingc} we will explain how this particular sequence $(\beta_n,K_n)$ was chosen. 

Depending on the signs of $b$ and $k$, the sequence $(\bn,\kn)$ in (\ref{eqn:betankn})
converges to $(\bc,\kcbc)$ from regions exhibiting markedly different physical behavior.  
For example, if $b > 0$ and $k > 0$, then $\bn < \bc$ and $K_n < K(\bn)$, 
and so $(\bn,\kn)$ converges to 
$(\beta,K)$ from the region $A$, corresponding to each point of which
there exists a unique equilibrium macrostate [Thm.\ \ref{thm:secondorder}(a)]. On the other
hand, if $k < 0$, then $K_n > K(\bn)$, and so $(\bn,\kn)$ converges to $(\beta,K)$ from a region
of points corresponding to each of which there exist two equilibrium macrostates.
If, in addition, $b > 0$, then this region lies above the curve $B$ of 
second-order points [Thm.\ \ref{thm:secondorder}(b)], while
if $b < 0$, then this region lies above the curve of first-order
points described in Theorem \ref{thm:firstorder}.  Despite the markedly different physical behavior associated with these various
regions, all the scaling limits in this paper are proved by a unified method,
regardless of the direction of approach of $(\bn,\kn)$ to $(\beta,K)$.  The situation with respect
to the MDPs is the same.  These remarks concerning the proofs of the scaling limits
and the MDPs will be amplified in section \ref{section:G} after we introduce the tools
that will be used in the proofs.

The occurrence of a particular one of the scaling limits enumerated in Table 2.1
depends on $\gamma$ and 
on the values of $\alpha$ and $\theta$ and thus
on the speed at which $(\beta_n,K_n) \goto (\beta_c,K_c(\beta_c))$ and on the direction of approach.
Only case 1 expresses the influence of $C$ alone, giving the same limit for
$(\beta_n,K_n)$ in (\ref{eqn:betankn}) as the limit in (\ref{eqn:6thorder}), which
holds for the constant sequence $(\beta_n,K_n) = (\beta_c,K_c(\beta_c))$.
Case 1 occurs if the convergence $(\beta_n,K_n) \goto (\beta_c,K_c(\beta_c))$ is sufficiently fast;
namely, $\alpha > 1/3$ and $\theta > 2/3$. 
Case 2, which expresses the influence of $B$ alone, occurs if the convergence 
of $(\beta_n,K_n) \goto (\beta_c,K_c(\beta_c))$is 
sufficiently slow but $\theta$ is relatively large compared to $\alpha$. 
Case 3, which expresses the influence of $A$ alone,
occurs if the convergence is 
sufficiently slow but, in contrast with case 2, $\alpha$ is relatively large compared to $\theta$.
Finally, cases 4--13, which express the influence of more than one set
$A$, $B$, and $C$, occur if the convergence of $(\beta_n,K_n) \goto (\beta_c,K_c(\beta_c))$
occurs at an appropriate critical rate.  For example,
cases 10--13 express the influence of all three sets $A$, $B$, and $C$ 
and so correspond to the most complicated form of the limiting density.
This case occurs when $\alpha={1}/{3}$,
$\theta = {2}/{3}$, and $\gamma = {1}/{6}$.   

The scaling limits for $S_n/n^{1-\gamma}$ listed in Table 2.1
are analyzed in Theorem \ref{thm:betakinc}, where we determine 
the values of $\alpha$, $\theta$, and $\gamma$ 
leading to the 13 different cases.
The dependence of $(\beta_n,K_n)$ in (\ref{eqn:betankn}) upon $\alpha$ and $\theta$ is complicated; 
because $\beta_n$ is a function of $\alpha$, $K_n$ is both a function of $\theta$ and,
through $\beta_n$, a function of $\alpha$.  However, as we will see, for the 
appropriate choice of $\gamma \in (0,{1}/{2})$, 
in the expression for the scaling limit of $S_n/n^{1-\gamma}$ the $\alpha$ and the 
$\theta$ decouple in such a way that the limits given in Theorem \ref{thm:betakinc}
can be read off in a systematic way.  

In Figure 2 in the introduction we indicate the subsets of the positive quadrant of the 
$\alpha$-$\theta$ plane leading to all the cases in Table 2.1.
The subsets labeled $C$, $B$, and $A$ correspond to cases 1, 2, and 3, respectively, and the subsets labeled
$B+C$, $A+C$, $A+B$, and $A+B+C$ correspond to cases 4--5, 6--7, 8--9, and 
10--13, respectively.  
The relationship between the $\alpha$-$\theta$ 
plane exhibited in Figure 2 and the $\beta$-$K$ plane, inside which lies the tricritical point,
is that each point in the $\alpha$-$\theta$ plane corresponds, through the formulas for $\beta_n$ and $K_n$
given in (\ref{eqn:betankn}), to a curve in the $\beta$-$K$ plane.  

In Figure 3 we exhibit three different curves in the $\beta$-$K$ plane, labeled (a), (b), and (abc).
These curves correspond to three different choices of $\alpha$ and $\theta$, three different choices
of $(\beta_n, K_n)$ in (\ref{eqn:betankn}), and three different limits in Table 2.1.  
The curve labeled (a) corresponds to $\alpha = 1$ and $\theta = {1}/{3}$, which in turn
corresponds to case 3 of the scaling limit; this case shows 
the influence only of region A.  The curve labeled (b) corresponds to $\alpha = {1}/{4}$
and $\theta = 1$, which in turn corresponds to case 2 of the scaling limit;
this case shows the influence only of region B. 
Finally, the curve labeled (abc) corresponds to $\alpha = {1}/{3}$, $b > 0$,
$\theta = {2}/{3}$, and $k > 0$; the associated scaling limit in case 10 
shows the influence of all three sets $A$, $B$, and $C$.  

\begin{figure}[h]
\begin{center}
\epsfig{file=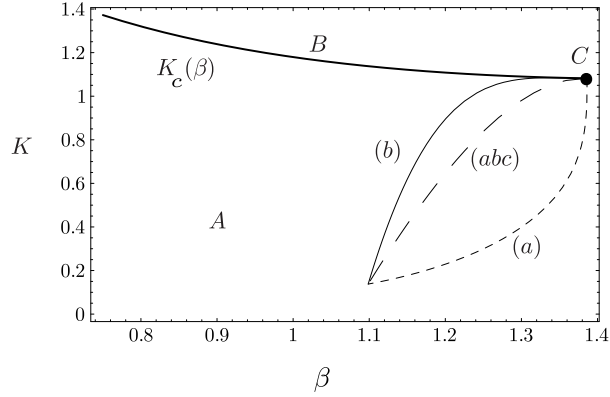,width=8cm}
\caption{\small Three choices of $(\beta_n, K_n)$ that show the influence of $A$, of $B$, 
and of $A$, $B$, and $C$ in (\ref{eqn:13cases})}
\end{center}
\end{figure}

It is worth noting a contrast between the scaling limits in (\ref{eqn:c2c4c6}) 
and those in Table 2.1.  In (\ref{eqn:c2c4c6}) the
three scaling limits for $S_n/n^{1-\gamma}$ hold with respect to $P_{n,\beta,K}$ 
for fixed $(\beta,K) \in A$, $(\beta,K) \in B$, and $(\beta,K) \in C$.  In each of these
three cases the value of $\gamma$ is fixed to be, respectively, ${1}/{2}$, ${1}/{4}$,
and ${1}/{6}$.  By contrast, we will see in Theorem \ref{thm:betakinc}
that in 4 of the 13 cases of the scaling limits for $S_n/n^{1-\gamma}$
stated in Table 2.1, the limit theorems hold for a range of values of $\gamma$. 
These are cases 2, 3, 8, and 9.  In the other cases, each of which includes the influence
of the tricritical point $(\beta_c,K_c(\beta_c))$, $\gamma$ equals the fixed value ${1}/{6}$. 

We now make a transition from the scaling limits to the MDPs. 
As we have seen, the scaling limits state that for appropriate choices of $(\bn,\kn)$ and of $\gamma = \gamma_0$
\be
\label{eqn:scalingG}
\pnbnkn\{{S_n}/{n^{1-\gamma_0}} \in dx\} \Longrightarrow \exp[-G(x)] \, dx,
\ee
where $G$ takes one of the 13 forms in Table 2.1.  For any $\gamma \in (0,\gamma_0)$,
one can show that if $D$ is any Borel set whose closure does not contain 0, then
\[
\lim_{n \goto \infty} \pnbnkn\{{S_n}/{n^{1-\gamma}} \in D\} = 0.
\]
A natural question is to determine the rate at which these 
and related probabilities converge to 0 
when $(\beta_n,K_n)$ is defined in (\ref{eqn:betankn}).
In Theorem \ref{thm:mdpc} we define a quantity $w$ in terms of $\alpha$, 
$\theta$, and $\gamma$ having the property that
when $w < 0$, $S_n/n^{1-\gamma}$ satisfies an MDP with exponential speed
$\nw$ and rate function $G(x) - \bar{G}$, where $G$ is the same function
appearing in (\ref{eqn:scalingG}) and $\bar{G} = \inf_{y \in \R} G(y)$. 
This MDP implies that for suitable sets $D$
\[
\pnbnkn\{{S_n}/{n^{1-\gamma}} \in D\} \goto 0 \ \mbox{ like } 
\exp[-\nw \inf_{x \in D} (G(x) - \bar{G})].
\]
In order to emphasize the similarity with the scaling limits, we summarize this class
of MDPs by the formal notation 
\be
\label{eqn:mdpformal}
\pnbnkn\{{S_n}{n^{1-\gamma}} \in dx\} \asymp \exp[-\nw G(x)],
\ee
in which the constant $\bar{G}$ is not shown.

The situation with the MDPs
is completely analogous to the situation for the scaling limits. 
Specifically, as we exhibit in Table 2.2,
there are 13 cases of the MDP (\ref{eqn:mdpformal}), 
each of which shows the influence of one or more of the sets
$C$, $B$, and $A$ depending on the speed at which 
the sequence $(\bn,\kn)$ defined in (\ref{eqn:betankn}) converges to $(\bc,\kc(\bc))$
and on its direction of approach.
The coefficient $c_6 = 9/40$ is the same as
in (\ref{eqn:6thorder}), $\bar{c}_4 = 3/16$, and $b$ and $k$ are
the nonzero real numbers appearing in (\ref{eqn:betankn}) and subject only to the 
requirement that $G(x) \goto \infty$ as $|x| \goto \infty$.
The MDPs for $S_n/n^{1-\gamma}$ listed in Table 2.2 are analyzed in Theorem
\ref{thm:mdpc}, where we determine the values of $\alpha$, $\theta$, and $\gamma$ 
that lead to each of the cases.

\begin{center}
% \label{table:13casesmdp}
\begin{tabular}{||c|c|l||} \hline \hline
{\em {\bf case}} & {\em {\bf influence}} & \boldmath $\pnbnkn\{S_n/n^{1-\gamma} \in dx\} \asymp
\exp[-\nw G(x)] \, dx$ \unboldmath \\ 
\hline \hline 
1 & $C$ & $G(x) = c_6 x^6$, $\ c_6 > 0$ \\ \hline
2 & $B$ & $G(x) = b \bar{c}_4 x^4$, $\ b > 0$, $\bar{c}_4 > 0$ \\ \hline
3 & $A$ & $G(x) = k \bc x^2$, $\ k > 0$ \\ \hline
4--5 & $B+C$ & $G(x) = b \bar{c}_4 x^4 + c_6 x^6$, $\ b \not = 0$ \\ \hline
6--7 & $A+C$ & $G(x) = k \bc x^2 + c_6 x^6$, $\ k \not = 0$ \\ \hline
8--9 & $A+B$ & $G(x) = k \bc x^2 + b \bar{c}_4 x^4$, $\ k \not = 0$, $b > 0$ \\ \hline
10--13 & $A+B+C$ & $G(x) = k \bc x_2 + b \bar{c}_4 x^4 + c_6 x^6$,
$k \not = 0$, $b \not = 0$ \\ \hline
\end{tabular}

\vspace{.05in}
Table 2.2: {\small 13 cases of the MDPs
in (\ref{eqn:mdpformal}) for $(\bn,\kn)$ 
in (\ref{eqn:betankn}) and 
$\gamma \in (0,{1}/{2})$}
\end{center}

The MDPs for $S_n/n^{1-\gamma}$ have an unexpected consequence concerning a new
class of distribution limits for $S_n/n^{1-\gamma}$ that give deeper insight
into the fine structure of the phase transitions in a neighborhood of the tricritical
point.  In an effort to understand the physical significance of these new limits, analogs
of them are now being investigated for a class of non-mean-field models, including the 
Blume-Emery-Griffiths model \cite{EllMac}.  In order to appreciate these
new results, we first consider a consequence of the large deviation principle
stated in part (a) of Theorem \ref{thm:ldppnbetak}.  Since
$\ebetak = \{0\}$ for $(\beta,K) \in A \cup B \cup C$, it follows that for
any positive sequence $(\bn,\kn) \goto (\beta,K) \in A \cup B \cup C$
\[
\pnbnkn\{{S_n}/{n} \in dx\} \Longrightarrow \delta_0.
\]

The MDPs for $S_n/n^{1-\gamma}$ listed in Table 2.2 lead to refinements of this
limit for $(\bc,\kcbc) \in C$ in those cases in which the set of 
global minimum points of $G$ contains nonzero points.  These are precisely the cases
in which the coefficients of $G$ are not all positive: cases 5 ($b < 0$), 7 ($k < 0$),
9 ($k < 0$), 11 ($k < 0$, $b > 0$), 12 ($k > 0$, $b < 0$), and 13 ($k < 0$, $b < 0$).
In all these cases except for case 12, the set of global minimum points of $G$ 
consists of two symmetric, nonzero points $\pm x(b,k)$.  Hence, using the appropriate
value of $\gamma$ and the appropriate sequence $(\bn,\kn)$ given in Theorem \ref{thm:mdpc}, 
we deduce from the corresponding MDP the limit
\be
\label{eqn:refinemdp}
\pnbnkn\{{S_n}/{n^{1-\gamma}} \in dx\} \Longrightarrow 
\ts \frac{1}{2}\!\left(\delta_{x(b,k)} + \delta_{-x(b,k)}\right)\! .
\ee
In each of these cases $(\bn,\kn)$ approaches $(\bc,\kcbc)$ from a region of points 
$(\beta,K)$ corresponding
to each of which there exist two equilibrium macrostates $\pm z(\beta,k)$ [Thms.\ 
\ref{thm:secondorder}(b), \ref{thm:firstorder}(c)].  As we have already seen, 
for each
$(\beta,K)$ in this region the limit (\ref{eqn:largek}) holds.  The new limit
(\ref{eqn:refinemdp}) shows that as $(\bn,\kn) \goto (\bc,\kcbc)$
from this two-phase region, the model retains a trace of the 
two equilibrium macrostates $\pm z(\beta,K)$, replacing them
by the quantities $\pm x(b,k)$.  The physical significance of this limit
as well as the limit (\ref{eqn:refine3mdp}) to be stated in the next paragraph 
is currently under investigation \cite{EllMac}.
A similar phenomenon occurs in case 4 of Theorem \ref{thm:mdpb},
which proves MDPs for $S_n/n^{1-\gamma}$ for appropriate sequences
$(\bn,\kn)$ converging to $(\beta,K)$ lying in the curve $B$ of
second-order points.

The situation in case 12 in Table 2.2 ($k > 0$, $b < 0$) is even more fascinating
than in the other cases.  For fixed $b < 0$, fixed $n \in \N$, and decreasing $k > 0$, the set
of global minimum points of $G$ undergoes a discontinuous bifurcation, changing 
from a unique global minimum point at 0 for $k$ large, to three global
minimum points at $0, \pm x(b,k)$ for a critical value of $k = \mbox{const}\cdot b^2$,
to two global minimum points at $\pm x(b,k)$ for $k$ small. 
As $k$ decreases, $(\bn,\kn)$ crosses the first-order critical curve from below;
the changing forms of the sets of global minimum points of $G$ replicate
the changing forms of $\ebetak$ for fixed $\beta > \bc$ and increasing $K > 0$
[Thm.\ \ref{thm:firstorder}].  In particular, when the set of global
minimum points of $G$ equals $\{0, \pm x(b,k)\}$, the MDP corresponding to 
case 12 together with other information yields the limit
\be
\label{eqn:refine3mdp}
\pnbnkn\{{S_n}/{n^{1 - \gamma}} \in dx\}
\Longrightarrow \bar{\lambda}_0 \delta_0 + \bar{\lambda}_1 \!
\left(\delta_{x(b,k)} + \delta_{-x(b,k)}\right)\! ,
\ee
where $\bar{\lambda}_0$ and $\bar{\lambda}_1$ are positive numbers
satisfying $\bar{\lambda_0} + 2 \bar{\lambda}_1 = 1$.
This limit is reminiscent of the limit (\ref{eqn:largebetakcbeta}),
in which the equilibrium macrostates $\pm z(\beta,K)$ 
are replaced by their traces $\pm x(b,k)$. 

Although in general the values of $\alpha$, $\theta$, and $\gamma$
leading to each of the 13 cases of the MDPs in Table 2.2 differ from the values of these
parameters leading to the corresponding case of the scaling limit in Table 2.1, the tables have
a number of obvious similarities.
This resemblance between the two tables
reaches deeper.  In fact, both sets of results are proved by a unified method.  
In order to explain this, let $f$ be any bounded, continuous function mapping $\R$ into $\R$
and let $(\bn,\kn)$ be any positive sequence.
The starting point of the proofs of both the scaling limits and the MDPs 
[see Lem.\ \ref{lem:G}] is that whenever
$\gamma \in (0,{1}/{2})$, we have
\be
\label{eqn:whyscalinglimits}
E\{f({S_n}/{n^{1-\gamma}} + \ve_n)\} 
= \frac{1}{Z_n} \cdot \int_{\R} f(x) \, \exp[-nG_{\beta_n, K_n}(x/n^{\gamma})] \, dx.
\ee
The function $G_{\beta,K}$ in this display is defined in (\ref{eqn:introgbetak}); 
its global minimum value equals the canonical
free energy for the model.
In addition, $\ve_n$ represents a sequence of random variables that converges to 0 as $n \goto \infty$, 
and $Z_n$ is a normalizating constant.

The quantity $w$ in the MDP (\ref{eqn:mdpformal}) is defined
by $w = \min\{2\gamma + \theta -1,4\gamma + \alpha -1, 6\gamma -1 \}$.
This quantity also plays a key role in the scaling limits for $S_n/n^{1-\gamma}$, 
which like the MDPs arise from the choice of $(\bn,\kn)$ in (\ref{eqn:betankn}).  
When $w = 0$, the scaling limits listed in Table 2.1 follow at least formally
from (\ref{eqn:whyscalinglimits}) and 
the fact that for each $x \in \R$
\[
\lim_{n \goto \infty} nG_{\bn,\kn}(x/n^\gamma) = G(x).
\]
The proof of this limit relies on an analysis of the Taylor expansion of $G_{\bn,\kn}$
at 0, which 
has 13 different forms depending on the choices of $\gamma$ and of the parameters
$\alpha$ and $\theta$ appearing in the definition (\ref{eqn:betankn}) 
of $(\bn,\kn)$.  Details are given in Theorem \ref{thm:betakinc}.

We now assume that $w < 0$.  
Given $\psi$ be any bounded, continuous function, we substitute $f = \exp(\nw \psi)$ into 
(\ref{eqn:whyscalinglimits}), obtaining
\beas
\label{eqn:whymdp}
\lefteqn{
E\{\exp[\nw \psi({S_n}/{n^{1-\gamma}} + \ve_n)]\}} \\
&& \nonumber = \frac{1}{Z_n} \cdot
\int_{\R} \exp\!\left[\nw \left\{\psi(x) - n^{1+w}G_{\beta_n, K_n}(x/n^\gamma)\right\}\right] 
 dx.
\eeas
When $w < 0$, the last display, the fact that for each $x \in \R$
\[
\lim_{n \goto \infty} n^{1+w} G_{\bn,\kn}(x/n^\gamma) = G(x),
\]
and the fact that $\ve_n \goto 0$ in probability at a rate faster than $\exp(-\nw)$ give the formal
asymptotics
\beas
\lefteqn{
E\{\exp[\nw \psi({S_n}/{n^{1-\gamma}} + \ve_n)]\}} \\
&& \approx 
\int_{\R} \exp\!\left[\nw \left\{\psi(x) - (G(x) - \bar{G})\right\}\right] 
dx \\
&& \approx\exp\!\left[\nw \ts \sup_{x \in \R}\left\{\psi(x) - (G(x) - \bar{G})\right\}\right]\! ,
\eeas
where $\bar{G} = \inf_{y \in \R} G(y)$.  In section \ref{section:mdp} we show to convert
this formal calculation into a limit known as the Laplace principle, which is
equivalent to the MDPs for $S_n/n^{1-\gamma}$ listed in Table 2.2.
As in the proof of the scaling limits, the proof of the Laplace limit relies
on an analysis of the Taylor expansion of $G_{\bn,\kn}$ at 0.
Despite the similarity in the proofs
of the scaling limits and the Laplace principles, the proof of the latter is much more delicate,
requiring additional estimates not needed in the proof of the former.

We start our analysis of the BEG model in the next section.

\section{Phase-Transition Structure of the BEG Model}
\beginsec
\label{section:phasetr}

After defining the BEG model, we summarize its phase-transition structure in Theorems
\ref{thm:secondorder} and \ref{thm:firstorder}.  In (\ref{eqn:introgbetak})
we introduce the function
$G_{\beta,K}$, in terms of which the scaling limits and the
MDPs for $S_n/n^{1-\gamma}$ will be deduced
later in the paper.

The BEG model is a lattice-spin model defined on the complete graph on $n$ vertices $1,2, \ldots, n$.  The spin at site $j \in \{1,2,\ldots, n\}$ is denoted by $\omega_j$, a quantity taking values in $\Lambda=\{-1,0,1\}$. 
The configuration space for the model is the set $\Lambda^n$ containing all 
sequences $\omega = (\omega_1,\omega_2,
\ldots, \omega_n)$ with each $\omega_j \in \Lambda$.  In terms of a positive parameter $K$
representing the interaction strength, the Hamiltonian is defined by
\[
H_{n, K}(\omega)=\sum_{j=1}^n \omega_j^2 - \frac{K}{n}\!\left( \sum_{j=1}^n 
\omega_j \right)^2
\]
for each $\omega \in \Lambda^n$.  
For $n \in \N$, inverse temperature $\beta > 0$, and $K>0$,
the canonical ensemble for the BEG model is the sequence of probability measures that assign to each
subset $B$ of $\Lambda^n$ the probability
\be
\label{eqn:pnbetak}
P_{n, \beta, K}(B)= \frac{1}{Z_n(\beta, K)} \cdot \int_B \exp[-\beta H_{n,K}] \, dP_n.
\ee
In this formula $P_n$ is the product measure on $\Lambda^n$ with identical one-dimensional marginals 
\[
\rho = \ts \frac{1}{3}(\delta_{-1}+\delta_0+\delta_1),
\]
and $Z_n(\beta, K)$ is the normalizing constant $\int_{\Lambda^n} \exp[-\beta H_{n,K}] dP_n$.

In \cite{EllOttTou} the analysis of the 
canonical ensemble $P_{n, \beta, K}$ was facilitated by expressing it
 in the form of a Curie-Weiss-type model.  This is done by absorbing 
the noninteracting component of the Hamiltonian into the product measure $P_n$, obtaining
\be
\label{eqn:rewritecanon}
P_{n,\beta,K}(d\omega) = 
\frac{1}{\tilde{Z}_n(\beta,K)} \cdot \exp\!\left[ n \beta K 
\!\left(\frac{S_n(\omega)}{n} \right)^2 \right] P_{n,\beta}(d\omega).
\ee
In this formula $S_n(\omega)$ equals
the total spin $\sum_{j=1}^n \omega_j$,
$P_{n,\beta}$ is the product measure on $\Lambda^n$ with identical one-dimensional marginals
\be
\label{eqn:rhobeta}
\rho_\beta(d \omega_j) = 
\frac{1}{Z(\beta)} \cdot \exp(-\beta \omega_j^2) \, \rho(d \omega_j),
\ee
$Z(\beta)$ is the normalizing constant 
$\int_\Lambda \exp(-\beta \omega_j^2) \rho(d \omega_j) = 1 + 2 e^{-\beta}$,
and $\tilde{Z}_n(\beta,K)$ is the normalizing constant
$[Z(\beta)]^n/Z_n(\beta,K)$. 

Although $P_{n,\beta,K}$ has the form of a Curie-Weiss model
when rewritten as in (\ref{eqn:rewritecanon}), 
it is much more complicated because of the $\beta$-dependent
product measure $P_{n,\beta}$ and the presence of the parameter $K$. 
These complications introduce new features not present in the Curie-Weiss model \cite[\S IV.4, \S V.9]{Ellis}; 
these features include the existence of a second-order
phase transition for all sufficiently small $\beta > 0$ and all sufficiently large $K > 0$
and a first-order phase transition for all sufficiently large $\beta > 0$ 
and all sufficiently large $K > 0$.
The existence of a second-order phase transition and a first-order phase transition
also implies the existence of
 a tricritical point, which separates the two phase transitions and is one of the main
focuses of the present paper.  

The starting point of the analysis of the 
phase-transition structure of the BEG model is the large deviation
principle (LDP) satisfied by the spin per site $S_n/n$ with respect to $P_{n,\beta,K}$.  
In order to state the form of the rate function, we introduce the cumulant generating function $c_\beta$ of the measure
$\rho_\beta$ defined in (\ref{eqn:rhobeta}); for $t \in \R$ this function is defined by 
\beas
\label{eqn:cbeta}
c_\beta(t) & = & \log \int_\Lambda \exp (t\omega_1) \, \rho_\beta (d\omega_1)
 \\ 
& = & 
\log \!\left[ \frac{1+e^{-\beta}(e^t+e^{-t})}{1+2e^{-\beta}} \right]. \nonumber 
\eeas
We also introduce the Legendre-Fenchel transform of $c_\beta$, which is defined for $z \in [-1,1]$ by
\[
J_\beta(z) = \sup_{t \in \R} \{tz - c_\beta(t)\};
\]
$J_\beta(z)$ is finite for $z \in [-1,1]$.
$J_\beta$ is the rate function in Cram\'{e}r's theorem, which 
is the LDP for $S_n/n$ with respect to the product measures
$P_{n,\beta}$ \cite[Thm.\ II.4.1]{Ellis} and is one of the components of
the proof of the LDP for $S_n/n$ with respect to $P_{n,\beta,K}$.
This LDP and a related limit are stated in parts (a) and (b)
of the next theorem.  Part (a) is proved in Theorem 3.3 in \cite{EllOttTou},
and part (b) in Theorem 2.4 in \cite{EllHavTur}.

\begin{thm}
\label{thm:ldppnbetak}  For all $\beta > 0$ and $K > 0$ the following conclusions hold.

{\em (a)} With respect to the canonical ensemble $P_{n,\beta,K}$,
$S_n/n$ satisfies the LDP on $[-1,1]$ with exponential speed $n$ and rate function
\[
I_{\beta,K}(z) = J_\beta(z) - \beta K z^2 - \inf_{y \in \R}
\{J_\beta(y) - \beta K y^2 \}.
\]

{\em (b)}  We define the canonical free energy
\[
\varphi(\beta,K) = - \lim_{n \goto \infty} \frac{1}{n} \log Z_n(\beta,K),
\]
where $Z_n(\beta,K)$ is the normalizing constant in {\em (\ref{eqn:pnbetak})}.
Then $\varphi(\beta,K) = \inf_{y \in \R}\{J_\beta(y) - \beta K y^2\}$.
\end{thm}

The LDP in part (a) of the theorem implies that those $z \in [-1,1]$
satisfying $I_{\beta,K}(z) > 0$ have an exponentially small probability
of being observed in the canonical ensemble. Hence we define 
the set of equilibrium macrostates by
\[
\ebetak = \{z \in [-1,1] : I_{\beta,K}(z) = 0\}.
\]
In \cite{EllOttTou} we used the notation $\tilde{\cal E}_{\beta,K}$ to describe
this set, using the notation $\ebetak$ to describe a different but related
set of equilibrium macrostates.  In the present paper we write $\ebetak$ 
instead of $\tilde{\cal E}_{\beta,K}$ in order to simplify the notation.

For $z \in \R$ we define
\be
\label{eqn:introgbetak} 
G_{\beta,K}(z) = \beta K z^2 - c_\beta(2\beta K z).
\ee
The calculation of the zeroes of $I_{\beta,K}$ --- equivalently, the global minimum points
of $J_{\beta,K}(z) - \beta K z^2$ --- is greatly facilitated by the following observations
made in Proposition 3.4 in \cite{EllOttTou}: 
\begin{enumerate}
\item The global minimum points of 
$J_{\beta,K}(z) - \beta K z^2$ coincide with the global minimum points of $\gbk$,
which are much easier to calculate. 
\item The minimum values $\min_{z \in \R}\{J_{\beta,K}(z) - \beta K z^2\}$ 
and $\min_{z \in \R}G_{\beta,K}(z)$ coincide and both equal 
the canonical free energy $\varphi(\beta,K)$
defined in part (b) of Theorem \ref{thm:ldppnbetak}.
\end{enumerate}
Item 1 gives the alternate characterization that 
\be
\label{eqn:ebetak}
\ebetak = \{z \in [-1,1] : z \mbox{ minimizes } G_{\beta,K}(z)\}.
\ee
In the context of Curie-Weiss-type models, the form of $G_{\beta,K}$ is explained
on page 2247 of \cite{EllOttTou}.

As shown in the next two theorems, the 
structure of $\ebetak$ depends on the relationship between $\beta$ and the
critical value $\beta_c = \log 4$.
We first describe $\ebetak$ for 
$0 < \beta \leq \beta_c$ and then for $\beta > \beta_c$.  
In the first case $\ebetak$ undergoes a continuous
bifurcation as $K$ increases through the critical value $K_c(\beta)$
defined in (\ref{eqn:kcbeta}); physically, this bifurcation
corresponds to a second-order phase transition.  The following theorem 
is proved in Theorem 3.6 in \cite{EllOttTou}.

\begin{thm}
\label{thm:secondorder} 
For $0 < \beta \leq \beta_c$, we define
\be
\label{eqn:kcbeta}
K_c(\beta) = \frac{1}{2\beta c''_\beta(0)} =
\frac{e^\beta + 2}{4\beta}.
\iffalse
\frac{1}{4\beta e^{-\beta}} + \frac{1}{2\beta}.
\fi
\ee
For these values of $\beta$, $\ebetak$ has the following structure.

{\em(a)} For $0 < K \leq K_c(\beta)$,
${\mathcal{E}}_{\beta,K} = \{0\}$.

{\em(b)} For $K > K_c(\beta)$, there exists 
${z}(\beta,K) > 0$ such that
${\mathcal{E}}_{\beta,K} = \{\pm z(\beta,K) \}$.

{\em(c)} ${z}(\beta,K)$ is
a positive, increasing, continuous function for $K > K_c(\beta)$, and
as $K \goto (K_c(\beta))^+$, $z(\beta,K) \goto 0$. 
Therefore, ${\mathcal{E}}_{\beta,K}$ exhibits a continuous bifurcation
at $K_c(\beta)$.
\end{thm}

For $\beta \in (0,\bc)$, the curve $(\beta,K_c(\beta))$ is the curve of second-order 
points.   As we will see in a moment, for $\beta \in (\bc,\infty)$
the BEG model also has a curve
of first-order points, which we denote by the same notation $(\beta,K_c(\beta))$.
In order to simplify the notation, we do not follow the convention
in \cite{EllOttTou}, where we distinguished between the second-order phase transition and 
the first-order phase transition by
writing $K_c(\beta)$ for $0 < \beta \leq \beta_c$ as $K^{(2)}_c(\beta)$ and writing $K_c(\beta)$ for
$\beta > \beta_c$ as $K^{(1)}_c(\beta)$.  

We now describe $\ebetak$ for 
$\beta > \beta_c$.  In this case $\ebetak$ undergoes a discontinuous
bifurcation as $K$ increases through an implicitly defined critical value.
Physically, this bifurcation
corresponds to a first-order phase transition.  The following theorem 
is proved in Theorem 3.8 in \cite{EllOttTou}.  

\begin{thm}
\label{thm:firstorder} 
For all $\beta > \beta_c $, $\ebetak$ has
the following structure in terms of the quantity
$K_c(\beta)$, denoted by $K_c^{(1)}(\beta)$ in {\em \cite{EllOttTou}}
and defined implicitly for $\beta > \beta_c$ on page {\em 2231} of
{\em \cite{EllOttTou}}.

{\em(a)} For $0 < K < K_c(\beta)$,
${\mathcal{E}}_{\beta,K} = \{0\}$.

{\em(b)} There exists $z(\beta,K_c(\beta)) > 0$
such that ${\mathcal{E}}_{\beta,K_c(\beta)} =
\{0,\pm z(\beta,K_c(\beta))\}$.

{\em(c)} For $K > K_c(\beta)$ 
there exists $z(\beta,K) > 0$
such that ${\mathcal{E}}_{\beta,K} =
\{\pm z(\beta,K)\}$.

{\em(d)} $z(\beta,K)$
is a positive, increasing, continuous function for $K \geq K_c(\beta)$, and 
as $K \goto K_c(\beta)^+$, $z(\beta,K) \goto 
z(\beta,K_c(\beta)) > 0$.  Therefore,
${\mathcal{E}}_{\beta,K}$ exhibits a discontinuous bifurcation
at $K_c(\beta)$.
\end{thm}

We end this section by outlining the proofs of the laws of large numbers in (\ref{eqn:smallk})
and (\ref{eqn:smallbetakcbeta}) and its breakdown in (\ref{eqn:largek}).  
The upper large deviation bound in
the LDP stated in part (a) of Theorem \ref{thm:ldppnbetak} implies that 
for any $\beta > 0$ and $K > 0$ the limiting mass of $S_n/n$ with respect to $P_{n,\beta,K}$
concentrates on the elements of $\ebetak$.  According to Theorems \ref{thm:secondorder}(a)
and \ref{thm:firstorder}(a), $\ebetak = \{0\}$
when $0 < \beta \leq \beta_c$ and $0 < K \leq K_c(\beta)$ and when 
$\beta > \beta_c$ and $K < K_c(\beta)$.  For these values of $\beta$ and $K$, 
the laws of large numbers in (\ref{eqn:smallk})
and (\ref{eqn:smallbetakcbeta}) follow immediately.  For $\beta > 0$ and $K > K_c(\beta)$,
we have $\ebetak = \{\pm z(\beta,K)\}$, and so by symmetry the limit (\ref{eqn:largek})
follows.  The proof of the limit (\ref{eqn:largebetakcbeta}) is postponed until after
Theorem \ref{thm:secondordertype}
because it requires more detailed information about the elements 
of $\ebetak$ when $\beta > \beta_c$ and $K = K_c(\beta)$.  

In the next section we present additional properties of the function $G_{\beta,K}$
introduced in (\ref{eqn:introgbetak}).  These properties will be used in later sections
to prove the scaling limits and the MDPs for $S_n/n^{1-\gamma}$.

\section{Properties of \boldmath $G_{\beta,K}$ \unboldmath}
\beginsec
\label{section:G}

As we saw in (\ref{eqn:ebetak}), the global minimum points of
\beas
\label{eqn:gbetak} 
G_{\beta,K}(z) & = & \beta K z^2 - c_{\beta}(2\beta K z) \\ \nonumber
& = & \beta K z^2 - \log \!\left[ \frac{1+e^{-\beta}(e^{2\beta K z} + e^{-2\beta K z})}{1+2e^{-\beta}} \right]
\eeas
coincide with the elements of $\ebetak$, the set of equilibrium
macrostates for the BEG model.  In this section we study further properties
of $G_{\beta,K}$ that will be used in later sections
to prove the scaling limits and the MDPs for $S_n/n^{1-\gamma}$ with respect to $P_{n,\beta,K}$
and with respect to $P_{n,\beta_n,K_n}$ for appropriate sequence $(\beta_n,K_n)$
and for appropriate choices of $\gamma$.   

We first show that for any $\gamma \in [0,1)$ the $\pnbnkn$-distribution of $S_n/n^{1-\gamma}$ can
be expressed in terms of $G_{\beta_n,K_n}$ and an independent normal random variable.  The
next lemma can be proved like Lemma 3.3 in \cite{EllNew}, which applies to the Curie-Weiss
model, or like Lemma 3.2 in \cite{EllWan}, which applies to the Curie-Weiss-Potts model.
In an equivalent form, the next lemma is well known in the literature as the 
Hubbard-Stratonovich transformation, 
where it is invoked to analyze models with quadratic Hamiltonians
(see, e.g., \cite[p.\ 2363]{AntRuf}). 
After the statement of the lemma, we outline how we will use it in order to 
deduce the scaling limits of $S_n/n^{1-\gamma}$.

\begin{lemma}
\label{lem:G}
Given a positive sequence $(\bn,\kn)$, let 
$W_n$ be a sequence of $N(0,(2\beta_n K_n)^{-1})$ random variables defined 
on a probability space $(\Omega,\mathcal{F},Q)$.
Then for any $\gamma \in [0,1)$ and any bounded, measurable function $f$
\bea
\label{eqn:G}
\lefteqn{
\int_{\Lambda^n \times \Omega} f\!\left(\frac{S_n}{n^{1-\gamma}} + \frac{W_n}{n^{1/2-\gamma}} \right)
d(P_{n,\beta_n,K_n} \times Q)} 
\\ \nonumber && = \frac{1}{\int_{\R} \exp [-nG_{\beta_n,K_n}(x/n^{\gamma})] \, dx} \cdot
\int_{\R} f(x) \, \exp[-nG_{\beta_n, K_n}(x/n^{\gamma})] \, dx.
\eea
\end{lemma}

As we will see in Theorems \ref{thm:betakina}, \ref{thm:betakinb}, and
\ref{thm:betakinc}, the scaling limits have different forms depending on which of the
following three sets $(\beta,K)$ lies in: the singleton set $C$ containing the
tricritical point $(\beta_c,K_c(\beta_c))$,
the curve $B$ of second-order points
\[
B = \{(\beta,K) \in \R^2 : 0 < \beta < \beta_c, K = K_c(\beta)\},
\]
and the single-phase region 
\[
A = \{(\beta,K) \in \R^2 : 0 < \beta \leq \beta_c, 0 < K < K_c(\beta)\}.
\]
These sets are shown in Figure 1 in the introduction.

We now indicate how we will use Lemma \ref{lem:G} to prove the scaling limits of $S_n/n^{1-\gamma}$ for
$\gamma \in (0,{1}/{2}]$.
Let $(\bn,\kn)$ be a suitable positive sequence converging to 
$(\beta,K) \in A \cup B \cup C$.  
Assume that $(\bn,\kn)$ and $\gamma$ are chosen so that the
limit of the right hand side of (\ref{eqn:G}) exists as 
$n \goto \infty$.  We first consider $\gamma < {1}/{2}$.  
Since $\beta_n$ and $K_n$ are bounded and 
uniformly positive over $n$, rewriting the limit of the left hand side
in terms of characteristic functions shows that $W_n/n^{1/2-\gamma}$ 
does not contribute.  Hence it follows that 
\bea
\label{eqn:limit}
\lefteqn{
\lim_{n \goto \infty}
\int_{\Lambda^n} f(S_n/n^{1-\gamma}) \, dP_{n,\beta_n,K_n}} \\
\nonumber 
&& = \lim_{n \goto \infty}
\frac{1}{\int_{\R} \exp [-nG_{\beta_n,K_n}(x/n^{\gamma})] \, dx} \cdot
\int_{\R} f(x) \, \exp[-nG_{\beta_n, K_n}(x/n^{\gamma})] \, dx.
\eea
From this formula we will be able to determine the scaling limits of $S_n/n^{1-\gamma}$
when $(\bn,\kn) \goto (\beta,K) \in B \cup C$ [Thms.\ \ref{thm:betakinb}, \ref{thm:betakinc}]. 
Using an analogous formula, we 
will be able to determine the MDPs of $S_n/n^{1-\gamma}$
when $(\bn,\kn) \goto (\beta,K) \in B \cup C$ [Thms.\ \ref{thm:mdpb}, \ref{thm:mdpc}].

Now consider $\gamma = {1}/{2}$, which corresponds
to the central-limit-type scaling for $S_n$ in (\ref{eqn:2ndorder}).
In this case (\ref{eqn:G}) yields
\bea
\label{eqn:limit2}
\lefteqn{
\lim_{n \goto \infty}
\int_{\Lambda^n \times \Omega} f(S_n/n^{1/2} + W_n) \, d(P_{n,\bn,\kn} \times Q)} \\
\nonumber 
&& = \lim_{n \goto \infty}
\frac{1}{\int_{\R} \exp [-nG_{\beta_n,K_n}(x/n^{1/2})] \, dx} \cdot
\int_{\R} f(x) \, \exp[-nG_{\beta_n, K_n}(x/n^{1/2})] \, dx.
\eea
In contrast to when $\gamma \in (0,1/2)$, $W_n$ now contributes to the limit.
Again the use of characteristic
functions enables one to determine the scaling limit of $S_n/n^{1/2}$ when
$(\bn,\kn) \goto (\beta,K) \in A$ [Thm.\ \ref{thm:betakina}].

\iffalse
In subsequent sections of this paper we will use Lemma \ref{lem:G} and in particular
(\ref{eqn:limit}), (\ref{eqn:limit2}), and related formulas to deduce the forms
of the scaling limits and the MDPs of $S_n/n^{1-\gamma}$ with respect to 
$P_{n,\beta_n,K_n}$, where $(\beta_n,K_n)$ is an appropriate positive
sequence converging to $(\beta,K)
\in A \cup B \cup C$ and $\gamma$ is appropriately chosen.  For any $\beta > 0$ and $K > 0$
the set $\ebetak$ of equilibrium macrostates
is characterized in (\ref{eqn:ebetak}) as the set of global 
minimum points of $G_{\beta,K}$.  
\fi
Formulas (\ref{eqn:limit}) and (\ref{eqn:limit2}) suggest how to proceed in proving
the scaling limits of $S_n/n^{1-\gamma}$. 
First consider $(\bn,\kn)$ for which $G_{\bn,\kn}$ has a unique
global minimum point at 0 [Thms.\ \ref{thm:secondorder}(a), \ref{thm:firstorder}(a)].  
As (\ref{eqn:limit}) and (\ref{eqn:limit2}) suggest,
the forms of the scaling limits of 
$S_n/n^{1-\gamma}$ with respect to $P_{n,\beta_n,K_n}$
depend on the forms of the Taylor expansions
of $G_{\beta_n,K_n}$ in the neighborhood of the global minimum point 0.
One of the attractive features of our analysis is that
the same Taylor expansions can be used to handle sequences $(\bn,\kn)$ 
for which $G_{\bn,\kn}$ has nonunique global minimum points.  Such 
sequences arise naturally
in the scaling limits and the MDPs to be proved later in the paper;
in fact, it is precisely such sequences for which the MDPs 
yield the new class of distribution limits of the form (\ref{eqn:refinemdp})
and (\ref{eqn:refine3mdp}).
What makes it possible to use the same Taylor expansions regardless
of the nature of the global minimum points of $G_{\bn,\kn}$ is Lemma
\ref{lem:awayfrom0}, the main technical innovation in this paper.

\iffalse
According to part (a) of Theorem \ref{thm:secondorder},
for all $(\bn,\kn)$ satisfying $\bn > 0$ and $0 < \kn < K(\bn)$ 
the set of global minimum points of $G_{\beta,K}$ equals $\{0\}$.
Hence when $(\bn,\kn)$ approaches $(\beta,K)$ from inside $A \cup B \cup C$,
(\ref{eqn:limit})--(\ref{eqn:limit2}) suggest that the forms of the scaling limits of 
$S_n/n^{1-\gamma}$ with respect to $P_{n,\beta_n,K_n}$
depend on the forms of the Taylor expansion
of $G_{\beta_n,K_n}$ in the neighborhood of the global minimum point 0.  This is in fact 
the case.  The surprise is that the same Taylor expansion can be used to deduce the forms
of the scaling limits when $(\bn,\kn)$ approaches $(\beta,K)$ from outside $A \cup B \cup C$;
although for such values of $(\bn,\kn)$, $\gn$ has nonzero global minimum points, these
minimum points converge to 0.
\fi

Preliminary information on the forms of the relevant Taylor expansions
is presented in Theorems \ref{thm:secondordertype} and \ref{thm:taylor}.  
In the proofs of the scaling limits, in order to
justify replacing $nG_{\beta_n, K_n}(x/n^{\gamma})$ in (\ref{eqn:limit})
by $n$ times the Taylor expansion evaluated at $x/n^\gamma$ and taking 
limits under the integral, one invokes the dominated convergence theorem,
for which the appropriate bounding function depends on the particular
sequence $(\beta_n,K_n)$. This will be handled on a case-by-case basis
in subsequent sections.  Finally, one must show that the contributions 
to the limit in (\ref{eqn:limit}) and (\ref{eqn:limit2}) by all $x$ for which $x/n^\gamma$ lies
in the complement of a neighborhood of 0 is exponentially small.  The relevant error
estimate is given in part (c) of Lemma \ref{lem:awayfrom0}.  Similar
considerations apply to the proofs of the MDPs in section \ref{section:mdp}, for 
which the relevant error estimate is given in part (d) of Lemma \ref{lem:awayfrom0}.

The steps outlined
in the preceding paragraph for deducing the scaling limits of $S_n/n^{1-\gamma}$
from (\ref{eqn:limit}) and (\ref{eqn:limit2}) are well known in the related contexts of the
Curie-Weiss model and the Curie-Weiss-Potts model.  Scaling limits for 
these models are studied
in \cite{EllNew,EllNewRos} and in \cite{EllWan} for fixed values of 
the inverse temperature defining the corresponding canonical ensemble.  
In contrast to those earlier papers, our study of scaling limits for the BEG model 
necessitates a considerably more careful analysis
because we work with the canonical ensemble $\pnbnkn$,
allowing sequences $(\beta_n,K_n)$ rather than only fixed values of $(\beta,K)$.

The analysis of the Taylor expansions of $G_{\beta,K}$ in the neighborhood
of a global minimum point involves the notion of the type of a global minimum point,
which we next introduce.  We temporarily consider any $\beta > 0$
and any $K > 0$ and then specialize to $(\beta,K) \in A \cup B \cup C$.
Let $\tilde{z}$ be an element of $\ebetak$.  Since
$G_{\beta,K}$ is real analytic and $\tilde{z}$ is a global
minimum point, there exists a positive integer $r = r(\tilde{z})$ 
such that $G_{\beta,K}^{(2r)}(\tilde{z}) > 0$ and
\[
G_{\beta,K}(z) = G_{\beta}(\tilde{z}) +
\frac{G_{\beta,K}^{(2r)}(\tilde{z})}{(2r)!} (z-\tilde{z})^{2r} +
O((z-\tilde{z})^{2r+1}) \ \mbox{ as } z \longrightarrow \
\tilde{z}.
\]
We call $r(\tz)$ the type of the global minimum point $\tz$.
If $r = 1$, then $G_{\beta,K}^{(2)}(\tz) = 2\beta K - (2\beta K)^2 (c_\beta)''(2\beta K\tz)$, and if $r \geq 2$,
then $G_{\beta}^{(2r)}(\tilde{z}) = -(2\beta K)^{2r} c_\beta^{(2r)}(\tz)$.  

In Theorem 6.3 in \cite{EllOttTou} 
the types of the elements of $\ebetak$ are determined for all $\beta > 0$
and $K > 0$.
In our study of scaling limits of $S_n/n^{1-\gamma}$ in the present paper, 
we focus on $(\beta,K) \in A \cup B \cup C$,
for which $\ebetak = \{0\}$ [Thm.\ \ref{thm:secondorder}(a)].
Although the conclusion in \cite{EllOttTou} that for $(\beta,K) \in B$ the type of 0 equals 2
is correct, the formula for $G^{(4)}_{\beta,K}(0)$ given in (6.6) in that paper has a small error.
The correct formula for $G^{(4)}_{\beta,K}(0)$ is given in (\ref{eqn:G4}) with
$(\bn,\kn) = (\beta,K)$.  

\begin{thm}
\label{thm:secondordertype}
For all $(\beta,K) \in A \cup B \cup C$, $\ebetak = \{0\}$.

{\em (a)} For all $(\beta,K) \in A$, $\tz = 0$ has type $r = 1$.

{\em (b)} For all $(\beta,K_c(\beta)) \in B$, $\tz = 0$ has type $r = 2$.

{\em (c)} For $(\beta,K) = (\beta_c,K_c(\beta_c)) \in C$, 
$\tz = 0$ has type $r = 3$.
\end{thm}

For all other values of $\beta > 0$ and $K > 0$ not considered in
Theorem \ref{thm:secondordertype}, the elements of $\ebetak$ all
have type $r = 1$.  This includes the values $0 < \beta \leq \beta_c$ and $K > K_c(\beta)$
[Thm.\ \ref{thm:secondorder}(b)] and the values $\beta > \beta_c$, $K > 0$
[Thm.\ \ref{thm:firstorder}].  In these two cases the fact that the elements
of $\ebetak$ all have type $r = 1$ is proved in \cite{EllOttTou}
in part (c) of Theorem 6.3 and in Theorem 6.4.

We now point out how to prove the breakdown of the law of large numbers stated in (\ref{eqn:largebetakcbeta}),
which holds for $\beta > \beta_c$ and $K = K_c(\beta)$.  In this case, 
$\mathcal{E}_{\beta,K_c(\beta)} = \{0,\pm z(\beta,K)\}$.  Since each of the elements
of $\mathcal{E}_{\beta,K_c(\beta)}$ has type $r = 1$, the limit in (\ref{eqn:largebetakcbeta})
is proved exactly as in part (c) of Theorem 2.3 in \cite{EllWan}, which treats
the breakdown of the law of large numbers for the Curie-Weiss-Potts model at $\beta
= \beta_c$.  In (\ref{eqn:largebetakcbeta}), 
\be
\label{eqn:lambdakappa}
\lambda_0 = \frac{\kappa_0}{\kappa_0 + 2\kappa_1} \ \mbox{ and } \ 
\lambda_1 = \frac{\kappa_1}{\kappa_0 + 2\kappa_1},
\ee
where $\kappa_0 = [G^{(2)}_{\beta,K_c(\beta)}(0)]^{-1/2}$ and 
$\kappa_1 = [G^{(2)}_{\beta,K_c(\beta)}(z(\beta,K_c(\beta)))]^{-1/2}$.

We return to Lemma \ref{lem:G} and in particular 
to (\ref{eqn:limit})--(\ref{eqn:limit2}), which express the scaling limit of $S_n/n^{1-\gamma}$
in terms of the function $nG_{\beta_n,K_n}(x/n^\gamma)$.  Using the information 
about the three different types of the global minimum point of $G_{\beta,K}$ at 0
for $(\beta,K) \in A$, $(\beta,K) \in B$, and $(\beta,K) \in C$, 
we now indicate the three different forms of the Taylor
expansion of $nG_{\beta_n,K_n}(x/n^\gamma)$ needed to deduce the scaling limits
of $S_n/n^{1-\gamma}$.  These involve the quantities $G_{\beta_n,K_n}^{(2)}(0)$, 
$G_{\beta_n,K_n}^{(4)}(0)$, and $G_{\beta_n,K_n}^{(6)}(0)$, for the first two of which explicit formulas  
in terms of $\beta_n$ and $K_n$ are given.  As we will see in later sections,
these formulas will guide us into how we should choose the sequences $(\beta_n,K_n)$ so that
all the different scaling limits of $S_n/n^{1-\gamma}$ appear.
Since $G_{\beta_n,K_n}$ is symmetric around 0, all odd-order derivatives
of this function evaluated at 0 vanish; in addition, $G_{\beta_n,K_n}(0) = 0$.

In order to state part (d) of the theorem, we define for $\beta > 0$
\be
\label{eqn:kbeta}
K(\beta) = \frac{1}{2 c_\beta''(0)} = \frac{e^\beta + 2}{4\beta}.
\ee
For $0 < \beta \leq \bc$ this function coincides with the function $K_c(\beta)$ defined
in (\ref{eqn:kcbeta}), while for $\beta > \bc$, $K(\beta) > K_c(\beta)$ \cite[Thm.\ 3.8]{EllOttTou}.
Thus for $(\beta,K) \in B$ we have $K = K_c(\beta) = K(\beta)$ while for $(\beta,K) \in C$
we have $\beta = \bc$ and $K = \kcbc = K(\bc)$. 

\begin{thm}
\label{thm:taylor}
Let $(\beta_n,K_n)$ be any positive bounded sequence and $\gamma$ any positive
number. The following conclusions hold.

{\em (a)}  Assume that $(\beta_n,K_n) \goto (\beta,K) \in A$.  
Then the type of $0 \in \ebetak$ equals $1$.  In addition, for any 
$R > 0$ and for all $x \in \R$ satisfying $|x/n^\gamma| < R$ there exists 
$\xi = \xi(\xng)
\in [-\xng,\xng]$ such that
\be
\label{eqn:TaylorG1}
nG_{\beta_n, K_n}(x/n^\gamma) = 
\frac{1}{n^{2\gamma-1}} \frac{G_{\beta_n, K_n}^{(2)}(0)}{2!} x^2 + 
\frac{1}{n^{3\gamma - 1}}A_n(\xi(\xng))  x^3.
\ee
The error terms $A_n(\xi(\xng))$ are uniformly bounded over $n \in \N$ and 
$x \in (-\rng,\rng)$.
Furthermore, as $n \goto \infty$, $G_{\beta_n, K_n}^{(2)}(0) \goto G_{\beta, K}^{(2)}(0) > 0$.

{\em (b)} Assume that $(\beta_n,K_n) \goto (\beta,K_c(\beta)) \in B$.  Then the type of $0 \in 
\mathcal{E}_{\beta,K_c(\beta)}$ is $2$.  In addition, for any
$R > 0$ and for all $x \in \R$ satisfying $|x/n^\gamma| < R$ there exists $\xi
= \xixng \in [-\xng,\xng]$ such that 
\be
\label{eqn:TaylorG2}
nG_{\beta_n, K_n}(x/n^\gamma) = 
\frac{1}{n^{2\gamma-1}}\frac{G_{\beta_n, K_n}^{(2)}(0)}{2!} x^2 + 
\frac{1}{n^{4\gamma-1}}\frac{G_{\beta_n, K_n}^{(4)}(0)}{4!} x^4 + 
\frac{1}{n^{5\gamma - 1}}B_n(\xixng)  x^5.
\ee
The error terms $B_n(\xixng)$ are uniformly bounded over $n \in \N$ and 
$x \in (-\rng,\rng)$.
Furthermore, as $n \goto \infty$, 
$G_{\beta_n, K_n}^{(2)}(0) \goto G_{\beta, K_c(\beta)}^{(2)}(0) = 0$ while
$G_{\beta_n, K_n}^{(4)}(0) \goto G_{\beta, K_c(\beta)}^{(4)}(0) > 0$.

{\em (c)} Assume that $(\beta_n,K_n) \goto (\beta_c,K_c(\beta_c))$.  
Then the type of $0 \in \mathcal{E}_{\beta_c,K_c(\beta_c)}$ is $3$.
In addition, for any
$R > 0$ and for all $x \in \R$ satisfying $|x/n^\gamma| < R$
there exists $\xi = \xixng \in [-\xng,\xng]$ such that
\bea
\lefteqn{
\label{eqn:TaylorG3}
nG_{\beta_n, K_n}(x/n^\gamma) = } \\
\nonumber
&& \frac{1}{n^{2\gamma-1}}\frac{G_{\beta_n, K_n}^{(2)}(0)}{2!} x^2 + 
\frac{1}{n^{4\gamma-1}}\frac{G_{\beta_n, K_n}^{(4)}(0)}{4!} x^4 + 
\frac{1}{n^{6\gamma-1}}\frac{G_{\beta_n, K_n}^{(6)}(0)}{6!} x^6 + 
\frac{1}{n^{7\gamma - 1}} C_n(\xixng)  x^7.
\eea
The error terms $C_n(\xixng)$ are uniformly bounded over $n \in \N$ and 
$x \in (-\rng,\rng)$.
Furthermore, as $n \goto \infty$, $G_{\beta_n, K_n}^{(2)}(0) \goto G_{\beta_c,K_c(\beta_c)}^{(2)}(0) = 0$ and
$G_{\beta_n, K_n}^{(4)}(0) \goto G_{\beta_c,K_c(\beta_c)}^{(4)}(0) = 0$ while
$G_{\beta_n, K_n}^{(6)}(0) \goto G_{\beta_c,K_c(\beta_c)}^{(6)}(0) = 2 \cdot 3^4$.

{\em (d)} For $\beta > 0$ we define $K(\beta)$
in {\em (\ref{eqn:kbeta})}.  Then in {\em (\ref{eqn:TaylorG1})}--{\em (\ref{eqn:TaylorG3})}
\be
\label{eqn:G2}
G_{\beta_n, K_n}^{(2)}(0)=\frac{2\beta_n K_n  (e^{\beta_n} + 2 - 4\beta_n K_n)}{e^{\beta_n} + 2}
= \frac{2\beta_n K_n [K(\beta_n) - K_n]}{K(\beta_n)}
\ee
and
\be
\label{eqn:G4}
G_{\beta_n, K_n}^{(4)}(0)= \frac{2 (2\beta_n K_n)^4 (4-e^{\beta_n})  }{(e^{\beta_n} + 2)^2}.
\ee
\end{thm}

\noi
{\bf Proof.}  In parts (a), (b), and (c) the type of the global minimum point at 0
is specified in Theorem \ref{thm:secondordertype}.
The formulas for $G_{\beta_n, K_n}^{(2)}(0)$ and $G_{\beta_n, K_n}^{(4)}(0)$
in part (d) follow
from an explicit calculation of the derivatives and from the formula for $K(\beta)$
given in (\ref{eqn:kbeta}).  In addition, one evaluates
the limits of the Taylor coefficients given in the last sentence of each
part (a), (b), and (c)
using the continuity of the derivatives $G^{(2j)}_{\beta,K}(0)$ with respect to $\beta$ and 
$K$ and the fact that the type of the global minimum point of $G_{\beta,K}$ at 0 is, respectively,
$r = 1$, $r = 2$, and $r = 3$.

We now prove the form of the Taylor expansion given in part (c);
the forms of the Taylor expansions given in
parts (a) and (b) are proved similarly.  By Taylor's Theorem, for any 
$R > 0$ and for all $u \in \R$ satisfying $|u| < R$ there exists $\xi
= \xi(u) \in [-u,u]$ such that
\be
\label{eqn:TaylorG3u}
G_{\beta_n, K_n}(u) = \frac{G_{\beta_n, K_n}^{(2)}(0)}{2!} u^2 + 
\frac{G_{\beta_n, K_n}^{(4)}(0)}{4!} u^4 + 
\frac{G_{\beta_n, K_n}^{(6)}(0)}{6!} u^6 + 
C_n(\xi(u)) u^7,
\ee
where $C_n(\xi(u)) = G^{(7)}_{\beta_n,K_n}(\xi(u))/7!$.  
Because the sequence $(\beta_n,K_n)$ is positive and bounded, 
there exists $b \in (0,\infty)$ such that $0 < \beta_n \leq b$ and $0 < K_n \leq b$ for all $n$.  
As a continuous function 
of $(\beta,K,x)$ on the compact set $[0,b] \times [0,b] \times [-R,R]$,
$G^{(7)}_{\beta,K}(x)$ is uniformly bounded.  It follows that 
$G^{(7)}_{\beta_n,K_n}(\xi(u))$, and thus $C_n(\xi(u))$, are uniformly bounded over $n \in \N$ 
and $u \in (-R,R)$.  Multiplying both sides of (\ref{eqn:TaylorG3u}) by $n$ 
and substituting $u = x/n^\gamma$ yields part (c).  \ \ink

\skp

This completes our preliminary discussion of the Taylor expansions of 
$nG_{\beta_n, K_n}(x/n^\gamma)$ as they are needed to deduce the scaling limits of 
$S_n/n^{1-\gamma}$ via Lemma \ref{lem:G}.
In order to finalize our analysis of these scaling limits,
we will have to prove that the
contributions to the integrals in (\ref{eqn:limit}) and (\ref{eqn:limit2}) 
by $x \in \R$ satisfying $|x/n^\gamma| \geq R$ converge to 0 as $n \goto \infty$.
In part (c) of the next lemma we prove that the convergence to 0 is
exponentially fast.  The technical hypothesis in part (c) is satisfied in 
each of the theorems that proves the scaling limits [Thms.\ \ref{thm:betakina},
\ref{thm:betakinb}, \ref{thm:betakinc}].  In part (d) of the next lemma 
we prove the exponentially fast convergence to 0 of a related integral
that arises in the proof of the MDPs.  As we verify in
the proof of Theorem \ref{thm:mdpb}, the technical hypothesis in part (d)
is satisfied in that setting.  The estimates in parts (c) and (d) are 
consequences of the LDP proved in part (b), which in turn follows 
from part (a) and the representation formula in Lemma \ref{lem:G}. 
\iffalse
Lemma \ref{lem:awayfrom0} is the main technical innovation in
this paper because it allows us to handle any positive 
sequence $(\bn,\kn)$ converging to $(\beta,K) \in A \cup B \cup C$,
regardless of its direction of approach to $(\beta,K)$.
\fi

Lemma \ref{lem:awayfrom0} is the main technical innovation in this paper.
When adapted to the BEG model, the precursors of Lemma \ref{lem:awayfrom0}
given in Lemma 3.5 in \cite{EllNew} and Lemma 3.3 in \cite{EllWan}
are able to handle only positive sequences $(\bn,\kn)$ converging to $(\beta,K) 
\in A \cup B \cup C$ for which $G_{\bn,\kn}$ has a unique global minimum
point at 0.  In order to handle sequences $(\bn,\kn)$ for which $G_{\bn,\kn}$ has 
nonunique global minimum points, the modifications that would be necessary 
in the precursors of Lemma \ref{lem:awayfrom0} 
would introduce serious technical complications 
in the proofs of the scaling limits and the MDPs.
\iffalse
On the other hand, sequences $(\bn,\kn)$ naturally arise in
the scaling limits and the MDPs for which $G_{\bn,\kn}$ has 
nonunique global minimum points.  The modifications that would be necessary 
in the precursors of Lemma \ref{lem:awayfrom0} to handle such sequences
would introduce complications in the Taylor expansions of 
$n G_{\bn,\kn}(x/n^\gamma)$ that are not covered by Theorem
\ref{thm:taylor} and that would result in nearly insurmountable
technical obstacles in the proofs of the scaling limits and the MDPs.
\fi
By allowing us to handle any positive sequence
$(\bn,\kn)$ converging to $(\beta,K) \in A \cup B \cup C$,
parts (c) and (d) of Lemma \ref{lem:awayfrom0} 
are universal bounds that enable us to avoid these technical
complications altogether.

\begin{lemma}
\label{lem:awayfrom0}
Let $(\bn,\kn)$ be any positive sequence converging
to $(\beta,K) \in A \cup B \cup C$ and as in Lemma {\em \ref{lem:G}}, let $W_n$
be a sequence of $N(0,(2\beta_n K_n)^{-1})$ random variables defined 
on a probability space $(\Omega,\mathcal{F},Q)$. The following conclusions hold.

{\em (a)} There exist $a_1 > 0$ and $a_2 > 0$ 
such that for all $n \in \N$ and
all $x \in \R$, $\gn(x) \geq a_1 (|x|-1)^2 - a_2$.

{\em (b)} With respect to $\pnbnkn \times Q$, $S_n/n + W_n/n^{1/2}$ satisfies the LDP on $\R$ 
with exponential speed $n$ and rate function $G_{\beta,K}$. 

{\em (c)} Given $\gamma > 0$ and $R > 0$, we define
\be
\label{eqn:yn}
y_n = \int_{\{|x| < Rn^\gamma\}} \exp[-n \gn(x/n^\gamma)] \, dx.
\ee
\iffalse
\[
\lim_{n \goto \infty} n \gn(x/n^\gamma) = G(x) \ \mbox{ for all } x \in \R,
\]
\[
n \gn(x/n^\gamma) \geq H(x) \ \mbox{ for all } x \in \R \mbox{ satisfying } |x/n^\gamma| < R,
\]
and $\int_{\R} \exp[-H(x)] dx < \infty$.
\fi
If the sequence $y_n$ is bounded, then
there exists $a_3 > 0$ and $a_4 > 0$ such that for all sufficiently large $n$
\[
\int_{\{|x| \geq R n^\gamma\}} \exp[-n \, G_{\bn,\kn}(x/n^\gamma)] \, dx 
\leq a_3 \exp({-n a_4}) \goto 0.
\]

{\em (d)}  Assume that there exist $\gamma > 0$, $R > 0$, $u \in (0,1)$, $a_5 > 0$, 
and $a_6 \in \R$ such that for all sufficiently large $n$
\[
y_n = \int_{\{|x| < Rn^\gamma\}} \exp[-n \gn(x/n^\gamma)] \, dx
\leq a_5 \exp(n^u a_6).
\]
Then there exists $a_7 > 0$ such that for all sufficiently large $n$
\[
\int_{\{|x| \geq R n^\gamma\}} \exp[-n \, G_{\bn,\kn}(x/n^\gamma)] \, dx 
\leq 2 a_5 \exp({-n a_7}) \goto 0.
\]
\end{lemma}

\noi
{\bf Proof.}  (a) Because the sequence $(\bn,\kn)$ is bounded and 
remains a positive distance from the origin and the 
coordinate axes, there exist $0 < b_1 < b_2 < \infty$
such that $b_1 \leq \beta_n \leq b_2$ and $b_1 \leq K_n \leq b_2$ for all $n \in \N$.
The conclusion of part (a) is a consequence of the elementary inequalities
\beas
G_{\beta_n,K_n}(x) & = & \beta_n K_n x^2 - c_{\beta_n}(2\beta_n K_n x) \\
& \geq & \beta_n K_n x^2 - 2\beta_n K_n |x| - \log 4 \: \geq \:
b_1^2 (|x|-1)^2 - b_2^2 - \log 4.
\eeas

(b) We prove that for any bounded, continuous function $\psi$
\be
\label{eqn:wantthislaplace}
\lim_{n \goto \infty} \frac{1}{n} \log 
\int_{\Lambda^n \times \Omega} \exp\!\left[n \psi\!\left(\frac{S_n}{n} + \frac{W_n}{n^{1/2}}\right)\right] 
d(\pnbnkn \times Q) = \sup_{x \in \R}\{\psi(x) - G_{\beta,K}(x)\}.
\ee
This Laplace principle implies the LDP stated in part (b) \cite[Thm.\ 1.2.3]{DupEll}.
$G_{\beta,K}$ is continuous, and by part (a) of this lemma applied to 
the constant sequence $(\bn,\kn) = (\beta,K)$, this
function has compact level sets.  Since $(\beta,K) \in A \cup B \cup C$, $\gbk$ has a unique
global minimum point at 0, and therefore 
$\inf_{x \in \R} G_{\beta,K}(x) = 0$.  It follows that $G_{\beta,K}$ is a rate function.
We now use Lemma \ref{lem:G} with $\gamma = 0$ to rewrite the integral in the last display as
\bea
\label{eqn:laplacelaplace}
\lefteqn{
\int_{\Lambda^n \times \Omega} \exp\!\left[n \psi\!\left(\frac{S_n}{n} + \frac{W_n}{n^{1/2}}
 \right)\right] d(P_{n,\beta_n,K_n} \times Q)} 
\\ \nonumber && = \frac{1}{\int_{\R} \exp [-nG_{\beta_n,K_n}(x)] \, dx} \cdot
\int_{\R} \exp[n\{\psi(x) - G_{\beta_n, K_n}(x)\}] \, dx.
\eea

By part (a) of this lemma, there exist 
$M > 0$ and $a_8 > 0$ having the following three properties: 
\begin{enumerate}
\item $G_{\bn,\kn}(x) \geq a_8 x^2$ for
all $n \in \N$ and all $x \in \R$ satisfying $|x| \geq M$. 
\item The supremum of $\psi - \gbk$
on $\R$ is attained on the interval $[-M,M]$. 
\item Let $\Delta = \sup_{x \in \R}\{\psi(x) - G_{\beta,K}(x)\}$. Then
$\|\psi\|_\infty - a_8 M^2 \leq - |\Delta| - 1$. 
\end{enumerate}
Since $\gn$ converges uniformly to $\gbk$ on $[-M,M]$, we have for any $\delta > 0$ and all
sufficiently large $n$
\beas
\lefteqn{
\exp(-n \delta) \int_{\{|x| \leq M\}} \exp[n \{\psi(x) - \gbk(x)\}] \, dx }
\\ \nonumber && \leq \int_{\{|x| \leq M\}} \exp[n\{\psi(x) - \gn(x)\}] \, dx 
\\ \nonumber && \leq \exp(n \delta) \int_{\{|x| \leq M\}} \exp[n\{\psi(x) - \gbk(x)\}] \, dx.
\eeas
In addition, by items 1 and 3
\beas
\lefteqn{
\int_{\{|x| > M\}} \exp[n\{\psi(x) - \gn(x)\}] \, dx } \\
& & \leq 
\exp[n \|\psi\|_\infty] \int_{\{|x| > M\}} \exp[-n a_8 x^2] \, dx \\
& & \leq  \frac{1}{n M a_8} \exp[n \|\psi\|_\infty - n a_8 M^2] \\
& & \leq   \frac{1}{n M a_8} \exp[-n (|\Delta| + 1)].
\eeas

We now put these estimates together. For all sufficiently large $n$ we have
\beas
\lefteqn{
\exp(-n \delta) \int_{\{|x| \leq M\}} \exp[n \{\psi(x) - \gbk(x)\}] \, dx }
\\ \nonumber && \leq \int_{\R} \exp[n\{\psi(x) - \gn(x)\}] \, dx 
\\ \nonumber && \leq \exp(n \delta) \int_{\{|x| \leq M\}} \exp[n\{\psi(x) - \gbk(x)\}] \, dx
	+ \frac{1}{n M a_8} \exp[-n (|\Delta| + 1)].
\eeas
Since by item 2
\beas
\lefteqn{
\lim_{n \goto \infty} \frac{1}{n} \log 
\int_{\{|x| \leq M\}} \exp[n\{\psi(x) - \gbk(x)\}] \, dx}\\
&& = \sup_{\{|x| \leq M\}} \{\psi(x) - \gbk(x)\} = \sup_{x \in \R} \{\psi(x) - \gbk(x)\},
\eeas
we see that
\beas
\lefteqn{\sup_{x \in \R} \{\psi(x) - \gbk(x)\} - \delta } \\
&& \leq \liminf_{n \goto \infty}
\frac{1}{n} \log \int_{\R} \exp[n\{\psi(x) - G_{\beta_n, K_n}(x)\}] \, dx \\
&& \leq \limsup_{n \goto \infty}
\frac{1}{n} \log \int_{\R} \exp[n\{\psi(x) - G_{\beta_n, K_n}(x)\}] \, dx \\
&& \leq \sup_{x \in \R} \{\psi(x) - \gbk(x)\} + \delta,
\eeas
and since $\delta > 0$ is arbitrary, it follows that 
\[
\lim_{n \goto \infty}
\frac{1}{n} \log \int_{\R} \exp[n\{\psi(x) - G_{\beta_n, K_n}(x)\}] \, dx = 
\sup_{x \in \R} \{\psi(x) - \gbk(x)\}.
\]
We combine this limit with the same limit for $\psi = 0$ and use (\ref{eqn:laplacelaplace})
together with the fact that $\inf_{x \in \R} G_{\beta,K}(x) = \gbk(0) = 0$,
concluding that 
\beas
\lefteqn{
\lim_{n \goto \infty} \frac{1}{n} \log 
\int_{\Lambda^n \times \Omega} \exp\!\left[n \psi\!\left(\frac{S_n}{n} + \frac{W_n}{n^{1/2}}\right)\right] d(\pnbnkn \times Q)} \\
&& = \sup_{x \in \R}\{\psi(x) - G_{\beta,K}(x)\} - \inf_{x \in R}\gbk(x) 
= \sup_{x \in \R}\{\psi(x) - G_{\beta,K}(x)\}.
\eeas
This is the Laplace principle (\ref{eqn:wantthislaplace}).  
The proof of part (b) is complete.

(c)  Since $\gbk$ has a unique global minimum point at 0, the LDP proved in part (b) implies 
the existence of $a_{9} > 0$ such that for all $n \in \N$
\be
\label{eqn:kenb}
\pnbnkn \times Q \!\left\{\frac{S_n}{n} + \frac{W_n}{n^{1/2}}
\not \in (-R,R) \right\} \leq \exp({-n a_{9}}).
\ee
Using Lemma \ref{lem:G}, we rewrite the probability in the last display as
\bea
\label{eqn:znyn}
\lefteqn{
\pnbnkn \times Q \!\left\{\frac{S_n}{n} + \frac{W_n}{n^{1/2}} 
\not \in (-R,R)\right\} }  \\ \nonumber
&& = \pnbnkn \times Q \!\left\{\frac{S_n}{n^{1-\gamma}} + \frac{W_n}{n^{1/2 - \gamma}}
\not \in (-R n^\gamma, R n^\gamma) \right\} \\ \nonumber
&& = \frac{1}{\int_{\R} \exp [-nG_{\beta_n,K_n}(x/n^{\gamma})] \, dx} \cdot
\int_{\{|x| \geq R n^\gamma\}} \exp[-nG_{\beta_n, K_n}(x/n^{\gamma})] \, dx \\ \nonumber
&& = \frac{z_n}{y_n + z_n},
\eea
where $y_n$ is defined in (\ref{eqn:yn}) and 
\[
z_n = \int_{\{|x| \geq Rn^\gamma\}} \exp[-n \, G_{\bn,\kn}(x/n^\gamma)] \, dx.
\]
Since by hypothesis the sequence $y_n$ is bounded, there exists $y > 0$ such that
$y_n \leq y$ for all $n$. 
It follows from (\ref{eqn:kenb}) and (\ref{eqn:znyn}) that for all sufficiently large $n$
\[
\ts \frac{1}{2} z_n \leq z_n (1 - \exp({-n a_{9}})) \leq y_n \exp({-n a_{9}}) 
\leq y \exp({-n a_{9}})
\]
and thus for all sufficiently large $n$, 
$z_n \leq 2y \exp({-n a_{9}})$.  This
completes the proof of part (c). 

(d) Exactly as in the proof of part (c), we have for all sufficiently large $n$
\[
\ts \frac{1}{2} z_n \leq z_n (1 - \exp({-n a_{9}})) \leq y_n \exp({-n a_{9}}).
\]
Since by hypothesis $y_n \leq a_5 \exp(n^u a_6)$ and $u \in (0,1)$, it follows that for 
all sufficiently large $n$
\[
z_n \leq 2 a_5 \exp(-n a_{9} + n^u a_6) \leq 2 a_{5} \exp(-n a_{9}/2).
\]
This completes the proof of part (d).   \ \ink 

\skp

In the next section we begin our analysis of the scaling limits of $S_n/n^{1-\gamma}$
in the simplest case by considering $(\beta_n,K_n) \goto (\beta,K) \in A$.  In the two sections
following the next one, we will uncover
a wider variety of scaling limits by considering sequences $(\beta_n,K_n)$ converging
to $(\beta,K_c(\beta)) \in B$ and to $(\beta_c,K_c(\beta_c)) \in C$.

\section{1 Scaling Limit for \boldmath $(\beta_n, K_n) \!\goto \!(\beta,K) \in A$ \unboldmath}
\beginsec
\label{section:scalinga}

In this short section, we deduce the unique scaling limit of $S_n/n^{1-\gamma}$
when $(\beta_n,K_n)$ is any positive sequence converging
to $(\beta,K) \in A$.  The unique global minimum point of $\gbk$ at 0 
has type $r = 1$ [Thm.\ \ref{thm:secondordertype}(a)]. As the next theorem shows, the scaling limit 
with respect to $P_{n,\beta_n,K_n}$ has the form of a central limit-type
theorem that is independent of the particular sequence chosen. 
In addition, the only value of $\gamma$ for which $S_n/n^{1-\gamma}$ has a nontrivial limit
is $\gamma = {1}/{2}$.  
We are including this scaling limit in order to highlight the much more complicated behavior
of the scaling limits of $S_n/n^{1-\gamma}$ in the subsequent two sections, in which $(\beta_n,K_n)
\goto (\beta,K_c(\beta)) \in B$ and $(\beta_n,K_n) \goto (\bc,\kcbc) \in C$ 
and in which different forms of the limit can be obtained by choosing different sequences.

The following theorem, stated for $0 < \beta \leq \beta_c$ and $0 < K < K_c(\beta)$,
is also valid for $\beta > \beta_c$ and $0 < K < K_c(\beta)$, and the proof is essentially the same.
The key observation is that for $\beta > \bc$, we have $K(\beta) = (e^\beta + 2)/(4\beta)
> K_c(\beta)$ \cite[Thm.\ 3.8]{EllOttTou}.  Hence if $K < K_c(\beta)$, then
also $K < K(\beta)$ and thus $G_{\beta,K}^{(2)}(0)$ in (\ref{eqn:gbetak20}) is positive.

\begin{thm}
\label{thm:betakina}
Let $(\beta_n,K_n)$ be an arbitrary positive sequence that converges
to $(\beta,K) \in A$; thus $\beta$ and $K$ satisfy
$0 < \beta \leq \beta_c$ and $0 < K < K_c(\beta)$.  Then 
\[
P_{n,\beta_n,K_n}\{{S_n}/{n^{1/2}} \in dx\} \Longrightarrow 
\exp(-c_2 x^2) \, dx,
\]
where $c_2 > 0$ is defined by
\be
\label{eqn:c2betak}
c_2 = \frac{1}{2} \cdot \frac{1}{[G_{\beta,K}^{(2)}(0)]^{-1} - [2\beta K]^{-1}} = \beta[K(\beta) - K].
\ee
Thus the limit is independent of the particular sequence $(\bn,\kn)$ that is chosen.
\end{thm}

\noi
{\bf Proof.}  We use the Taylor expansion in 
part (a) of Theorem \ref{thm:taylor} with $\gamma = {1}/{2}$.
By continuity, $G_{\beta_n, K_n}^{(2)}(0)$ given in (\ref{eqn:G2}) converges
to 
\be
\label{eqn:gbetak20}
G_{\beta,K}^{(2)}(0) = \frac{2\beta K [K(\beta) - K]}{K(\beta)},
\ee
which is positive since $0 < K < K_c(\beta) = K(\beta)$.  
For any $R > 0$ the error terms $A_n(x/n^{1/2})$ in the Taylor expansion are uniformly bounded
over $n \in \N$ and $x \in (-R n^{1/2},R n^{1/2})$.  It follows that for all $x \in \R$ 
\[
\lim_{n \goto \infty} 
nG_{\beta_n, K_n}(x/n^{1/2}) = \ts \frac{1}{2} G_{\beta,K}^{(2)}(0)x^2
\]
and that $R > 0$ can be chosen to be sufficiently small so that for all
sufficiently large $n$ and all $x \in \R$ satisfying $|x/n^{1/2}| < R$
\[
nG_{\beta_n, K_n}(x/n^{1/2}) \geq \ts \frac{1}{4} G_{\beta,K}^{(2)}(0) x^2.
\] 
Since $\int_{\R} \exp[-G_{\beta,K}^{(2)}(0) x^2/4] dx < \infty$,
the dominated convergence theorem implies
that for any bounded, continuous function $f$ 
\[
\lim_{n \goto \infty} \int_{\{|x| < R n^{1/2}\}} f(x) \, \exp[-nG_{\beta_n, K_n}(x/n^{1/2})] \, dx
= \int_{\R} f(x) \, \exp[- G^{(2)}_{\beta,K}(0) x^2/2] \, dx.
\]
The existence of this limit implies that the sequence 
$y_n = \int_{\{|x| < R n^{1/2}\}} \exp[-nG_{\beta_n, K_n}(x/n^{1/2})] dx$ is bounded.
Hence, combining this limit with part (c) of Lemma \ref{lem:awayfrom0} yields
\[
\lim_{n \goto \infty} \int_{\R} f(x) \, \exp[-nG_{\beta_n, K_n}(x/n^{1/2})] \, dx
= \int_{\R} f(x) \, \exp[- G^{(2)}_{\beta,K}(0) x^2/2] \, dx.
\]
We now augment this limit with the same limit for $f = 1$ and use (\ref{eqn:limit2}) to obtain
\beas
\lefteqn{
\lim_{n \goto \infty}
\int_{\Lambda^n \times \Omega} f(S_n/n^{1/2} + W_n) \, d(P_{n,\beta,K_c(\beta)} \times Q)} \\
&& = \frac{1}{\int_{\R} \exp[- G^{(2)}_{\beta,K}(0) x^2/2] \, dx} 
\cdot \int_{\R} f(x) \exp[- G^{(2)}_{\beta,K}(0) x^2/2] \, dx.
\eeas
We omit the straightforward argument using characteristic functions 
that enables one to deduce from the last display
that
\[
\pnbnkn\{S_n/n^{1/2} \in dx\} \Longrightarrow \exp(- c_2 x^2) \, dx,
\]
where $c_2$ is given by the first equality in (\ref{eqn:c2betak}).  A similar argument involving moment generating
functions is given on pages 70--71 of \cite{EllWan}.  
The positivity of $c_2$ and the second formula for $c_2$ given in (\ref{eqn:c2betak})
follow from (\ref{eqn:gbetak20}).
This completes the proof of the theorem. \ \ink

\skp
In Theorem \ref{thm:mdpa} we prove an MDP for $S_n/n^{1-\gamma}$
that is related to the scaling limit proved in Theorem \ref{thm:betakina}.
As in the latter theorem, the form of the MDP is independent of the particular
sequence $(\bn,\kn)$ converging to $(\beta,K) \in A$.
In the next section we see the first example of scaling limits 
for $S_n/n^{1 - \gamma}$ where different forms of the limit 
can be obtained by choosing different sequences $(\beta_n,K_n)
\goto (\beta,K_c(\beta)) \in B$.

\section{4 Scaling Limits for \boldmath 
$(\beta_n, K_n) \! \goto \! (\beta,K_c(\beta)) \in B$ \unboldmath}
\beginsec
\label{section:scalingb}

In this section we determine the scaling limits of $S_n/n^{1-\gamma}$ with
respect to $P_{n,\beta_n,K_n}$, where $(\beta_n,K_n)$ is an appropriate positive sequence
converging to $(\beta,K_c(\beta)) \in B$ and $\gamma \in (0,{1}/{2})$ is appropriately
chosen.  We recall that $B$ is the curve of second-order 
points for the BEG model.  For any $(\beta,K) \in B$, we have $0 < \beta < \beta_c = \log 4$ and 
\[
K = K_c(\beta) = \frac{1}{2\beta c_\beta''(0)} = \frac{e^\beta + 2}{4\beta}.
\]

The scaling limits that we obtain involve limiting densities proportional to $\exp[-G(x)]$,
where $G$ takes one of the 4 forms of an even polynomial of degree 4 or 2
satisfying $G(0) = 0$ and $G(x) \goto \infty$ as $|x| \goto \infty$.  There are 3 such $G$'s 
of degree 4; namely, $G(x) = c_4 x^4$, where $c_4 > 0$
and $G(x) = k \beta x^2 + c_4 x^4$, where $c_4 > 0$ and either $k > 0$ or $k < 0$.  There is also 1 such
$G$ of degree 2; namely, $G(x) = k \beta x^2$, where $k > 0$. 
These 4 cases are all obtained in Theorem \ref{thm:betakinb};
the forms of the limits depend on the choice of $K_n \goto K_c(\beta)$ but are independent
of the choice of $\beta_n \goto \beta$. 

In order to determine the forms of the scaling limits of $S_n/n^{1-\gamma}$ with
respect to $P_{n,\beta_n,K_n}$, we start by recalling the Taylor expansion given 
in part (b) of Theorem \ref{thm:taylor}.  For any $\gamma > 0$ and $R > 0$ and for all $x \in \R$
satisfying $|x/n^\gamma| < R$ there exists $\xi \in [-x/n^\gamma,x/n^\gamma]$ such that
\be
\label{eqn:TaylorG2again}
nG_{\beta_n, K_n}(x/n^\gamma) = 
\frac{1}{n^{2\gamma-1}}\frac{G_{\beta_n, K_n}^{(2)}(0)}{2!} x^2 + 
\frac{1}{n^{4\gamma-1}}\frac{G_{\beta_n, K_n}^{(4)}(0)}{4!} x^4 + 
\frac{1}{n^{5\gamma - 1}}B_n(\xi(x/n^\gamma))  x^5.
\ee
The error terms $B_n(\xi(x/n^\gamma))$ are uniformly bounded over $n \in \N$ and 
$x \in (-\rng,\rng)$.
According to part (b) of Theorem \ref{thm:secondordertype}, the unique global
minimum point of $G_{\beta,K_c(\beta)}$ at 0 has type 2.  Hence
by continuity, as $n \goto \infty$, 
\[
G_{\beta_n, K_n}^{(2)}(0) = 
\frac{2 \beta_n K_n [K(\beta_n) - K_n]}{K(\beta_n)} \goto G_{\beta, K_c(\beta)}^{(2)}(0) = 0
\]
while $G_{\beta_n, K_n}^{(4)}(0) \goto G_{\beta, K_c(\beta)}^{(4)}(0) > 0$.  We recall that in the last display
$K(\beta) = (e^\beta + 2)/(4\beta)$ for $\beta > 0$.

Fixing $\beta \in (0,\beta_c)$, we let $\beta_n$ be an arbitrary positive sequence that
converges to $\beta$, and we let $\theta$ be a positive number.
The key insight is to choose $K_n$ so that $G_{\beta_n, K_n}^{(2)}(0) \goto 0$ 
at a rate $1/n^\theta$, where $1/n^\theta$ 
counterbalances the term $1/n^{2\gamma-1}$ appearing in (\ref{eqn:TaylorG2again}).
Since $2 \beta_n K_n/K(\beta_n)$ has the positive limit $2\beta$ as $n \goto \infty$, 
we achieve this by choosing $k \not = 0$ and defining
\be
\label{eqn:choosekn}
K_n = K(\beta_n) - {k}/{n^\theta}.
\ee
Since $\bn \goto \beta$ and $K(\cdot)$ is continuous, 
it follows that $K_n \goto K(\beta) = K_c(\beta)$.  Hence
\[
G_{\beta_n, K_n}^{(2)}(0) = 
\frac{k}{n^\theta} \cdot \frac{2 \beta_n K_n}{K(\beta_n)} = \frac{k}{n^\theta} \cdot C_n^{(2)},
\ \mbox{ where } \ C_n^{(2)} > 0 \ \mbox{ and } \ C_n^{(2)} \goto 2\beta.
\]
With these choices (\ref{eqn:TaylorG2again}) becomes
\be
\label{eqn:5point2'}
nG_{\beta_n, K_n}(x/n^\gamma) = 
\frac{1}{n^{2\gamma+\theta-1}}\frac{k C_n^{(2)}}{2!} x^2 + 
\frac{1}{n^{4\gamma-1}}\frac{G_{\beta_n, K_n}^{(4)}(0)}{4!} x^4 + 
\frac{1}{n^{5\gamma - 1}}B_n(\xi(x/n^\gamma))  x^5.
\ee
As we will see in Theorem \ref{thm:betakinb}, 
the scaling limits depend on the value of $\gamma$ and on $K_n$
through the value of $\theta$, but are independent of the sequence $\beta_n \goto \beta$.

In the last display we assume that the coefficients multiplying $x^2$ 
and $x^4$ both appear
with nonnegative powers of $n$ and that at least one of these two coefficients
has $n$ to the power $0$.  Then in the limit $n \goto \infty$ any coefficient 
including the error term that has
a positive power of $n$ will vanish while any coefficient 
that has $n$ to the power $0$ will converge to a positive constant.  This preliminary 
analysis shows the possibility of multiple scaling limits for different
choices of $\gamma$ and $\theta$.  In order to confirm this possibility, we define
\[
v = \min\{2\gamma + \theta -1, 4\gamma -1\}
\]
and focus on the cases in which $v = 0$. 
As we will see in the final section of the paper, $v < 0$ corresponds 
to 4 different MDPs for $S_n/n^{1 - \gamma}$.  On the other hand, if $v > 0$, then one obtains
neither scaling limits nor MDPs.

\iffalse
If $\gamma \geq {1}/{2}$, then both terms in the definition of $v$
are positive.  Hence if $v = 0$, then in all cases we have $\gamma < {1}/{2}$, 
and so the term $W_n/n^{1/2 - \gamma}$ in (\ref{eqn:G})
does not contribute to the limit $n \goto \infty$.
Hence we can determine the scaling limits of $S_n/n^{1-\gamma}$
by using (\ref{eqn:limit}).
As we will see in the proof of Theorem \ref{thm:betakinb},
in all cases we have $\gamma \in [{1}/{4},{1}/{2})$. 
\fi

In the next theorem we show that $v = 0$ corresponds to 3 different
choices of $\gamma$ and $\theta$, which in turn correspond to 4 different
sequences $K_n$ in (\ref{eqn:choosekn}). The additional sequence arises
because when $x^2$ is not the highest order term in the scaling limit (cases 3--4),
$k$ can be chosen to be
either positive or negative.  As shown in Table 6.1 in part (b) 
of the theorem, for each of these 4 different sequences
we obtain 4 different scaling limits of $S_n/n^{1-\gamma}$.  
In case 1 we can also choose $k$ to be any real number;
this affects only the definition of the sequence $K_n$, not
the form of the scaling limit.  
\iffalse
In case 2 we obtain a Gaussian limit, but this
is for a different value of $\gamma$ from the value $\gamma = 1/2$ in Theorem
\ref{thm:betakina}, where we consider $(\bn,\kn) \goto (\beta,K) \in A$.
\fi

\begin{figure}[h]
\begin{center}
\epsfig{file=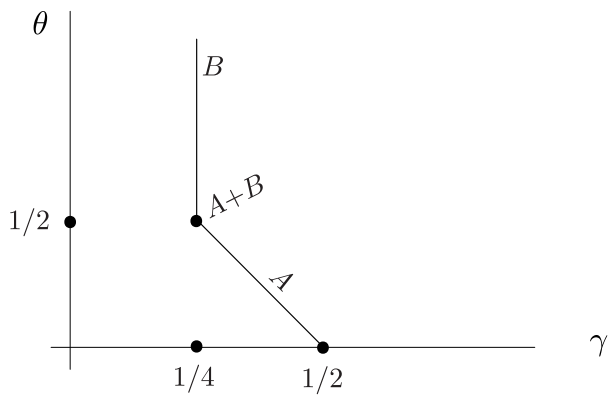,width=8cm}
\caption{\small Influence of $A$ and $B$ on scaling limits when $(\bn,\kn)
\goto (\beta,K_c(\beta)) \in B$}
\end{center}
\end{figure} 

The results of the theorem confirm one's intuition concerning the influence of the regions on the scaling limits.
Of the 4 cases, case 1 corresponds to the largest values of $\theta$ --- namely,
$\theta > {1}/{2}$ --- and thus the most rapid
convergence of $K_n \goto K_c(\beta)$. In this case only $B$ influences the form of the limiting
density, which is proportional to $\exp(-c_4 x^4)$; $c_4$ defined in (\ref{eqn:c4betak})
is positive since $e^\beta < e^{\beta_c} = 4$.
By contrast, case 2 corresponds to the smallest values of $\theta$ --- namely,
$\theta \in (0,{1}/{2})$ --- and thus the slowest
convergence of $K_n \goto K_c(\beta)$. In this case only $A$ influences the form of the limiting density,
which is proportional to $\exp(-\beta x^2)$; thus we have $S_n/n^{1 - \gamma}$ converging in distribution
to a normal random variable even though the non-classical scaling is given by $n^{1-\gamma}$,  
where $\gamma = (1-\theta)/2 \in ({1}/{4},{1}/{2})$.  Finally, cases 3 and 4 correspond to the critical
speed $\theta = {1}/{2}$.  In this case both $A$ and $B$ influence the form 
of the limiting density, which is proportional to $\exp(-k\beta x^2 - c_4 x^4)$ with $c_4 > 0$ and either $k > 0$ or $k < 0$.
In Figure 4 we indicate the subsets of the positive quadrant of the 
$\theta$-$\gamma$ plane leading to the 4 cases just discussed.  Using Table 5.1, one easily
checks that as $\theta$ increases through the critical value $1/2$, the values of $\gamma$ in
the scaling limit change continuously while the forms of the limiting densities change
discontinuously.

\begin{thm}
\label{thm:betakinb}
For fixed $\beta \in (0,\beta_c)$, let $\beta_n$ be an arbitrary positive
sequence that converges
to $\beta$.  Given $\theta > 0$ and $k \not = 0$, define
\[
K_n = K(\beta_n) - {k}/{n^\theta},
\]
where $K(\beta) = (e^\beta + 2)/(4\beta)$ for $\beta > 0$.
Then $(\bn,\kn) \goto (\beta,K_c(\beta)) \in B$.
Given $\gamma \in (0,1)$, we also define
\be
\label{eqn:Gforb}
G(x) = \delta{(v, 2\gamma+\theta-1)} k\beta x^2 +  \delta(v, 4\gamma-1) c_4 
x^4,
\ee
where $\delta(a,b)$ equals {\em 1} if $a = b$ and equals {\em 0} if $a \not = b$ 
and $c_4 > 0$ is given by
\be
\label{eqn:c4betak}
c_4 = \frac{G_{\beta,K_c(\beta)}^{(4)}(0)}{4!} =
\frac{2 [2\beta K_c(\beta)]^4 (4 - e^\beta)}{4!(e^\beta + 2)^2} 
= \frac{(e^\beta + 2)^2 (4 - e^\beta)}{2^3 \cdot 4!}.
\ee
The following conclusions hold.

{\em (a)}  Assume that $v = \min\{2\gamma + \theta -1, 4\gamma -1\}$ equals $0$.
Then 
\be
\label{eqn:scalingwithg}
\pnbnkn\!\left\{S_n/n^{1-\gamma} \in dx\right\} \Longrightarrow \exp[-G(x)] \, dx.
\ee

{\em (b)}  We have $v = 0$ if and only if one of the {\em 4} cases enumerated in Table {\em 6.1} holds.
Each of the {\em 4} cases corresponds to a set of values of $\theta$ and $\gamma$, to the influence of one or more
sets $B$ and $A$, 
and to a particular scaling limit in {\em (\ref{eqn:scalingwithg})}.  In 
case {\em 1} the choice of $k \in \R$ does not affect the form
of the scaling limit.
%
% \vspace{.1in}  
\begin{center}
\begin{tabular}{||c|l|l|l||} \hline \hline
{\em {\bf case}} & {\em {\bf values of}} $\theta$ & {\em {\bf values of}} \boldmath $\gamma$ \unboldmath
& {\em {\bf scaling limit of}} \boldmath $S_n/n^{1-\gamma}$ \unboldmath \\ \cline{1-1}
{\em {\bf influence}} & & & \\ \hline \hline
{\em 1} & $\theta > \frac{1}{2}$ & $\gamma = \frac{1}{4}$ &
% $S_n/n^{1-1/2} \Longrightarrow 
$\exp(-c_4 x^4) \, dx$ \\ \cline{1-1} 
$B$ & & & $c_4 > 0$, $k \in \R$\\ \hline \hline
{\em 2} & $\theta \in (0,\frac{1}{2})$ & 
$\gamma = \frac{1-\theta}{2} \in (\frac{1}{4},\frac{1}{2})$ &
% $S_n/n^{1-\gamma} \Longrightarrow 
$\exp(- k \beta x^2) \, dx $ \\ \cline{1-1}
$A$ & & & $k > 0$ \\ \hline \hline
{\em 3--4} & $\theta = \frac{1}{2}$ & $\gamma = \frac{1}{4}$ & 
% $S_n/n^{1-1/4} \Longrightarrow 
$\exp(-k\beta x^2 - c_4 x^4) \, dx$ \\ \cline{1-1}
$A + B$ & &  & $k \not = 0$ \\ \hline \hline
\end{tabular}

\vspace{.05in}
{\em Table 6.1}: {\small {\em Values of } $\theta$ {\em and} $\gamma$ {\em and scaling limits
in part (b) of Theorem \ref{thm:betakinb}}}
\end{center}
\end{thm}

\noi
{\bf Note.}  Let $\beta_n = \beta$ for all $n$.  The constant sequence $(\bn,\kn) = (\beta,K_c(\beta))$
for all $n$ corresponds to the choice $\theta = \infty$ in case 1.  As in the proof of case 1,
one shows that $P_{n,\beta,K_c(\beta)}\{S_n/n^{1-1/4} \in dx\} \Longrightarrow
\exp(-c_4 x^4) dx$.  This scaling limit was mentioned in (\ref{eqn:4thorder}).

\skp
\noi
{\bf Proof of Theorem \ref{thm:betakinb}.}  We first prove part (b) assuming part (a), and then we prove part (a).

(b)  $v = \min\{2\gamma + \theta -1, 4\gamma -1\}$
equals 0 if and only if each of the quantities in this minimum is nonnegative
and one or more of the quantities equals 0.
As (\ref{eqn:Gforb}) makes clear, $4\gamma - 1 = 0$ corresponds to the influence of $B$
and $2\gamma + \theta - 1 = 0$  to the influence of $A$.  
We have the following 4 mutually exclusive and exhaustive cases, which correspond to the 
4 cases in Table 6.1.
\begin{itemize}
\item {\bf Case 1: Influence of \boldmath $B$ \unboldmath alone.} 
$2 \gamma + \theta - 1 >0$, $4\gamma -1 = 0$, and $k \in \R$.  In this case 
$\gamma = {1}/{4}$ and $\theta > 1 - 2\gamma = {1}/{2}$, 
which corresponds to the second and third columns for case 1 in Table 6.1. 
\item {\bf Case 2: Influence of \boldmath $A$ \unboldmath alone.} 
$2\gamma + \theta - 1 =0$, $4 \gamma - 1 > 0$, and $k > 0$. In this case 
$\gamma > {1}/{4}$ and $\theta = 1 - 2\gamma < {1}/{2}$.
Since $\theta$ must be positive,
we have $\gamma = (1-\theta)/2 \in ({1}/{4},{1}/{2})$. 
Hence case 2 corresponds to the second and third columns for case 2
in Table 6.1. 
\item {\bf Cases 3--4: Influence of \boldmath $A$ \unboldmath and \boldmath $B$\unboldmath.} 
$2\gamma + \theta -1 = 0$, $4\gamma - 1 = 0$, $k > 0$ for case 3, and $k < 0$ for case 4.  In these 2 cases 
$\gamma = {1}/{4}$ and $\theta = 1 - 2\gamma =
{1}/{2}$, which corresponds to the second and third columns for cases 3 and 4 in Table 6.1. 
\end{itemize}
In cases 1, 2, 3, and 4 we have, respectively, $G(x) = c_4 x^4$,  
$G(x) = k\beta x^2$ with $k > 0$, $G(x) = k\beta x^2 + c_4 x^4$ with 
$k > 0$, and $G(x) = k\beta x^2 + c_4 x^4$ with 
$k < 0$. In combination
with part (a), we obtain the 4 forms of the scaling limits listed in the last column of Table 6.1.

(b) We prove the 4 scaling limits corresponding to the 4 cases listed in 
Table 6.1. As the discussion prior to the statement of the theorem indicates, the quantity 
$v = \min\{2\gamma + \theta -1, 4\gamma -1\}$ is defined in 
such a way that in each of the 4 cases
defined by the choices of $\theta$, $\gamma$, and $k$ in Table 6.1, we have for each $x \in \R$  
\[
\lim_{n \goto \infty} n G_{\beta_n,K_n}(x/n^\gamma) = G(x).
\]
Since in each case we have $\gamma \in [{1}/{4},{1}/{2})$, 
the term $W_n/n^{1/2 - \gamma}$ in (\ref{eqn:G})
does not contribute to the limit $n \goto \infty$.
Hence we can determine the scaling limits of $S_n/n^{1-\gamma}$
by using (\ref{eqn:limit}).
In order to justify taking the limit inside the integrals
on the right hand side of (\ref{eqn:limit}), we return to (\ref{eqn:5point2'})
and use the fact that for all sufficiently large $n$, $C_n^{(2)} > 0$
and $G_{\bn,\kn}^{(4)}(0) > 0$.  It follows that
$R > 0$ can be chosen to be sufficiently small so that for all sufficiently large $n$ and 
all $x \in \R$ satisfying $|x/n^\gamma| < R$ there exists 
a polynomial $H(x)$ satisfying 
\be
\label{eqn:H}
nG_{\beta_n,K_n}(x/n^\gamma) \geq H(x)
\ee
and $\int_\R \exp[-H(x)] dx < \infty$.
In case 1 when $k \geq 0$ as well as in cases 2 and 3, $H(x) = G(x)/2$; in case 1 when
$k < 0$ and in case 4, which corresponds to $k < 0$,
\be
\label{eqn:case4}
H(x) = - 2|k| \beta x^2 + c_4 x^4/2.
\ee
The last two displays in combination with the dominated convergence theorem imply that
for any bounded, continuous function $f$
\[
\lim_{n \goto \infty} \int_{\{|x| < Rn^\gamma\}} f(x) \, \exp[-n G_{\beta_n,K_n}(x/n^\gamma)] \, dx
= \int_{\R} f(x) \, \exp[-G(x)] \, dx.
\]
The existence of this limit implies that the sequence 
$y_n = \int_{\{|x| < Rn^\gamma\}} \exp[-n G_{\beta_n,K_n}(x/n^\gamma)] dx$ is bounded.
Hence, combining this limit with part (c) of Lemma \ref{lem:awayfrom0} yields
\[
\lim_{n \goto \infty} \int_{\R} f(x) \, \exp[-n G_{\beta_n,K_n}(x/n^\gamma)] \, dx
= \int_{\R} f(x) \, \exp[-G(x)] \, dx.
\]
If we augment this limit with the same limit for $f = 1$ and use (\ref{eqn:limit}), then
we conclude that in each of the 4 cases 
\[
\lim_{n \goto \infty}
\int_{\Lambda^n} f({S_n}/{n^{1-\gamma}}) \, dP_{n,\beta_n,K_n} 
= \frac{1}{\int_{\R} \exp [-G(x)] \, dx} \cdot
\int_{\R} f(x) \, \exp[-G(x)] \, dx.
\]
This yields the scaling limits in part (a).  The proof of the theorem is complete.  \ \ink

\skp
This finishes our analysis of scaling limits for $S_n/n^{1-\gamma}$ with respect
to $P_{n,\beta_n,K_n}$, where the sequence $(\beta_n,K_n)$ converging
to $(\beta,K_c(\beta)) \in B$ 
is defined in Theorem \ref{thm:betakinb}.  This analysis
is a warm-up for the even more interesting analysis of the scaling limits 
for sequences $(\beta_n,K_n)$ converging to the tricritical point.

\section{13 Scaling Limits for \boldmath $(\beta_n, K_n) \! \goto \! (\beta_c,K_c(\beta_c))$ \unboldmath}
\beginsec
\label{section:scalingc}

In Theorem \ref{thm:betakinb} we obtained 4 forms of scaling limits for $S_n/n^{1-\gamma}$
using sequences $(\beta_n,K_n)$ converging to a second-order 
point $(\beta,K_c(\beta)) \in B$.  The limiting 
densities are proportional to $\exp[-G(x)]$, where $G$ takes of the 4 forms of an even
polynomial of degree 4 or 2 satisfying $G(0) = 0$ and $G(x) \goto \infty$ as $|x| \goto \infty$. 
In each case the form of the limit
is independent of the choice of $\beta_n \goto \beta$ but depends on the choice
of  $K_n \goto K_c(\beta)$.  
Like the BEG model at $(\beta,K_c(\beta)) \in B$, the Curie-Weiss model
has a second-order phase transition at a critical inverse temperature $\bar{\beta}_c$.
The 4 scaling limits and the 4 MDPs analyzed
in Theorem \ref{thm:mdpb} are analogous to the scaling limits
and MDPs that hold in the Curie-Weiss model when the inverse temperature
converges to $\bar{\beta}_c$ along appropriate sequences $\beta_n$ \cite{EicLow}.
However, the 13 scaling limits proved in the present section and the 13 analogous
MDPs obtained in Theorem \ref{thm:mdpc} depend on the nature of the 
tricritical point, a feature not shared with the Curie-Weiss model.   

We now use the insights gained in the preceding section to study the more
complicated problem of scaling limits for $S_n/n^{1-\gamma}$
using sequences $(\beta_n,K_n)$ converging to the tricritical point 
$(\beta_c,K_c(\beta_c)) = (\log 4,3/[2 \log 4]) $.
As in the preceding section, we choose $\theta > 0$,
$k \not = 0$, and 
\be
\label{eqn:chooseknagain}
K_n = K(\beta_n) - {k}/{n^\theta},
\ee
where $K(\beta) = (e^\beta + 2)/(4\beta)$ for $\beta > 0$.
In contrast to the preceding section, we now also have 
to pick the sequence $\beta_n$ appropriately.  Theorem \ref{thm:betakinc}
shows that 13 scaling limits arise for different choices of $\theta$, $\gamma$, and the parameter appearing
in the definition of $\bn$.   The limiting 
densities are proportional to $\exp[-G(x)]$, where $G$ takes one of the 13 forms of an even
polynomial of degree 6, 4, or 2 satisfying $G(0) = 0$ and $G(x) \goto \infty$ as $|x| \goto \infty$. 

In order to determine the forms of the scaling limits for $S_n/n^{1-\gamma}$ with respect to 
$\pnbnkn$, we use the Taylor expansion given in part (c) of Theorem \ref{thm:taylor}.
For any $\gamma > 0$ and $R > 0$ and for all $x \in \R$ satisfying $|x/n^\gamma| < R$
there exists $\xi \in [-x/n^\gamma,x/n^\gamma]$ such that
\bea
\lefteqn{
\label{eqn:TaylorG3again}
nG_{\beta_n, K_n}(x/n^\gamma) = } \\
\nonumber
&& \frac{1}{n^{2\gamma-1}}\frac{G_{\beta_n, K_n}^{(2)}(0)}{2!} x^2 + 
\frac{1}{n^{4\gamma-1}}\frac{G_{\beta_n, K_n}^{(4)}(0)}{4!} x^4 + 
\frac{1}{n^{6\gamma-1}}\frac{G_{\beta_n, K_n}^{(6)}(0)}{6!} x^6 + 
\frac{1}{n^{7\gamma - 1}} C_n(\xi(x/n^\gamma))  x^7.
\eea
The error terms $C_n(\xi(x/n^\gamma))$ are uniformly bounded over $n \in \N$ and 
$x \in (-\rng,\rng)$.
According to part (c) of Theorem \ref{thm:secondordertype}, the unique
global minimum point of $G_{\beta_c,K_c(\beta_c)}$ at 0 has type 3.
Hence by continuity, as $n \goto \infty$, 
\[
G_{\beta_n, K_n}^{(2)}(0) = \frac{2\beta_n K_n [K(\beta_n) - K_n]}{K(\beta_n)}
\goto G_{\beta_c,K_c(\beta_c)}^{(2)}(0) = 0,
\]
\[
G_{\beta_n, K_n}^{(4)}(0) = \frac{2 (2\beta_n K_n)^4 (4-e^{\beta_n})}{(e^{\beta_n} + 2)^2}
\goto G_{\beta_c,K_c(\beta_c)}^{(4)}(0) = 0,
\] 
while $G_{\beta_n, K_n}^{(6)}(0) \goto G_{\beta_c,K_c(\beta_c)}^{(6)}(0) = 2 \cdot 3^4$.

As in the preceding section, we choose $K_n$ as in (\ref{eqn:chooseknagain}) so that
$G_{\beta_n, K_n}^{(2)}(0) \goto 0$ at a rate $1/n^\theta$, where $1/n^\theta$ counterbalances
the term $1/n^{2\gamma-1}$ appearing in (\ref{eqn:TaylorG3again}).  
We also choose $\beta_n$ so that $G_{\beta_n, K_n}^{(4)}(0) \goto 0$ at 
a rate $1/n^\alpha$, where $1/n^\alpha$ counterbalances the term $1/n^{4\gamma - 1}$ 
appearing in (\ref{eqn:TaylorG3again}).  This is achieved
by choosing $\alpha > 0$ and either $b > 0$ or $b < 0$ and then defining $\bn$
by the logarithmic formula
\be
\label{eqn:choosebetan} 
\beta_n = \log(4 - {b}/{n^{\alpha}}) = \log(e^{\beta_c} - {b}/{n^{\alpha}})\!;
\ee
if $b > 0$, then $\beta_n$ is well defined for all sufficiently large $n$.
Since $\bn \goto \beta$ and $K(\cdot)$ is continuous, 
it follows that 
$(\bn,K_n) \goto (\beta_c,K_c(\beta_c))$.   With this choice of $(\bn,\kn)$ we have
\be
\label{eqn:cn2again}
G_{\beta_n, K_n}^{(2)}(0) = 
\frac{k}{n^\theta} \cdot \frac{2 \beta_n K_n}{K(\beta_n)} = \frac{k}{n^\theta} \cdot C_n^{(2)},
\ \mbox{ where } \ C_n^{(2)} \goto 2\beta_c,
\ee
and 
\be
\label{eqn:cn4}
G_{\beta_n, K_n}^{(4)}(0) = \frac{b}{n^\alpha} \cdot \frac{2 (2\beta_n K_n)^4 }{(e^{\beta_n} + 2)^2}
= \frac{b}{n^\alpha} \cdot C_n^{(4)},
\ \mbox{ where } \ C_n^{(4)} \goto \frac{2 (2\beta_c K_c(\beta_c))^4}{(e^{\beta_c} + 2)^2} = \frac{9}{2} > 0.
\ee

The dependence of $(\beta_n,K_n)$ in (\ref{eqn:chooseknagain}) and (\ref{eqn:choosebetan}) upon 
$\alpha$ and $\theta$ is complicated; because $\beta_n$ is a function of $\alpha$, $K_n$
is both a function of $\theta$ and, through $\beta_n$, a function of $\alpha$. 
However, the $\alpha$ and $\theta$ decouple nicely when 
(\ref{eqn:cn2again}) and (\ref{eqn:cn4}) 
are substituted into (\ref{eqn:TaylorG3again}), yielding
\bea
\label{eqn:gammathetaalpha}
\lefteqn{
nG_{\beta_n, K_n}(x/n^\gamma) } \\
&& = \frac{1}{n^{2\gamma+\theta-1}} \frac{k C_n^{(2)}}{2!} x^2 + 
\frac{1}{n^{4\gamma+\alpha-1}} \frac{b C_n^{(4)}}{4!} x^4 + 
\frac{1}{n^{6\gamma-1}}\frac{G_{\beta_n, K_n}^{(6)}(0)}{6!} x^6 
+ \frac{1}{n^{7\gamma - 1}} C_n(\xi(x/n^\gamma)) x^7.
\nonumber
\eea

We continue the analysis as in the preceding section.
Let us suppose that in the last display the coefficients multiplying $x^2$, 
$x^4$, and $x^6$ all appear
with nonnegative powers of $n$ and that at least one of the coefficients
has $n$ to the power $0$.  Then in the limit $n \goto \infty$ any coefficient 
including the error term that has
a positive power of $n$ will vanish while any coefficient that has
$n$ to the power $0$ will converge to positive constants.  
In order to analyze the various cases, we define
\be
\label{eqn:w} 
w = \min\{2\gamma + \theta -1, 4\gamma + \alpha - 1, 6\gamma -1\},
\ee
and focus on the cases in which $w = 0$.  As we will see in the final section of the paper,
$w < 0$ corresponds to 13 different MDPs for $S_n/n^{1 - \gamma}$.  
On the other hand, if $w > 0$, then one obtains neither scaling limits nor MDPs.

In the next theorem we show that $w = 0$ corresponds to 7 different
choices of $\gamma$, $\theta$, and $\alpha$, which in turn correspond to 13 different
sequences $(\beta_n,K_n)$ defined in (\ref{eqn:chooseknagain}) and (\ref{eqn:choosebetan}). 
The additional sequences arise because when $x^4$ is not the highest order term in the scaling limit (cases 4--5, 8--13),
$b$ can be chosen to be either positive or negative;
similarly, when $x^2$ is not the highest order term
in the scaling limit (cases 6--13),
$k$ can be chosen to be either positive or negative.
As shown in Table 7.1 in 
part (b) of the theorem, for each of these 13 different sequences
we obtain a different scaling limit of $S_n/n^{1-\gamma}$.

The limiting densities in cases 1, 4--7, and 10--13 are new.  
In cases 2, 3b, 8, and 9 we obtain the same forms of the limiting
densities as in Theorem \ref{thm:betakinb}, where we considered
$(\bn,\kn) \goto (\beta,K) \in B$.  However, the values of $\gamma$ in the corresponding
scaling limits in the two theorems are different.  By contrast, the values
of $\gamma$ and $\theta$ as well as the forms of the limiting densities
are the same in case 3a in Theorem \ref{thm:betakinc} and in case 2 in
Theorem \ref{thm:betakinb}.

There are yet further possibilities concerning the sign of $b$ and $k$. In all the 
cases in which no $x^4$ term appears in the scaling limit (cases 1, 3, 6, 7),
we can choose $b$ to be any real number.  Similarly, in all the cases in which no $x^2$ term appears in the scaling limit (cases 1, 2, 4, 5),
we can choose $k$ to be any real number.  Although the choice
of $b$ or $k$ affects the definition of the sequence $(\bn,\kn)$, it
does not affect the form of the scaling limit.

Through the terms 
$x^6$, $x^4$, and $x^2$ appearing in the limiting densities, the scaling limits correspond to the influence
of one or more of the sets $C$, $B$, and $A$.  The influence of the various sets upon the form of the scaling
limits is shown in Figure 2 in the introduction, and details are given
in Table 7.1, which is included in part (b) of the next theorem. 
Case 3, which corresponds to the influence of $A$ alone, has two subcases, labeled 
3a and 3b in Table 7.1.  Case 3a corresponds to the lower region labeled $A$ in Figure 2
and case 3b to the upper region labeled $A$ in Figure 2.  Using Table 7.1, one easily
checks that as $(\alpha,\theta)$ crosses any of the lines in Figure 2 labeled $A+B$, $A+C$, or
$B+C$, the values of $\gamma$ in the scaling limits change continuously while the forms
of the limiting densities change discontinuously.

\begin{thm}
\label{thm:betakinc}
Given $\alpha > 0$, $\theta > 0$, $b \not = 0$, and
$k \not = 0$, define
\[
\beta_n = \log(4 - {b}/{n^\alpha}) 
= \log(e^{\beta_c} - {b}/{n^\alpha}) \ \mbox{ and } \ K_n = K(\beta_n) - {k}/{n^\theta},
\]
where $K(\beta) = (e^\beta + 2)/(4\beta)$ for $\beta > 0$.
Then $(\beta_n,K_n) \goto (\beta_c,K_c(\beta_c)) $.  Given $\gamma \in (0,1)$, we also define 
\be
\label{eqn:Gforc}
G(x) = \delta (w, 2\gamma+\theta-1) k \bc x^2 +   \delta (w, 4\gamma+\alpha-1) 
b \bar{c}_4 x^4 + \delta (w, 6\gamma-1)  c_6 x^6,
\ee
where $\bar{c}_4 = 3/16$ and $c_6 = 9/40$.
The following conclusions hold.

{\em (a)}  Assume that $w = \min\{2\gamma + \theta -1, 4\gamma + \alpha -1, 6\gamma -1\}$ equals 0.
Then 
\be
\label{eqn:scalingwithgagain}
P_{n,\beta_n,K_n}\{S_n/n^{1-\gamma} \in dx\} \Longrightarrow \exp[-G(x)] \, dx.
\ee

{\em (b)}  We have $w = 0$ if and only if one of the 
{\em 13} cases enumerated in Table {\em 7.1} holds.
Each of the {\em 13} 
cases corresponds to a set of values of $\theta$, $\alpha$, and $\gamma$, to the influence
of one or more sets $C$, $B$, $A$,  
and to a particular scaling limit in {\em (\ref{eqn:scalingwithgagain})}.  
The form of the scaling limit is not affected by the choice of $b \in \R$
in cases {\em 1}, {\em 3}, {\em 6}, and {\em 7} and 
by the choice of $k \in \R$ in cases {\em 1}, {\em 2}, {\em 4}, 
and {\em 5}.

\begin{center}
\begin{tabular}{||l|l|l|l||} \hline \hline
{\em {\bf case}} & {\em {\bf values of}} \boldmath $\alpha$ \unboldmath & 
{\em {\bf values of}} \boldmath $\gamma$ \unboldmath
& {\em {\bf scaling limit of}} \boldmath $S_n/n^{1-\gamma}$ \unboldmath \\ \cline{1-2}
{\em {\bf influence}} & {\em {\bf values of}} $\theta$ & & 
\\ \hline \hline
{\em 1} & $\alpha > \frac{1}{3}$ & $\gamma = \frac{1}{6}$ &
% $S_n/n^{1-1/6} \Longrightarrow 
$\exp(-c_6 x^6) \, dx$ \\ \cline{1-2}
 $C$ & $\theta > \frac{2}{3}$ &   &  $c_6 > 0$,
$b \in \R$, $k \in \R$ \\ \hline \hline
{\em 2} & $\alpha \in (0,\frac{1}{3})$ & $\gamma = \frac{1-\alpha}{4} \in 
(\frac{1}{6},\frac{1}{4}) $ &
% $S_n/n^{1-\gamma} \Longrightarrow 
$\exp(-b \bar{c}_4 x^4) \, dx$ \\ \cline{1-2}
 $B$ & $\theta > \frac{\alpha + 1}{2}$ &   &  $\bar{c}_4 > 0$, $b > 0$,
$k \in \R$ \\ \hline \hline
{\em 3a} & $\alpha > 0$ & $\gamma = \frac{1 - \theta}{2} \in (\frac{1}{4},\frac{1}{2})$ &
% $S_n/n^{1-\gamma} \Longrightarrow \
$\exp(-k \bc x^2) \, dx$ \\ \cline{1-2}
 $A$ & $\theta \in (0,\frac{1}{2})$ &   &  $k > 0$, $b \in \R$
\\ \hline \hline
{\em 3b} & $\theta \in [\frac{1}{2},\frac{2}{3})$ & $\gamma = \frac{1 - \theta}{2} \in (\frac{1}{6},\frac{1}{4}]$ &
% $S_n/n^{1-\gamma} \Longrightarrow 
$\exp(-k \bc x^2) \, dx$ \\ \cline{1-2}
$A$  & $\alpha > 2\theta - 1$ &   &  $k > 0$, $b \in \R$  \\ \hline \hline
{\em 4--5} & $\alpha = \frac{1}{3}$ & $\gamma = \frac{1}{6}$ &
% $S_n/n^{1-1/6} \Longrightarrow 
$\exp(-b \bar{c}_4 x^4 - c_6 x^6) \, dx$ \\ \cline{1-2}
 $B+C$ & $\theta > \frac{2}{3}$ &   & $b \not = 0$,
$k \in \R$  \\ \hline \hline
{\em 6--7} & $\alpha > \frac{1}{3}$ & $\gamma = \frac{1}{6}$ &
% $S_n/n^{1-1/6} \Longrightarrow 
$\exp(-k \bc x^2 - c_6 x^6) \, dx$ \\ \cline{1-2}
$A+C$  & $\theta = \frac{2}{3}$ &   & $k \not = 0$,
$b \in \R$  \\ \hline \hline
{\em 8--9} & $\alpha \in (0,\frac{1}{3})$ & $\gamma = \frac{1-\alpha}{4} 
\in (\frac{1}{6},\frac{1}{4})$ &
% $S_n/n^{1-1/6} \Longrightarrow 
$\exp(-k \bc x^2 - b \bar{c}_4 x^4) \, dx$ \\ \cline{1-2}
$A+B$  & $\theta = \frac{\alpha + 1}{2} \in (\frac{1}{2},\frac{2}{3})$ 
%\in (\frac{1}{2},\frac{2}{3})$ 
&   &  $k \not = 0$, $b > 0$ \\ \hline \hline
{\em 10--13} & $\alpha = \frac{1}{3}$ & $\gamma = \frac{1}{6}$ &
% $S_n/n^{1-1/6} \Longrightarrow 
$\exp(-k \bc x^2 - b \bar{c}_4 x^4 - c_6 x^6) \, dx$ \\ \cline{1-2}
$A+B+C$ & $\theta = \frac{2}{3}$ &   & $k \not = 0$, $b \not = 0$  \\ \hline \hline
\end{tabular}

\vspace{.05in}
{\em Table 7.1}: {\small {\em Values of } $\alpha$, $\theta$, {\em and} $\gamma$
{\em and scaling limits
in part (b) of Theorem \ref{thm:betakinc}}}  
\end{center}
\end{thm}

\skp
\noi
{\bf Note.}  The constant sequence $(\bn,\kn) = (\beta_c,K_c(\bc))$ for all $n$
corresponds to the choices $\alpha = \theta = \infty$ in case 1. 
As in the proof of case 1, one shows that $P_{n,\beta_c,K_c(\beta_c)}\{S_n/n^{1-1/6} \in dx\} \Longrightarrow
\exp(-c_6 x^6) dx$.  This scaling limit was mentioned in (\ref{eqn:6thorder}).  

\skp
\noi
{\bf Proof of Theorem \ref{thm:betakinc}.}  We first prove part (b) from part (a) and then prove part (a).

(b)  $w = \min\{2\gamma + \theta -1, 4\gamma + \alpha -1, 6\gamma -1\}$ equals 0 
if and only if each of the quantities in this minimum 
is nonnegative and one or more of the quantities equals 0.  
As (\ref{eqn:Gforc}) makes clear, $6\gamma -1 = 0$ corresponds to
the influence of $C$, $4\gamma + \alpha -1 = 0$  to the influence of $B$, 
and $2\gamma + \theta -1 = 0$ to the influence of $A$. 
We have the following 13 mutually exclusive and exhaustive cases, which 
correspond to the 13 cases in Table 7.1.  In each of the cases the equalities
and inequalities expressing the influence of one or more sets $C$, $B$, and $A$ are easily
verified to be equivalent to the equalities and inequalities involving
$\alpha$, $\theta$, and $\gamma$ given in the second and third columns of Table 7.1.
Case 3, the most complicated, divides into two subcases depending on the value of $\alpha$.

\begin{itemize}
\item {\bf Case 1: Influence of \boldmath $C$ \unboldmath alone.} 
$2\gamma + \theta -1 > 0$, $4\gamma + \alpha -1 > 0$, $6\gamma -1 = 0$, 
$b \in \R$, and $k \in \R$.
\item {\bf Case 2: Influence of \boldmath $B$ \unboldmath alone.} 
$2\gamma + \theta -1 > 0$, $4\gamma + \alpha -1 = 0$, $6\gamma -1 > 0$, $b > 0$,
and $k \in \R$.
\item {\bf Case 3: Influence of \boldmath $A$ \unboldmath alone.} 
$2\gamma + \theta -1 = 0$, $4\gamma + \alpha -1 > 0$, $6\gamma -1 > 0$, $k > 0$,
and $b \in \R$.
\item {\bf Cases 4--5: Influence of \boldmath $B$ \unboldmath and \boldmath $C$\unboldmath.}
$2\gamma + \theta -1 > 0$, $4\gamma + \alpha -1 = 0$, $6\gamma -1 = 0$, $b > 0$ for case 4
and $b < 0$ for case 5, and $k \in \R$.
\item {\bf Cases 6--7: Influence of \boldmath $A$ \unboldmath and \boldmath $C$\unboldmath.}
$2\gamma + \theta -1 = 0$, $4\gamma + \alpha -1 > 0$, $6\gamma -1 = 0$, $k > 0$ for case 6
and $k < 0$ for case 7, and $b \in \R$.
\item {\bf Cases 8--9: Influence of \boldmath $A$ \unboldmath and \boldmath $B$\unboldmath.}
$2\gamma + \theta -1 = 0$, $4\gamma + \alpha -1 = 0$, $6\gamma -1 > 0$, $k > 0$ for 
case 8, $k < 0$ for case 9, and $b > 0$.
\item {\bf Cases 10--13: Influence of \boldmath $A$ \unboldmath, \boldmath $B$\unboldmath, and
\boldmath $C$\unboldmath.}
$2\gamma + \theta -1 = 0$, $4\gamma + \alpha -1 = 0$, $6\gamma -1 = 0$, $k > 0$ and $b > 0$
for case 10, $k < 0$ and $b > 0$ for case 11, $k > 0$ and $b < 0$ for case 12, 
and $k < 0$ and $b < 0$ for case 13.
\end{itemize}

In each of the 13 cases the form of $G(x)$ follows from (\ref{eqn:Gforc}).
In combination
with part (a), we obtain the 13 forms of the scaling limits listed in the last column of Table 7.1.

(a) The proof of the 13 scaling limits follows precisely the pattern of the proof of the
4 scaling limits listed in part (b) of Theorem \ref{thm:betakinb}.  As
the discussion preceding the statement of Theorem \ref{thm:betakinc} indicates, the quantity
$w = \min\{2\gamma + \theta -1, 4\gamma + \alpha -1, 6\gamma -1\}$ is defined in such a way
that in each of the 13 cases defined by the choices of $\alpha$, $\theta$,
$\gamma$, $k$, and $b$ in Table 7.1, we have for each $x \in \R$
\[
\lim_{n \goto \infty} n G_{\beta_n,K_n}(x/n^\gamma) = G(x).
\]
Since in each case we have $\gamma \in [{1}/{6},{1}/{2})$, 
the term $W_n/n^{1/2 - \gamma}$ in (\ref{eqn:G})
does not contribute to the limit $n \goto \infty$.
Hence we can determine the scaling limits of $S_n/n^{1-\gamma}$
by using (\ref{eqn:limit}).
In order to justify taking the limit inside the integrals
on the right hand side of (\ref{eqn:limit}), we return to (\ref{eqn:gammathetaalpha})
and use the fact that for all sufficiently large 
$n$, $C_n^{(2)} > 0$, $C_n^{(4)} > 0$,
and $G_{\bn,\kn}^{(6)}(0) > 0$.  It follows that
$R > 0$ can be chosen to be sufficiently small so that for all sufficiently large $n$ and 
all $x \in \R$ satisfying $|x/n^\gamma| < R$ there exists 
a polynomial $H(x)$ satisfying
\be
\label{eqn:Hagain}
nG_{\beta_n,K_n}(x/n^\gamma) \geq H(x)
\ee
and $\int_{\R} \exp[- H(x)] < \infty$.
We define $H(x) = G(x)/2$ in all the cases in which both $b \geq 0$ and $k \geq 0$
(cases 1--4, 6, 8, 10).
Otherwise, a suitable polynomial $H$ can be found as in
(\ref{eqn:case4}); the details are omitted.
As in the proof of Theorem \ref{thm:betakinb},
the dominated convergence theorem and part (c) of Lemma \ref{lem:awayfrom0}
imply that for any bounded, continuous function $f$
\[
\lim_{n \goto \infty} \int_{\R} f(x) \, \exp[-n G_{\beta_n,K_n}(x/n^\gamma)] \, dx
= \int_{\R} f(x) \, \exp[-G(x)] \, dx.
\]
From (\ref{eqn:limit}) we conclude that in each of the 13 cases in part (b)
\[
\pnbnkn\!\left\{S_n/n^{1 - \gamma} \in dx\right\}
\Longrightarrow \exp[-G(x)] \, dx.
\]
This completes the proof of the theorem. \ \ink

\skp
Two special cases of the scaling limits in Theorem \ref{thm:betakinc} are worth
pointing out.  Given $\theta > 0$ and $k \not = 0$, the sequence
\[
\beta_n = \beta_c \ \mbox{ and } \ K_n = K(\beta_c) - {k}/{n^\theta}
\]
corresponds to the choice $\alpha = \infty$ in Theorem \ref{thm:betakinc}.
With this sequence and with the same proofs, one obtains
exactly the same limits as in cases 1, 3, 6, and 7 in this theorem with the same choices of $\theta$,
$\gamma$, and $k$.  Similarly, given $\alpha > 0$ and $b \not = 0$,
the sequence
\[
\beta_n = \log(4 - {b}/{n^\alpha}) \ \mbox{ and } \ K_n = K(\beta_c)
\]
corresponds to the choice $\theta = \infty$ in Theorem \ref{thm:betakinc}.
With this sequence and with the same proofs, one obtains
exactly the same limits as in cases 1, 2, 4, and 5 in this theorem with the same choices of $\alpha$,
$\gamma$, and $b$.

This completes our analysis of scaling limits for $S_n/n^{1-\gamma}$ with respect
to $P_{n,\beta_n,K_n}$, where the sequence
$(\beta_n,K_n)$ converging to $(\beta_c,K_c(\beta_c))$ 
is defined in Theorem \ref{thm:betakinc}.  In the next
section we study MDPs for $S_n/n^{1 - \gamma}$ for appropriate sequences
$(\bn,\kn)$ converging to $(\beta,K) \in A \cup B \cup C$ and for 
appropriate choices of $\gamma$. 
We obtain 1 MDP for $(\beta,K) \in A$, 4 MDPs for $(\beta,K_c(\beta)) \in B$, 
and 13 MDPs for $(\bc,\kcbc) \in C$.

\section{18 MDPs for
\boldmath $(\beta_n, K_n) \! \goto \! (\beta,K) \in A \cup B \cup C$ \unboldmath}
\beginsec
\label{section:mdp}

In this section we turn to a new problem, 
which is to formulate MDPs for $S_n/n^{1-\gamma}$ 
with respect to $P_{n,\beta_n,K_n}$, first for appropriate
sequences $(\beta_n,K_n)$
converging to $(\beta,K_c(\beta)) \in B$, then for $(\beta_n,K_n)$ 
converging to $(\beta,K) \in A$, and finally for $(\beta_n,K_n)$ 
converging to $(\beta_c,K_c(\beta_c)) \in C$.  These results are stated, respectively,
in Theorem \ref{thm:mdpb}, Theorem \ref{thm:mdpa}, and Theorem \ref{thm:mdpc}.  
In proving the first result, we introduce the methods that are also
used to prove the third.  The proof
of the MDP when $(\beta_n,K_n) \goto (\beta,K) \in A$ proceeds 
differently from the proofs of the other MDPs in this section, relying on the 
G\"{a}rtner-Ellis Theorem.  After the proof of that MDP, we will remark on why
the same method cannot be used to prove all the MDPs in this section.
Although an MDP is an LDP, we shall follow the example
of \cite{EicLow}, who in their study of Curie-Weiss-type models speak about
an MDP whenever the exponential speed $a_n$ of the large deviation probabilities
satisfies $a_n/n \goto 0$ as $n \goto \infty$.  Also see \cite[\S 3.7]{DemZei}.

When $(\beta_n,K_n) \goto (\beta,K_c(\beta)) \in B \cup C$,
we will prove the MDPs by a method that is closely related to the proofs
of the scaling limits earlier in this paper.  Thus, rather than focus on the large
deviation probabilities directly, we prove that $S_n/n^{1-\gamma}$ satisfies 
an equivalent Laplace principle.  
Despite the similarity in the proof of the scaling limits and the Laplace principles, 
the proof of the latter 
is much more delicate, requiring additional estimates not needed 
in the proof of the former.

We start by considering the MDPs when $(\beta_n,K_n)$ converges to 
$(\beta,K_c(\beta)) \in B$.
In order to formulate these limit theorems, 
we adapt the methods used in section \ref{section:scalingb}, 
where we proved scaling limits for such sequences $(\beta_n,K_n)$. 
For $\beta \in (0,\bc)$ let $\beta_n$ be an arbitrary positive
sequence that converges to $\beta$. 
Given $\theta > 0$ and $k \not = 0$, we then define $K_n \goto K_c(\beta)$ as in (\ref{eqn:choosekn}).  
With this choice,
part (b) of Theorem \ref{thm:taylor} implies that for any $\gamma > 0$ and $R > 0$
and for all $x \in \R$ satisfying $|x/n^\gamma| < R$ there exists
$\xi \in [-x/n^\gamma,x/n^\gamma]$ such that [see (\ref{eqn:5point2'})]
\be
\label{eqn:gammathetamdpb}
nG_{\beta_n, K_n}(x/n^\gamma) = 
\frac{1}{n^{2\gamma+\theta-1}}\frac{C_n^{(2)}}{2!} x^2 + 
\frac{1}{n^{4\gamma-1}}\frac{G_{\beta_n, K_n}^{(4)}(0)}{4!} x^4 + 
\frac{1}{n^{5\gamma - 1}}B_n(\xi(x/n^\gamma))  x^5.
\ee
The error terms $B_n(\xi(x/n^\gamma))$ are uniformly bounded over $n \in \N$ and 
$x \in (-\rng,\rng)$,
$C_n^{(2)} \goto 2\beta$, and $G_{\beta_n,K_n}^{(4)}(0) \goto G_{\beta,K}^{(4)}(0) > 0$.

Given $\gamma \in (0,1)$, we define 
\be
\label{eqn:vmdpb}
v = \min\{2\gamma + \theta -1, 4\gamma -1\}.
\ee
In Theorem \ref{thm:betakinb} we prove that when $v = 0$,
$S_n/n^{1-\gamma}$ satisfies the scaling limit 
\[
\pnbnkn\{S_n/n^{1-\gamma} \in dx\} \Longrightarrow \exp[-G(x)]dx,
\] 
where
\[
G(x) = \delta(v,2\gamma + \theta -1) k \beta x^2 + \delta(v,4\gamma-1) c_4x^4
\]
and $c_4$ is defined in (\ref{eqn:c4betak}).
As enumerated in Table 6.1, 
the 4 different forms of the limiting density depend on the values
of $\gamma$ and $\theta$ and the sign of $k$.

In Theorem \ref{thm:mdpb} 
we prove the analogous results on the level of MDPs.
Assume that the quantity $v$ defined in (\ref{eqn:vmdpb}) is 
negative.  Then, when $(\bn,\kn)$ is chosen
as in Theorem \ref{thm:betakinb}, $S_n/n^{1-\gamma}$ 
satisfies the MDP with exponential speed $\nv$ and rate function 
$\Gamma(x) = G(x) - \inf_{y \in \R}G(y)$, where $G$ is defined in 
the last display.
We prove the MDP in Theorem \ref{thm:mdpb} by showing that 
when $v < 0$, $S_n/n^{1-\gamma}$ 
satisfies the Laplace principle with speed $\nv$ and rate function $\Gamma$;
i.e., for any bounded, continuous function $\psi$
\[
\lim_{n \goto \infty} \frac{1}{\nv} \log \int_{\Lambda^n} \exp[\nv \, \psi(S_n/n^{1-\gamma})] \, dP_{n,\bn,\kn}
=  \sup_{x \in \R}\{\psi(x) - \Gamma(x)\}.
\]
By Theorem 1.2.3 in \cite{DupEll} 
the fact that $S_n/n^{1-\gamma}$ satisfies the Laplace principle 
implies that $S_n/n^{1-\gamma}$ satisfies the LDP
with the same speed $\nv$ and the same
rate function $\Gamma$; i.e., for any closed subset $F$ in $\R$
\[
\limsup_{n \goto \infty} \frac{1}{\nv} \log P_{n,\bn,\kn}\{S_n/n^{1-\gamma} \in F\}
\leq -\inf_{x \in F} \Gamma(x)
\]
and for any open subset $\Phi$ in $\R$
\[
\liminf_{n \goto \infty} \frac{1}{\nv} \log P_{n,\bn,\kn}\{S_n/n^{1-\gamma} \in \Phi\}
\geq -\inf_{x \in \Phi} \Gamma(x).
\]
$\Gamma$ is obviously a rate function.  One easily checks that in all 4 cases
given in part (b) of Theorem \ref{thm:mdpb} $-v < 1$.  Hence $\nv/n \goto 0$ as $n \goto \infty$, 
and so we have an MDP.  In cases 1, 2, and 3, we have $\inf_{y \in \R}G(y) = 0$ and
thus $\Gamma = G$; in case 4, $\inf_{y \in \R}G(y) < 0$.

As in the scaling limits in Theorem \ref{thm:betakinb},
the rate function in Theorem \ref{thm:mdpb} 
takes the 4 forms enumerated in cases 1, 2, 3, and 4
in Table 8.1. In case 2 the requirement that $G(x) \goto \infty$
as $|x| \goto \infty$ forces $k > 0$.  By contrast, in case 4,
$k < 0$ is allowed.  In case 1 we can also choose $k$ to be any real 
number; this affects only the definition
of the sequence $K_n$, not the form of the rate function.

The forms of the rate functions
reflect the influence, respectively, of $B$, of $A$, and of $A$ and $B$.  
In each case
the particular set or sets that influence the form of $G$ depend on
the speed at which $(\bn,\kn)$ approaches $(\beta,K_c(\beta))$ and 
the direction of approach.  Case 2, which corresponds to the influence 
of $A$ alone, has two subcases, labeled 2a and 2b in Table 8.1. 

In Figure 5 and in Table 8.1 we indicate
the subsets of the positive quadrant of the $\theta$-$\gamma$ plane leading 
to the 4 cases of the MDPs in Theorem \ref{thm:mdpb}.
Subcases 2a and 2b correspond, respectively, to the left half 
and the right half of the triangle labeled $A$ in Figure 5.
An interesting connection between the MDPs in Theorem \ref{thm:mdpb}
and the scaling limits 
in Theorem \ref{thm:betakinb} is revealed by comparing Figure 5
with Figure 4, which exhibits 
the subsets of the positive quadrant of the 
$\theta$-$\gamma$ plane leading 
to the 4 cases of the scaling limits in Theorem \ref{thm:betakinb}.
The subsets labeled $A$, $B$, and $A+B$ in Figure 4 are each a subset of the 
boundary of the set having the same label in Figure 5.  The relevant boundaries
in Figure 5 are labeled $\partial^+\!A$, $\partial^+\!B$, and $\partial^+(A+B)$,
the first two of which are indicated by dotted lines.  This relationship between
the two figures is not a surprise because the sets labeled $A$, $B$, and $A+B$
in Figure 4 are determined by solving $v = 0$ while the sets having the same labels
in Figure 5 are determined by solving $v < 0$.  

\begin{figure}[h]
\begin{center}
\epsfig{file=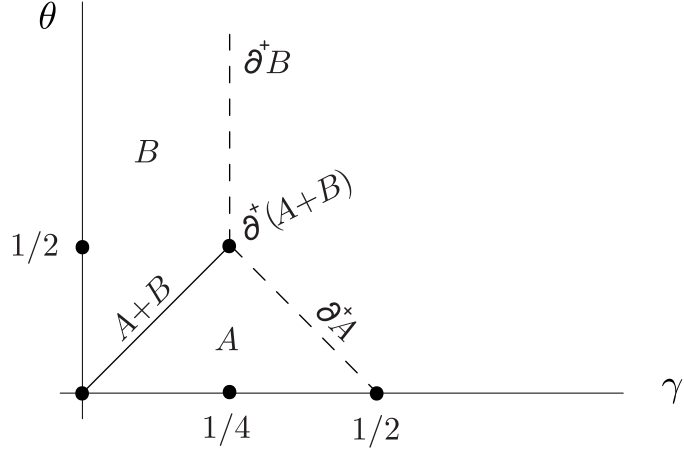,width=10cm}
\caption{\small Influence of $B$ and $A$ on MDPs when $(\bn,\kn)
\goto (\beta,K_c(\beta)) \in B$}
\end{center}
\end{figure} 

\begin{thm}
\label{thm:mdpb}
For fixed $\beta \in (0,\bc)$, let $\bn$ be an arbitrary positive sequence
that converges to $\beta$. 
Given $\theta > 0$ and $k \not = 0$, define
\[
K_n = K(\beta_n) - {k}/{n^\theta},
\]
where $K(\beta) = (e^\beta + 2)/(4\beta)$ for $\beta > 0$.
Then $(\bn,\kn) \goto (\beta,K_c(\beta)) \in B$.  Given $\gamma \in (0,1)$, we also define
\be
\label{eqn:gxforc}
G(x) = \delta{(v, 2\gamma+\theta-1)} k\beta x^2 +   \delta(v, 4\gamma-1) c_4 
x^4,
\ee
where $c_4 > 0$ is given by 
\[
c_4 = \frac{G_{\beta,K_c(\beta)}^{(4)}(0)}{4!} =
\frac{2[2 \beta K_c(\beta)]^4 (4 - e^\beta)}{4!(e^\beta + 2)^2}
= \frac{(e^\beta + 2)^2 (4 - e^\beta)}{2^3 \cdot 4!}.
\]
The following conclusions hold.

{\em (a)}  Assume that $v = \min\{2\gamma + \theta -1, 4\gamma -1\}$ satisfies $v < 0$.
Then with respect to $P_{n,\bn,\kn}$,
$S_n/n^{1-\gamma}$ satisfies the Laplace principle, and thus the MDP, with exponential speed $\nv$ and 
rate function $\Gamma(x) = G(x) - \inf_{y \in \R}G(y)$. 

{\em (b)}  We have $v < 0$ if and only if one of the 
{\em 4} cases enumerated in Table {\em 8.1} holds.
Each of the {\em 4} cases 
corresponds to a set of values of $\gamma$ and $\theta$, a choice of sign of $k$, 
the influence of one or more
sets $B$ and $A$, 
and a particular exponential speed and a particular form of the rate function in part {\em (a)}.
The function $G$ appearing in the definition of the 
rate function is shown in column {\em 5} in Table {\em 8.1}; in case {\em 4} the nonzero
constant $\inf_{y \in \R}G(y)$ in the definition of the rate function is not shown. 
In case {\em 1} the choice of $k \in \R$ does not affect the form of the
rate function.
% \vspace{.1in}  
\begin{center}
\begin{tabular}{||c|l|l|l|l||} \hline \hline
{\em {\bf case}} & {\em {\bf values of}} $\gamma$ & {\em {\bf values of}} \boldmath $\theta$ \unboldmath & 
{\em {\bf exp'l}}
& {\em {\bf function}} \boldmath $G$ \unboldmath {\em {\bf in}} 
\\ \cline{1-1}
{\em {\bf influence}} & & & {\em {\bf speed}} & {\em {\bf rate function}} \boldmath $\Gamma$ \unboldmath
\\ \hline \hline
{\em 1} & $\gamma \in (0,\frac{1}{4})$ & $\theta > 2 \gamma$ & $n^{1-4\gamma}$ &
% $S_n/n^{1-1/2} \Longrightarrow 
$c_4 x^4$ \\ \cline{1-1} 
$B$ & & & & $c_4 > 0$, $k \in \R$ \\ \hline \hline
{\em 2a} & $\gamma \in (0,\frac{1}{4}]$ & $\theta \in (0,2\gamma)$ & $n^{1-2\gamma-\theta}$ &
% $S_n/n^{1-\gamma} \Longrightarrow 
$k\beta x^2$ \\ \cline{1-1}
$A$ & & & & $k > 0$\\ \hline \hline
{\em 2b} & $\gamma \in (\frac{1}{4},\frac{1}{2})$ & $\theta \in (0,1-2\gamma)$ &
$n^{1-2\gamma-\theta}$ &
% $S_n/n^{1-\gamma} \Longrightarrow 
$k\beta x^2$ \\ \cline{1-1}
$A$ & & & & $k > 0$ \\ \hline \hline
{\em 3--4} & $\gamma \in (0,\frac{1}{4})$ & $\theta = 2\gamma$ & $n^{1-4\gamma}$ &
% $S_n/n^{1-1/4} \Longrightarrow 
$k\beta x^2 + c_4 x^4$ \\ \cline{1-1}
$A + B$ & & & & $k \not = 0$ \\ \hline \hline
\end{tabular}

\vspace{.05in}
{\em Table 8.1}: {\small {\em Values of } $\gamma$ {\em and} $\theta$, {\em exponential speeds, and rate functions
in part (b) of Theorem \ref{thm:mdpb}}}
\end{center}
\end{thm}

\skp
\noi
{\bf Proof.}  
We first prove part (b) from part (a) and then prove part (a).

(b) We have $v < 0$ in the following 4 mutually exclusive and exhaustive cases.
As (\ref{eqn:gxforc}) makes clear, $v = 4\gamma -1 < 0$ corresponds to the influence
of $B$ and $v = 2\gamma + \theta -1$ to the influence of $A$.
\begin{itemize}
\item {\bf Case 1: Influence of \boldmath $B$ \unboldmath alone.} 
$v = 4\gamma -1 < 0$, $4\gamma - 1 < 2\gamma + \theta -1$, and $k \in \R$.  In this case 
$\gamma \in (0,{1}/{4})$ and $\theta > 2\gamma$, 
which corresponds to the second and third columns for case 1 in Table 8.1. 
\item {\bf Case 2: Influence of \boldmath $A$ \unboldmath alone.} 
$v = 2\gamma + \theta - 1 < 0$, $2\gamma + \theta - 1 < 4 \gamma - 1$, and $k > 0$. 
In this case $0 < \theta < \min\{2\gamma, 1-2\gamma\}$.  Since
$0 < 2\gamma \leq 1 - 2\gamma \Leftrightarrow \gamma \in (0,{1}/{4}]$
and $0 < 1 - 2\gamma < 2\gamma \Leftrightarrow \gamma \in ({1}/{4},{1}/{2})$,
case 2 corresponds to the second and third columns for case 2a and case 2b in Table 8.1.
\item {\bf Cases 3--4: Influence of \boldmath $A$ \unboldmath and \boldmath $B$\unboldmath.} 
$v = 4\gamma -1 = 2\gamma + \theta -1 < 0$, $k > 0$ for case 3, and $k < 0$ for case 4. 
In these cases $0 < \gamma < {1}/{4}$ and $\theta = 2\gamma$.  Hence case 3--4 correspond
to the second and third columns for cases 3--4 in Table 8.1. 
\end{itemize}
In cases 1, 2, 3, and 4 we have, respectively, $G(x) = c_4 x^4$,  
$G(x) = k \beta x^2$ with $k > 0$, $G(x) = k \beta x^2 + c_4 x^4$ with $k > 0$,
and $G(x) = k \beta x^2 + c_4 x^4$ with $k < 0$.  In combination
with part (a), we obtain the 4 rate functions given in the last column of Table 8.1.

(a)  Our strategy is to prove that with respect to $P_{n,\bn,\kn} \times Q$,
$S_n/n^{1-\gamma} + W_n/n^{1/2 - \gamma}$
satisfies the Laplace principle with exponential speed $\nv$ and rate function $\Gamma$.  
In order to prove the Laplace principle for $S_n/n^{1-\gamma}$ alone, we need the following
estimate, which shows that $W_n/n^{1/2-\gamma}$ is superexponentially small relative to $\exp(\nv)$:
for any $\delta > 0$
\be
\label{eqn:superexpclose}
\limsup_{n \goto \infty} \frac{1}{\nv} \log Q\{|W_n/n^{1/2 - \gamma}| > \delta\} = -\infty.
\ee
According to Theorem 1.3.3 in \cite{DupEll}, if with respect to $P_{n,\bn,\kn} \times Q$,
$W_n/n^{1/2 - \gamma} + S_n/n^{1-\gamma}$
satisfies the Laplace principle with speed $\nv$ and rate function $\Gamma$,
then with respect to $P_{n,\bn,\kn}$, $S_n/n^{1-\gamma}$
satisfies the Laplace principle with speed $\nv$ and rate function $\Gamma$.
Since the Laplace principle implies the MDP \cite[Thm.\ 1.2.3]{DupEll}, part (a) of
the present theorem will be proved.

We now prove (\ref{eqn:superexpclose}). Denote the variance $(2\beta_n\kn)^{-1}$ of $W_n$ by $\sigma_n^2$.
Since $\bn$ and $\kn$ are bounded and uniformly positive over $n$, 
the sequence $\sigma_n^2$ is bounded
and uniformly positive over $n$.  We have the inequality
\beas
Q\{|W_n/n^{1/2 - \gamma}| > \delta\} & = & Q\{|N(0,\sigma_n^2)| > n^{1/2 - \gamma}\delta\} \\
& \leq & \frac{\sqrt{2} \sigma_n}{\sqrt{\pi} n^{1/2 - \gamma} \delta} \cdot\exp(-n^{1-2\gamma}\delta^2/[2\sigma_n^2]).
\eeas
Hence (\ref{eqn:superexpclose}) 
follows if $1 - 2\gamma > -v$.   Since $\gamma$ and $\theta$ are both positive, 
this is easily verified to hold
when either $v = 4\gamma - 1$ or $v = 2\gamma + \theta -1$.

We now turn to the Laplace principle for $S_n/n^{1-\gamma} + W_n/n^{1/2 - \gamma}$.
Let $\psi$ be an arbitrary bounded, continuous function.   
Choosing $f = \exp[\nv \psi]$ in Lemma \ref{lem:G} yields
\bea
\label{eqn:needthis}
\lefteqn{
\int_{\Lambda^n \times \Omega} \exp\!\left[\nv \, \psi\!\left(
\frac{S_n}{n^{1-\gamma}} + \frac{W_n}{n^{1/2-\gamma}}\right)\right] d(P_{n,\beta_n,K_n} \times Q)} \\ 
 \nonumber && = \frac{1}{\int_{\R} \exp [-nG_{\beta_n,K_n}(x/n^{\gamma})] \, dx} \cdot
\int_{\R} \exp[\nv \psi(x) - nG_{\beta_n, K_n}(x/n^{\gamma})] \, dx,
\eea
In order to obtain the appropriate expansion of $nG_{\beta_n, K_n}(x/n^{\gamma})$ in this display,
we multiply the numerator and denominator of 
the right hand side of (\ref{eqn:gammathetamdpb}) by $\nv$, obtaining
\[
nG_{\beta_n, K_n}(x/n^\gamma) = \nv G_n(x),
\]
where
\[
G_n(x) = 
\frac{1}{n^{2\gamma+\theta-1-v}}\frac{C_n^{(2)}}{2!} x^2 + 
\frac{1}{n^{4\gamma-1-v}}\frac{G_{\beta_n, K_n}^{(4)}(0)}{4!} x^4 + 
\frac{1}{n^{5\gamma-1-v}}B_n(\xi(x/n^\gamma))  x^5.
\]

The proof of the Laplace principle for $S_n/n^{1-\gamma} + W_n/n^{1/2 - \gamma}$ rests
on the following properties of $nG_{\beta_n, K_n}(x/n^{\gamma}) = \nv G_n(x)$,
which in turn are consequences of the Taylor expansion of $G_n(x)$ just given.
Because of the estimate (\ref{eqn:superexpclose}) on $W_n/n^{1/2 - \gamma}$,
the inequality in (\ref{eqn:uniform}), 
and the uniform convergence of $G_n$ to $G$
expressed in item 3 below, the proof of the MDPs, though analogous, is 
more delicate than the proof of the scaling 
limits in section \ref{section:scalingb}, for which the
a.s.\ convergence of $W_n/n^{1/2 - \gamma}$ to 0, the pointwise
convergence of $G_{\bn,\kn}(x/n^\gamma)$ to $G(x)$, and the lower bound (\ref{eqn:H}) suffice.
\begin{enumerate}
\item There exists $R > 0$ and a polynomial $H$
with the properties that $H(x) \goto \infty$ as $|x| \goto \infty$
and for all 
sufficiently large $n$ and all $x \in \R$ satisfying $|x/n^\gamma| < R$
\[
nG_{\bn,\kn}(x/n^\gamma) \geq  \nv H(x).
\]
In case 1 when $k \geq 0$ as well as in 
cases 2 and 3, $H(x) = G(x)/2$; in case 1 when $k < 0$ and in case 4,
which corresponds to $k < 0$, $H(x) = - 2|k| \beta x^2 + c_4 x^4/2$.
\item Let $\Delta = \sup_{x \in \R} \{\psi(x) - G(x)\}$.  
Since $H(x) \goto \infty$ and $G(x) \goto \infty$, 
there exists $M > 0$ with the properties that
\[
\sup_{|x| > M} \{\psi(x) - H(x)\} \leq -|\Delta| - 1,
\]
the supremum of $\psi - G$ on $\R$ is attained on the interval $[-M,M]$,
and the supremum of $-G$ on $\R$ is attained on the interval $[-M,M]$.
In combination with item 1, we see that for all $n \in \N$ satisfying $R n^\gamma > M$
\be
\label{eqn:uniform}
\sup_{M < |x| < Rn^\gamma} \{\nv\psi(x) - nG_{\bn,\kn}(x/n^\gamma)\} \leq -\nv(|\Delta| + 1).
\ee
\item Let $M$ be the number selected in item 2. 
Then for all $x \in \R$ satisfying $|x| \leq M$, $G_n(x) = n^{1+v}G_{\bn,\kn}(x/n^\gamma)$ 
converges uniformly to $G(x)$ as $n \goto \infty$.  
\end{enumerate}

Since $nG_{\beta_n, K_n}(x/n^{\gamma}) = \nv G_n(x)$, 
item 3 implies that for any $\delta > 0$ and all sufficiently large $n$
\beas
\label{eqn:gnversusg}
\lefteqn{
\exp(-\nv \delta) \int_{\{|x| \leq M\}} \exp[\nv (\psi(x) - G(x))] \, dx }
\\ \nonumber && \leq \int_{\{|x| \leq M\}} \exp[\nv \psi(x) - nG_{\beta_n, K_n}(x/n^{\gamma})] \, dx 
\\ \nonumber && \leq \exp(\nv \delta) \int_{\{|x| \leq M\}} \exp[\nv (\psi(x) - G(x))] \, dx.
\eeas
In addition, item 2 implies that 
\[
\int_{\{M < |x| < R n^\gamma\}} \exp[\nv \psi(x) - nG_{\beta_n, K_n}(x/n^{\gamma})] \, dx 
\leq 2 R n^\gamma \exp[-\nv(|\Delta| + 1)].
\]
Since $\psi$ is bounded, the last two displays show that there exist $a_5 > 0$ and $a_6 \in \R$
such that for all sufficiently large $n$
\[
\int_{\{|x| < Rn^\gamma\}} \exp[-n \gn(x/n^\gamma)] \, dx \leq a_5 \exp(\nv a_6).
\]
Since $-v \in (0,1)$, we conclude from part (d) of Lemma \ref{lem:awayfrom0} 
the existence of $a_7 > 0$ such that for all sufficiently large $n$
\[
\int_{\{|x| \geq Rn^\gamma\}} \exp[-n \gn(x/n^\gamma)] \, dx \leq 2 a_5 \exp(-n a_7).
\]

We now put these three estimates together.  
For all sufficiently large $n$ we have
\beas
\label{eqn:thisisitformdpb}
\lefteqn{\exp(-\nv \delta) \int_{\{|x| \leq M\}} \exp[\nv (\psi(x) - G(x))] \, dx }\\
\nonumber && \leq \int_{\R} \exp[\nv \psi(x) - nG_{\beta_n, K_n}(x/n^{\gamma})] \, dx \\
\nonumber && \leq
\exp(\nv \delta) \int_{\{|x| \leq M\}} \exp[\nv (\psi(x) - G(x))] \, dx + \delta_n,
\eeas
where
\[
\delta_n \leq 2R n^\gamma \exp[-\nv(|\Delta| + 1)] + 
2 a_5 \exp(-n a_7 + \nv \|\psi\|_\infty).
\]
Since $-v < 1$ and since by item 2
\beas
\lefteqn{
\lim_{n \goto \infty} \frac{1}{\nv} \log \int_{\{|x| \leq M\}} \exp[\nv (\psi(x) - G(x))] \, dx } \\
&& = \sup_{|x| \leq M} \{\psi(x) - G(x)\} = \sup_{x \in \R} \{\psi(x) - G(x)\},
\eeas
we have
\beas
\lefteqn{
\sup_{x \in \R} \{\psi(x) - G(x)\} - \delta } 
\\ && \leq
\liminf_{n \goto \infty} \frac{1}{\nv} \log \int_{\R} \exp[\nv \psi(x) - nG_{\beta_n, K_n}(x/n^{\gamma})] \, dx
\\ && \leq 
\limsup_{n \goto \infty} \frac{1}{\nv} \log \int_{\R} \exp[\nv \psi(x) - nG_{\beta_n, K_n}(x/n^{\gamma})] \, dx
\\ && \leq \sup_{x \in \R} \{\psi(x) - G(x)\} + \delta,
\eeas
and because $\delta > 0$ is arbitrary, it follows that
\[
\lim_{n \goto \infty} \frac{1}{\nv} \log \int_{\R} \exp[\nv \psi(x) - nG_{\beta_n, K_n}(x/n^{\gamma})] \, dx
= \sup_{x \in \R} \{\psi(x) - G(x)\}.
\]
Combining this limit with the same limit for $\psi = 0$, we conclude from 
(\ref{eqn:needthis}) that
\beas
\lefteqn{
\lim_{n \goto \infty} \frac{1}{\nv}
\log \int_{\Lambda^n \times \Omega} \exp\!\left[\nv \, \psi\!\left( 
\frac{S_n}{n^{1-\gamma}} + \frac{W_n}{n^{1/2-\gamma}}\right)\right] d(P_{n,\beta_n,K_n} \times Q)} \\
&& = \sup_{x \in \R} \{\psi(x) - G(x)\} + \inf_{y \in \R} G(y) 
= \sup_{x \in \R} \{\psi(x) - \Gamma(x)\}.
\eeas
This completes the proof that with respect to $P_{n,\beta_n,K_n} \times Q$,
$S_n/n^{1-\gamma} + W_n/n^{1/2-\gamma}$ satisfies the Laplace principle with 
exponential speed $\nv$ and rate function $\Gamma$.  Since $W_n/n^{1/2-\gamma}$ is superexponentially
small, we obtain the desired Laplace
principle for $S_n/n^{1-\gamma}$ with respect to $P_{n,\bn,\kn}$.  
The proof of the theorem is complete.  \ \ink

\skp

We next formulate the MDP for $S_n/n^{1-\gamma}$ when $(\bn,\kn)$ is an arbitrary positive
sequence that converges to $(\beta,K) \in A$; thus $\beta$ and $K$ 
satisfy $0 < \beta \leq \beta_c$ and $0 < K < K_c(\beta)$.  
Because in this case the normal random variable
$W_n$ contributes to the limit, we are not able to prove the MDP as we proved Theorem \ref{thm:mdpb}.
Instead we use the G\"{a}rtner-Ellis Theorem.   The following theorem is also valid for
$\beta > \beta_c$ and $0 < K < K_c(\beta)$, and the proof is essentially the same.
The key observation is that for $\beta > \bc$, we have $K(\beta) = (e^\beta + 2)/(4\beta)
> K_c(\beta)$ \cite[Thm.\ 3.8]{EllOttTou}.  Hence if $K < K_c(\beta)$, then
also $K < K(\beta)$ and thus $G_{\beta,K}^{(2)}(0)$ in (\ref{eqn:lastoneihope}) is positive.

\begin{thm}
\label{thm:mdpa}
Let $(\beta_n,K_n)$ be an arbitrary positive sequence that converges
to $(\beta,K) \in A$.  Let $\gamma$ 
be any number in $(0,{1}/{2})$.  Then with respect to
$P_{n,\beta_n,K_n}$, $S_n/n^{1-\gamma}$ satisfies the MDP with exponential speed $n^{1-2\gamma}$ 
and rate function $\beta[K(\beta) - K]x^2$.  
Thus the limit is independent of the particular sequence $(\bn,\kn)$ that is chosen.
\end{thm}

\noi
{\bf Proof.}  For $n \in \N$ and $t \in \R$ 
we use the monotone convergence theorem to replace 
$f$ in Lemma \ref{lem:G} by $\exp(n^{1-2\gamma}tx)$.  We then use
the Taylor expansion in part (a) of Theorem \ref{thm:taylor} and the fact that $G^{(2)}_{\bn,\kn}(0)$ given
in (\ref{eqn:G2}) converges to 
\be
\label{eqn:lastoneihope}
G^{(2)}_{\beta,K}(0) = \frac{2\beta K [K(\beta) - K]}{K(\beta)},
\ee
which is positive since $0 < K < K_c(\beta) = K(\beta)$.
As in the proof of part (a) of Theorem \ref{thm:mdpb}, 
there exists $M > 0$ such that the supremum of 
$tx - G^{(2)}_{\beta,K}(0) x^2/2$ is attained on the interval $[-M,M]$
and the following calculation is valid:
\beas
\lefteqn{
\lim_{n \goto \infty} \frac{1}{n^{1-2\gamma}}
\log \int_{\Lambda^n \times \Omega} \exp\!\left[n^{1-2\gamma} t \!
\left(\frac{S_n}{n^{1-\gamma}} + \frac{W_n}{n^{1/2-\gamma}}\right)\right] d(P_{n,\beta_n,K_n} \times Q)} \\
&& = \lim_{n \goto \infty} \frac{1}{n^{1-2\gamma}} \log 
\int_{\R} \exp\!\left[n^{1-2\gamma} tx - nG_{\beta_n, K_n}(x/n^{\gamma})\right] dx \\
&& \hspace{.25in} - \lim_{n \goto \infty} \frac{1}{n^{1-2\gamma}} \log 
\int_{\R} \exp\!\left[-nG_{\beta_n,K_n}(x/n^{\gamma})\right] dx \\
&& = \lim_{n \goto \infty} \frac{1}{n^{1-2\gamma}} \log 
\int_{\{|x| \leq M\}} 
\exp\!\left[n^{1-2\gamma}\!\left(tx - G^{(2)}_{\beta,K}(0) x^2/2\right)\right] dx \\
&& \hspace{.25in} - \lim_{n \goto \infty} \frac{1}{n^{1-2\gamma}} \log 
\int_{\{|x| \leq M\}} \exp\!\left[-G^{(2)}_{\beta,K}(0) x^2/2\right] dx \\
&& = \sup_{\{|x| \leq M\}} \!\left\{tx - G^{(2)}_{\beta,K}(0) x^2/2\right\} + 
\inf_{\{|x| \leq M\}} \!\left\{G^{(2)}_{\beta,K}(0) x^2/2\right\}
\\ && = \frac{t^2}{2G^{(2)}_{\beta,K}(0)}.
\eeas
Since $W_n$ is an $N(0,(2\beta_n K_n)^{-1})$ random variable and is independent of $S_n$, 
\beas
\lefteqn{
\lim_{n \goto \infty} \frac{1}{n^{1-2\gamma}}
\log \int_{\Lambda^n} \exp\!\left[n^{1-2\gamma} t \cdot
\frac{S_n}{n^{1-\gamma}}\right] dP_{n,\beta_n,K_n}} \\
&& = \lim_{n \goto \infty} \frac{1}{n^{1-2\gamma}}
\log \int_{\Lambda^n \times \Omega} \exp\!\left[n^{1-2\gamma} t
\left(\frac{S_n}{n^{1-\gamma}} + \frac{W_n}{n^{1/2-\gamma}}\right)\right] d(P_{n,\beta_n,K_n} \times Q) \\
&& \hspace{.25in} - \lim_{n \goto \infty} \frac{1}{n^{1-2\gamma}}
\log \int_{\Omega} \exp\!\left[n^{1/2-\gamma}t W_n\right] dQ \\
&& =  \frac{t^2}{2G^{(2)}_{\beta,K}(0)} - \frac{t^2}{4 \beta K} 
= \frac{t^2}{2} \cdot \frac{1}{2\beta[K(\beta) - K]}.
\eeas
The G\"{a}rtner-Ellis Theorem \cite{Ellis2} now implies that $S_n/n^{1-\gamma}$ 
satisfies the MDP with exponential speed $n^{1-2\gamma}$ and rate function
\[
I(x) = \sup_{t \in \R} \!\left\{tx - \frac{t^2}{2} \cdot \frac{1}{2\beta[K(\beta) - K]}\right\}
= \beta[K(\beta) - K] x^2.
\]
This completes the proof.  \ink

\skp
In the context of the proof of the preceding theorem, it is worthwhile pointing out
that the G\"{a}rtner-Ellis Theorem cannot be used to prove all the other MDPs 
for $S_n/n^{1-\gamma}$ in this section.  
For example, consider the MDPs in Theorem \ref{thm:mdpb}.
For any $t \in \R$ one calculates
\beas
g(t) & = &
\lim_{n \goto \infty} \frac{1}{\nv}
\log \int_{\Lambda^n \times \Omega} \exp\!\left[\nv \, t \! \left( 
\frac{S_n}{n^{1-\gamma}} + 
\frac{W_n}{n^{1/2-\gamma}}\right)\right] d(P_{n,\beta_n,K_n} \times Q)\\
& = & \sup_{x \in \R} \{tx - G(x)\} + \inf_{y \in \R} G(y) 
= \sup_{x \in \R} \{tx - [G(x) - \bar{G}]\},
\eeas
where $\bar{G} = \inf_{y \in \R}G(y)$. 
Thus $g$ equals the Legendre-Fenchel transform of $G - \bar{G}$.  If $G -\bar{G}$
is strictly convex on $\R$, as it is in cases 1, 2, and 3 in Theorem \ref{thm:mdpb}, then
$g$ is differentiable on $\R$ \cite[p.\ 253]{Roc}.  Hence by the G\"{a}rtner-Ellis
Theorem, $S_n/n^{1-\gamma}$ satisfies the MDP with exponential speed $\nv$ and rate function
given by the Legendre-Fenchel transform of $g$, which is $G -\bar{G}$.  In cases
1, 2, and 3 in Theorem \ref{thm:mdpb}, $\bar{G}$ equals 0, and we recover the 
form of the rate function in column 4 of Table 8.1. 
However, the situation is different in the MDP in case 4, in which
$G(x) = k \beta x^2 + c_4 x^4$ with $k < 0$.
Here $\bar{G} < 0$, $G$ is not convex on all of $\R$, and $g$ is not differentiable
on $\R$.  As a result, the G\"{a}rtner-Ellis Theorem cannot be applied
to obtain the lower large deviation bound for all open sets and thus to obtain the MDP.
In addition, the Legendre-Fenchel transform of $g$ 
equals 0 on a symmetric interval containing the origin, and thus it does not 
coincide with $G - \bar{G}$ on this interval. 
A similar situation holds in Theorem \ref{thm:mdpc}, in which we derive 13 MDPs for 
suitable sequences $(\bn,\kn) \goto (\bc,\kcbc)$.  In cases 1--4, 6, 8, and 10 in that theorem,
the coefficients in the polynomial $G$ are all positive, and so $G$ is strictly convex
and $\bar{G} = 0$.  Hence the corresponding MDPs can be derived via
the G\"{a}rtner-Ellis Theorem.  However, in all the other cases except for
case 12 with $k$ sufficiently large, the polynomial
$G$ is not convex on all of $\R$; as in case 4 in Theorem \ref{thm:mdpb},
the G\"{a}rtner-Ellis Theorem cannot be applied to obtain the MDP.

We now consider the final class of MDPs in this section.
This class arises when $(\beta_n,K_n)$ converges to $(\beta_c,K_c(\beta_c))$
along the same sequences considered in Theorem \ref{thm:betakinc},
where we proved scaling limits for $S_n/n^{1-\gamma}$ for $\gamma \in (0,1/2)$. 
Given $\alpha > 0$, $\theta > 0$, $b \not = 0$, 
and $k \not = 0$, these sequences are defined by 
\be
\label{eqn:betanknmdpc}
\bn = \log(4 - {b}/{n^\alpha}) = 
\log(e^{\bc} - {b}/{n^\alpha}) \ \mbox{ and } \ 
\kn = K(\bn) - {k}/{n^\theta}.
\ee
For these sequences
the parameter that plays the role of $v$ in Theorem \ref{thm:mdpb} is
\[
w = \min\{2\gamma + \theta -1, 4\gamma + \alpha - 1, 6\gamma -1\}.
\]
The 13 forms of the scaling limits of $S_n/n^{1-\gamma}$ 
 are proved in Theorem \ref{thm:betakinc}
under the assumption that $w = 0$.  
We now assume that $w < 0$.  Using the same Taylor expansion
that was used to deduce these scaling limits
[Thm.\ \ref{thm:taylor}(c)], one deduces the 13 forms of the Laplace principles
for $S_n/n^{1-\gamma}$.  These Laplace principles and the equivalent MDPs
are stated in the next theorem along with the choices of $\gamma$,
$\alpha$, $b$, $\theta$, and $k$ leading to the 13 forms of the rate function.
The only requirement on $b$ and $k$ is that $G(x) \goto \infty$ as $|x| \goto \infty$.
This requirement forces $b > 0$ in case 2 and $k > 0$ in case 3. 
The proof of the MDPs in the next theorem is omitted because it follows the same pattern of
proof of Theorem \ref{thm:mdpb}.

As in Theorem \ref{thm:betakinc}, there are further possibilities
concerning the sign of $b$ and $k$.  In all the 
cases in which no $x^4$ term appears in the scaling limit (cases 1, 3, 6, 7),
we can choose either $b$ to be any real number.  
Similarly, in all the cases in which no $x^2$ term appears in the scaling limit (cases 1, 2, 4, 5), we can choose either $k$ to be any real number.  Although the choice
of $b$ or $k$ affects the definition of the sequence $(\bn,\kn)$, it
does not affect the form of the rate function.

\begin{thm}
\label{thm:mdpc}
Given $\alpha > 0$, $\theta > 0$, $b \not = 0$, 
and $k \not =0$,
consider the sequence $(\bn,\kn)$ defined in {\em (\ref{eqn:betanknmdpc})}. 
\iffalse
\[
\beta_n = \log\!\left(4 - \frac{b}{n^\alpha}\right) \ \mbox{ and } \ K_n = K(\beta) - \frac{k}{n^\theta}
\]
\fi
Then $(\bn,\kn) \goto (\beta_c,K_c(\beta_c))$.  Given $\gamma \in (0,1)$, we also define 
\[
G(x) = \delta (w, 2\gamma+\theta-1) k \beta_c x^2 +   \delta (w, 4\gamma+\alpha-1) 
b \bar{c}_4 x^4 + \delta (w, 6\gamma-1)  c_6 x^6,
\]
where $\bar{c}_4 = 3/16$ and $c_6 = 9/40$.
The following conclusions hold.

{\em (a)}  Assume that $w = \min\{2\gamma + \theta -1, 4\gamma + \alpha - 1, 6\gamma -1\}$ 
satisfies $w < 0$.  Then with respect to $P_{n,\bn,\kn}$,
$S_n/n^{1-\gamma}$ satisfies the Laplace principle, and thus the MDP, with exponential speed $\nw$
and rate function $\Gamma(x) = G(x) - \inf_{y \in \R} G(y)$.  

{\em (b)}  We have $w < 0$ if and only if one of the 
{\em 13} cases enumerated in Table {\em 8.2} holds.
Each of the {\em 13} cases corresponds to a set of values of $\gamma$, $\alpha$, and $\theta$; 
a choice of signs of $b$ and $k$; the influence
of one or more sets $C$, $B$, $A$;  
and a particular exponential speed and a particular form of 
the rate function in part {\em (a)}.  
The function $G$ appearing in the definition of the 
rate function is shown in column {\em 5} in Table {\em 8.2};  
when $\inf_{y \in \R} G(y) \not = 0$, this additive constant in the definition of the 
rate function is not shown.  The form of the rate function
is not affected by the choice of $b \in \R$
in cases {\em 1}, {\em 3}, {\em 6}, and {\em 7} and 
by the choice of $k \in \R$ in cases {\em 1}, {\em 2}, {\em 4}, 
and {\em 5}.

% \vspace{.1in}  
\begin{center}
\begin{tabular}{||l|l|l|l|l||} \hline \hline
{\em {\bf case}} & {\em {\bf values of}} \boldmath $\gamma$ \unboldmath
& {\em {\bf values of}} \boldmath $\alpha$ \unboldmath & {\em {\bf exp'l}}
& {\em {\bf function}} \boldmath $G$ \unboldmath {\em {\bf in}}
\\ \cline{1-1} \cline{3-3}
{\em {\bf influence}} & & {\em {\bf values of}} \boldmath $\theta$ \unboldmath & {\em {\bf speed}} &
{\em {\bf rate function}} \boldmath $\Gamma$ \unboldmath  
 \\ \hline \hline
{\em 1} & $\gamma \in (0,\frac{1}{6})$ & $\alpha > 2\gamma$  & $n^{1-6\gamma}$&
% $S_n/n^{1-1/6} \Longrightarrow 
$c_6 x^6$\\ \cline{1-1} \cline{3-3}  
$C$ & & $\theta > 4 \gamma$ &   &  $c_6 > 0$, 
$b \in \R$, $k \in \R$  \\ \hline \hline
{\em 2} & $\gamma \in (0,\frac{1}{4}) $  & $\alpha \in (0,\min\{2\gamma,1-4\gamma\})$ & $n^{1-4\gamma-\alpha}$ &
% $S_n/n^{1-\gamma} \Longrightarrow 
$b \bar{c}_4 x^4$ \\ \cline{1-1} \cline{3-3}
 $B$ & & $\theta > 2\gamma + \alpha $ &   &  $b > 0$, $\bar{c}_4 > 0$,
$k \in \R$ \\ \hline \hline
{\em 3} & $\gamma \in (0,\frac{1}{2})$ & $\theta \in (0,\min\{4\gamma,1-2\gamma\})$  & $n^{1-2\gamma-\theta}$ &
% $S_n/n^{1-\gamma} \Longrightarrow \
$k \bc x^2$ \\ \cline{1-1} \cline{3-3}
 $A$ & & $\alpha > \max(\theta - 2\gamma,0)$ &   & $k > 0$, $b \in \R$  \\ \hline \hline
{\em 4--5} & $\gamma \in (0,\frac{1}{6})$ & $\alpha = 2\gamma$  & $n^{1-6\gamma}$ &
% $S_n/n^{1-1/6} \Longrightarrow 
$b \bar{c}_4 x^4 + c_6 x^6$ \\ \cline{1-1} \cline{3-3}
 $B+C$ & & $\theta > 4\gamma$ &   &  $b \not = 0$, $k \in \R$ 
\\ \hline \hline
{\em 6--7} & $\gamma \in (0,\frac{1}{6})$  & $\alpha > 2\gamma$ & $n^{1-6\gamma}$ &
% $S_n/n^{1-1/6} \Longrightarrow 
$k \bc x^2 + c_6 x^6$ \\ \cline{1-1} \cline{3-3}
$A+C$  & & $\theta = 4\gamma$ &   &  
$k \not = 0$, $b \in \R$ \\ \hline \hline
{\em 8--9} & $\gamma = (0,\frac{1}{4})$  & $\alpha \in (0,\min\{2\gamma,1-4\gamma\})$ & $n^{1-4\gamma-\alpha}$ &
% $S_n/n^{1-1/6} \Longrightarrow 
$k \bc x^2 + b \bar{c}_4 x^4$ \\ \cline{1-1} \cline{3-3}
$A+B$  & & $\theta = 2\gamma + \alpha$ 
%\in (\frac{1}{2},\frac{2}{3})$ 
&   &  $k \not = 0$, $b > 0$ \\ \hline \hline
{\em 10--13} & $\gamma \in (0,\frac{1}{6})$  & $\alpha = 2\gamma$ & $n^{1-6\gamma}$  &
% $S_n/n^{1-1/6} \Longrightarrow 
$k \bc x^2 + b \bar{c}_4 x^4 + c_6 x^6$ \\ \cline{1-1} \cline{3-3}
$A+B+C$ & & $\theta = 4\gamma$ &   &  $k \not = 0$, $b \not = 0$ \\ \hline \hline
\end{tabular}

\vspace{.05in}
{\em Table 8.2}: {\small {\em Values of } $\gamma$, $\alpha$, {\em and} $\theta$,
{\em exponential speeds, and rate functions in part (b) of Theorem \ref{thm:mdpc}}}  
\end{center}
\end{thm}

\iffalse
*Influence C alone
$\gamma \in (0, 1/6), \theta > 4\gamma, \alpha > 2\gamma$
MY COMMENT. I GET THE SAME ANSWER.

*Influence B alone
$\gamma \in (0, 1/4), 0 < \alpha < \min(2\gamma, 1-4\gamma), \theta >
\alpha + 2\gamma$
MY COMMENT. I GET THE SAME ANSWER.

*Influence A alone
$\gamma \in (0, 1/2), 0 < \theta < \min(4\gamma, 1-2\gamma), \alpha >
\theta - 2\gamma$
MY COMMENT.  HOW DOES ONE KNOW THAT $\theta-2\gamma >0$.  THUS ONE
MUST REPLACE $\alpha > \theta-2\gamma$ BY  
$\alpha > \max\{\theta-2\gamma, 0\}$.  DO YOU AGREE?
Marius: On region A you are correct, we should
include the zero.  I did not include it there.  

*Influence A+B alone
$\gamma \in (0, 1/4), 0 < \theta < \min(4\gamma, 1-2\gamma), \alpha =
\theta - 2\gamma$
MY COMMENT.  I DID THIS A DIFFERENT WAY, OBTAINING THE FOLLOWING: 
$\gamma \in (0, 1/4)$, $0 < \alpha < \min\{2\gamma, 1-4\gamma\}$,
$\theta = 2\gamma + \alpha$.  COULD YOU PLEASE VERIFY THIS?  WHICH ONE
IS BETTER, MINE OR YOURS?  Marius: On region A+B both our
answers are correct since they are equivalent from $\theta = 2\gamma +
\alpha$.

*Influence A+C alone
$\gamma \in (0, 1/6), \theta = 4\gamma, \alpha > 2\gamma$
MY COMMENT. I GET THE SAME ANSWER.

*Influence B+C alone
$\gamma \in (0, 1/6), \alpha = 2\gamma, \theta > 4\gamma$
MY COMMENT. I GET THE SAME ANSWER.

*Influence A+B+C alone
$\gamma \in (0, 1/6), \alpha = 2\gamma, \theta = 4\gamma.$
MY COMMENT. I GET THE SAME ANSWER.
\fi

\skp
As discussed in section \ref{section:overview}, the MDPs listed in Table 8.2 yield a new
class of distribution limits for $S_n/n^{1-\gamma}$ in those cases in which the set of 
global minimum points of $G$ contains nonzero points.  These are the cases
in which the coefficients of $G$ are not all positive: cases 5 ($b < 0$), 7 ($k < 0$),
9 ($k < 0$), 11 ($k < 0$, $b > 0$), 12 ($k > 0$, $b < 0$), and 13 ($k < 0$, $b < 0$).
In all these cases except for case 12, we obtain the limit (\ref{eqn:refinemdp}).
Case 12 exhibits the most complicated behavior, giving rise to the limit (\ref{eqn:refine3mdp})
for the critical value $k = 5 b^2/[2^7 \bc]$.  These limits and the underlying
physical phenomena are now being investigated for a class of non-mean-field models, including the 
Blume-Emery-Griffiths model \cite{EllMac}.  

This completes our study of limit theorems for the BEG model in 
the neighborhood of the tricritical point $(\bc,\kcbc) \in C$, 
in the neighborhood of second-order points $(\beta,K_c(\beta))
\in B$, and in the neighborhood of single-phase points $(\beta,K) \in A$.  It is an
unexpectedly rich and fruitful area of research, one that we hope
will inspire similar investigations for other statistical mechanical models.

\end{document}